# Color superconductivity in the Nambu Jona-Lasinio model complemented by a Polyakov loop

Eric Blanquier

12 Boulevard des capucines, F-11 800 Trèbes, France

E-mail: ericblanquier@hotmail.com

**Abstract**
The color superconductivity was studied with an adaptation of the Nambu and Jona-Lasinio model (NJL). This one was coupled to a Polyakov loop, to form the PNJL model. A $\mu$-dependent Polyakov loop potential was considered. An objective of the presented works was to describe the analytical calculations required to establish the equations to be solved, in all of the treated cases. They concerned the normal quark phase, the 2-flavor color-superconducting and the color-flavor-locked phases, in an $SU(3)_f \times SU(3)_c$ description. The calculations were performed according to the temperature $T$, the chemical potentials $\mu_f$ or according to the densities $\rho_f$, with or without the isospin symmetry. Among the obtained results, it was found that at low $T$, the restoration of the chiral symmetry, the deconfinement transition and the transition between the normal quark and 2-flavor color-superconducting phases occur via first order phase transitions at the same chemical potential. Moreover, an sSC phase was identified in the $\rho_q, \rho_s$ plane.

## 1. Introduction

The quark physics appears to be a very rich topic, which motivated various works to try to understand it better. It is possible to quote the recent results found in the framework of the RHIC and LHC programs. It notably concerns the observed behavior of the quark matter at high temperatures, assimilated to the one of a perfect fluid [1, 2]. From a theoretical point of view, the quark physics is also a fascinating topic of investigations. In the study of the phase transition hadronic matter/quark gluon plasma, several approaches have been performed in order to understand how this transition occurs. In this framework, the Quantum Chromodynamics (QCD) plays a central role. Results found with Lattice QCD (LQCD) [3-8] can be mentioned. However, the description of the quark physics is not limited to its behavior at high temperatures. Indeed, if it was imagined a rather simple phase diagram of the quarks matter at low temperatures and high densities, it seems to be well admitted that this vision has to be corrected. Indeed, in these conditions, the quarks are expected to form pairs. They are comparable to the Cooper pairs of electrons in the framework of the superconductivity phenomenon [9-11] due to electrons/phonons interactions. With the quarks, it is customary to speak about color superconductivity, because the quark pairs are caused by the strong force [12-20]. Since it exits several flavors/colors of quarks, several pairs are possible. Consequently, the color superconductivity is not associated with one unique phase, but with several ones. In an $SU(3)_{flavor}$ description, one mainly mentions the 2-flavor color-superconducting (2SC) phase, in which $u$ and $d$ quarks form pairs, and the color-flavor-locked (CFL) phase that involves $ud$, $us$ and $ds$ pairs. Other phases were also imagined: recent works related to this topic [19-24] show rather complex phase diagrams to represent them. In Nature, some of these ones are expected to constitute the core of neutron stars or, more generally, of compact stars. Future accelerator programs plan to explore the high densities regions of the phase diagram. Color superconductivity may be probably accessible to them [25]. To describe the color superconductivity theoretically, the QCD should be the best tool to be considered. However, at high densities, some difficulties appear in a QCD description involving the three quark colors, caused by the fermion sign problem [26]. Even if some approaches have been developed to take this problem into account [5], it constitutes a serious limitation in the study of the dense matter with the QCD.

The Nambu and Jona-Lasinio (NJL) model [27, 28] constitutes a reliable alternative. The main idea of this approach consists in gathering the various interactions between the quarks/antiquarks in punctual ones. It leads to consider massive gluons, described by constant terms. So, the gluon degrees of freedom are frozen in this description [29]. This model was progressively improved [29-42], in order to obtain an interesting approach to describe the quark physics at low energies. With the use of Matsubara's formalism [43], the NJL approach is fully able to work at finite temperature. The model is also usable at finite chemical potentials or finite densities. As a consequence, the NJL model appears as a serious candidate to study the color superconductivity. In fact, even if other models may also be considered, like the Dyson-Schwinger approach [44, 45] or the instanton model [46-49], the NJL model and its various versions, e.g. [50], already proved its reliability to describe the color superconductivity, e.g. [16, 19, 51, 52]. However, the major feature of the NJL approach, i.e. the absence of gluons as dynamical degrees of freedom, leads to its major limitation, i.e. the absence of color quark confinement. To try to correct this aspect, it was recently proposed to couple the quarks to a Polyakov loop [53], which is used in LQCD. It allows including an effective Polyakov loop potential associated with the gluons, and a coupling between the quarks/antiquarks and a static gluons field. It forms a version of the model known as the Polyakov Nambu Jona-Lasinio (PNJL) model [54-63]. Among the advantages of this approach, it can be mentioned, e.g., the suppression of colored states in the "confined phase" [64]. Furthermore, it was frequently reported that the PNJL description is more efficient than the pure NJL one to reproduce the results found with LQCD [65-68]. The inclusion of the loop has another consequence, as visible in [61]: it leads to a shifting of the quark masses towards higher temperatures. In previous publications [69-72], I studied again the differences between the NJL and PNJL at several levels. It concerned the modeling of particles (quarks/antiquarks) and composite ones (mesons, diquarks and baryons). This shifting of the masses was confirmed, and it also concerned the cross-sections involving these quoted particles. In [71], I investigated how the shifting can influence and boost the hadronization of a hot quark/antiquark plasma. More recently, improvements of models that use a Polyakov loop were proposed, via the inclusion of a quark back-reaction to the gluonic sector. One objective is notably to adjust the results to the recent LQCD data [73]. It can be done by a rescale in temperature in the Polyakov loop potential [74]. Other modifications consist in including an $N_f$ correction and a $\mu$-dependence in this potential [75-81].

Some adaptations must be performed to include the color superconductivity in the (P)NJL models. It was shown that such a work can be performed in the Nambu-Gorkov formalism [15], in which the quark-quark pairing is described by energy gaps $\Delta$. These gaps are used to identify and to characterize the various color superconducting phases. The (P)NJL works devoted to the color superconductivity can be divided into two groups [19]. In the first group, the calculations are performed as in other (P)NJL studies, i.e. according to the chemical potentials $\mu_f$ and/or according to the temperature $T$. It allows building phase diagrams in the $T, \mu_f$ plane or in the $\mu_f, \mu_{f'}$ ones [21]. The finality of the second group is to take into account the conditions in the interior of compact stars. It leads to consider the neutrality of the quark matter, according to the electric and color charges. In this framework, one also considers the $\beta$ equilibrium, associated with the quark $\beta^\pm$ disintegrations caused by the weak interaction. These constraints lead to establish relations between the chemical potentials of the quarks and the one of the electrons [15]. For both groups, since the theoretical treatment of the color superconductivity involves more difficult calculations than in pure (P)NJL ones, several approximations are proposed. In the literature, the most encountered ones concern the quark masses $m_f$. In some approaches, massless quarks [82, 83] are considered. In other cases, $m_u = m_d \equiv m_q$ is very frequently used. When electric or color neutrality conditions are not applied, this simplification can also involve the chemical potentials of these light quarks, i.e. $\mu_u = \mu_d \equiv \mu_q$. It corresponds to the isospin symmetry, which is often encountered in (P)NJL approaches. In idealized calculations, these simplifications are sometimes extended to the strange quarks. Among the other approximations, one can mention [84], in which the terms that mix quarks and antiquarks are neglected. Other works avoid performing some analytical developments thanks to numerical calculations. It notably concerns the estimation of the eigenvalues of the matrix that describes the involved physics [21-24].

Nevertheless, it should be relevant to investigate in details the mathematical developments that are required to model the color superconductivity in the (P)NJL descriptions. In particular, the reference [85] proposes a method to calculate the determinant of block matrices, which constitutes an important stage of the analytical calculations. It appears important to estimate if this method can help us to avoid some of the approximations mentioned upstream. Also, because of the various phases that exist in the framework of the color superconductivity, it seems useful to underline the relations and equations that are valid in all of the treated cases. Moreover, except in some papers [25, 52, 86], the calculations are performed according to the temperature and the chemical potentials, and rarely according to the densities $\rho_f$. The use of the baryonic $\rho_B$ or the strange $\rho_s$ densities should certainly facilitate comparisons of the obtained data with nuclear or astrophysical ones. Also, the analysis of the results obtained with $\mu_f$ and $\rho_f$ should give some interesting information. This remark particularly concerns the phase transitions. It includes the phase transition associated with the restoration of the chiral symmetry, but also the ones that occur between the different color superconducting phases. In the case of a first order one, how can we model the metastable and unstable states that are expected to be present near such a transition? Moreover, even if $m_u \approx m_d$ in most of the studied configurations, it should be relevant to study asymmetric matter [19]. It implies to see how to perform the calculations without the simplification $m_u = m_d \equiv m_q$, and with $\mu_u \neq \mu_d$. In studies that do not apply neutrality conditions, it leads to work beyond the isospin symmetry and to treat $\mu_u$ and $\mu_d$ as independent variables. Furthermore, an important aspect of this work could be to show the influence of the color superconductivity on various observables. The effect of the Polyakov loop should also be analyzed. In this framework, the differences in the results between the standard Polyakov loop potential and the $\mu$-dependent one should be underlined, notably in the superconducting regime.

In order to investigate the points formulated upstream, I propose to describe in section 2 the model that is used in this study, and its various versions. More precisely, it concerns the NJL model, the PNJL one, and the adaptations to be done to perform the description of the color superconductivity. The vocation of this section is to gather the general equations to be solved to work at finite temperature and finite densities/chemical potentials. The specific relations and results related to each case are then proposed in the next sections. As in [21], electric or color neutrality conditions and $\beta$ equilibrium are not included in this paper. In fact, I focus on descriptions in which the temperature, the densities or the chemical potentials are used as parameters, in order to allow a direct comparison with results that include or not the color superconductivity. In section 3, I recall the developments and results found with the (P)NJL models that do not consider this phenomenon. An objective is to detail and to explain the link between the results obtained at finite chemical potentials and at finite densities. At this occasion, the *problem of matter stability* is evoked, and the stability of the states observable in the $T, \rho_B$ plane is discussed. In sections 4 and 5, the 2SC phase is treated. In these two sections, the influence of the Polyakov loop on the results is mentioned. In the section 4, the calculations are performed in the framework of the isospin symmetry. The effect of the 2SC phase on various physical quantities is underlined: it concerns the masses of the treated particles, etc. In this paper, I focus on the quarks/antiquarks. In the section 5, the isospin symmetry is not used. A motivation is to see the theoretical limits of the 2SC phase according to the temperature, the two chemical potentials $\mu_{u,d}$ and the two densities $\rho_{u,d}$. Then, in section 6, the CFL phase is treated. Some results in the $\mu_q, \mu_s$ plane are produced in order to allow a comparison with the ones already published. Other data focus on a description of the CFL phase according to the densities. These works are done by a study of the two gaps $\Delta_{ud}$ and $\Delta_{qs}$ associated with this phase. Phase diagrams according to the densities are also proposed, for several temperatures.

## 2. Global description of the model

*2.1 Gap equation and grand potential*

In the mean field approximation, the masses $m_f$ of the dressed quarks are found by the gap equations written in the first line of (1), where $m_{0f}$ are the masses of the naked quarks, and $\langle\langle\bar{\psi}_f\psi_f\rangle\rangle$ is the chiral condensate of the flavor $f$ quarks. The chiral condensates are used as order parameters to describe the restoration of the chiral symmetry, at high temperatures or high densities/chemical potentials, when they converge towards zero. To work at finite temperature $T$ and finite densities $\rho_f$, the system of equations to be solved to estimate $m_f$ is written in the general form [19, 22, 29-31]

$$\begin{cases} m_f = m_{0f} - 4G\langle\langle\bar{\psi}_f\psi_f\rangle\rangle + 2K\langle\langle\bar{\psi}_j\psi_j\rangle\rangle\langle\langle\bar{\psi}_k\psi_k\rangle\rangle\Big|_{\substack{f=u,d,s \\ f\neq j \text{ and } f\neq k}} \\ \rho_f = \langle\langle\psi_f^+\psi_f\rangle\rangle\Big|_{f=u,d,s} \end{cases}. \quad (1)$$

More precisely, in an $SU(3)_f$ description, the masses $m_f$ of the dressed quarks and their associated chemical potentials $\mu_f$ are the six unknowns of this six-equation system, in which the temperature and the densities are used as parameters [69-72]. The system (1) stays valid in all the configurations treated in this paper. However, the expressions of $\langle\langle\bar{\psi}_f\psi_f\rangle\rangle$ and $\langle\langle\psi_f^+\psi_f\rangle\rangle = \langle\langle\bar{\psi}_f\gamma_0\psi_f\rangle\rangle$ depend on the treated case. Such expressions are obtainable from the grand potential $\Omega$, as done in the subsection 2.4. In this work, its most general expression is written as [19, 87]

$$\begin{aligned} \Omega = &\mathcal{U}(T,\Phi,\bar{\Phi}) + \Omega_M \\ &+ 2G\sum_{f=u,d,s}\langle\langle\bar{\psi}_f\psi_f\rangle\rangle^2 - 4K\langle\langle\bar{\psi}_u\psi_u\rangle\rangle\langle\langle\bar{\psi}_d\psi_d\rangle\rangle\langle\langle\bar{\psi}_s\psi_s\rangle\rangle \\ &- 2G_V\sum_{f=u,d,s}\langle\langle\psi_f^+\psi_f\rangle\rangle^2 \\ &+ G_{DIQ}\left(\langle\langle\psi_u\psi_d\rangle\rangle^2 + \langle\langle\psi_u\psi_s\rangle\rangle^2 + \langle\langle\psi_d\psi_s\rangle\rangle^2\right) \end{aligned}. \quad (2)$$

It includes a vector interaction (term associated with $G_V$) to be able to work at non-null densities. As recalled in [22], since this expression of $\Omega$ uses terms found in the framework of the Hartree approximation, the terms involving both $\langle\langle\bar{\psi}_f\psi_f\rangle\rangle$ and $\langle\langle\psi_f\psi_{f'}\rangle\rangle$ contributions are neglected. Moreover, the 't Hooft instanton-induced interaction [88] is taken into account. It corresponds to the terms associated with the constant $K$ in (1) and (2). As with the other constants that appear in these equations, its value can vary in the literature. In my previous papers [69-71] and in this one, I consider two sets of parameters, named P1 and EB, which gather the used values, Table 1. The P1 parameter set respects the isospin symmetry, i.e. $m_u = m_d \equiv m_q$, whereas $m_u \neq m_d$ for the EB parameter set. In fact, P1 and EB give rather comparable results, especially for the values of the chiral condensates [71]. The relations between the constants $G$, $G_V$ and $G_{DIQ}$ should be fixed by Fierz transformations, leading to $G_V = G/2$ and $G_{DIQ} = 3G/4$ [29, 60, 89]. However, it is not uncommon to treat these constants as more or less independent ones [69, 90]. Moreover, the effective Polyakov loop potential $\mathcal{U}(T,\Phi,\bar{\Phi})$ is related to the inclusion (or not) of this loop. This aspect is described in subsection 2.2.

The $\Omega_M$ term is the thermodynamical potential of a fermion/antifermion gas, which corresponds here to the $u, d, s$ quarks and the associated antiquarks. In subsection 2.3, this term is detailed.

**Table 1.** Parameter sets used for the NJL and PNJL models. The masses $m_{0f}$ of the naked quarks and the cutoff are expressed in MeV. Also, $G$, $G_V$ and $G_{DIQ}$ are in MeV$^{-2}$ and $K$ is in MeV$^{-5}$.

| Parameter set | $m_{0u}$ | $m_{0d}$ | $m_{0s}$ | cutoff $\Lambda$ | $G \Lambda^2$ | $K \Lambda^5$ | $G_V$ | $G_{DIQ}$ |
|---|---|---|---|---|---|---|---|---|
| P1 | 4.75 | 4.75 | 147.0 | 708.0 | 1.922 | 10.00 | 0.310 $G$ | 0.705 $G$ |
| EB | 4.00 | 6.00 | 120.0 | 708.0 | 1.922 | 10.00 | 0.295 $G$ | 0.705 $G$ |

*2.2 Inclusion of the Polyakov loop*

In the NJL model, the potential $\mathcal{U}$ is equal to zero. Otherwise, in the PNJL model, several forms of this potential are proposed in the literature. As in [59-62, 91-94], I propose to consider

$$\frac{\mathcal{U}(T, \Phi, \bar{\Phi})}{T^4} = -\frac{a(T)}{2} \Phi \bar{\Phi} + b(T) \ln\left[1 - 6\Phi\bar{\Phi} + 4\left(\Phi^3 + \bar{\Phi}^3\right) - 3\left(\Phi\bar{\Phi}\right)^2\right], \quad (3)$$

where the $a$, $b$ terms depend on the temperature $T$. They are written as

$$a(T) = a_0 + a_1\left(\frac{T_0}{T}\right) + a_2\left(\frac{T_0}{T}\right)^2 \text{ and } b(T) = b_3\left(\frac{T_0}{T}\right)^3. \quad (4)$$

The coefficients used in (4) are gathered in Table 2. [60].

**Table 2.** Parameters of the PNJL model.

| $a_0$ | $a_1$ | $a_2$ | $b_3$ | $T_0$ |
|---|---|---|---|---|
| 3.51 | −2.47 | 15.2 | −1.75 | 270 MeV |

In the standard PNJL model, the temperature $T_0$ is a constant, Table 2. At the opposite, $T_0$ can become a function of $N_f$ and $\mu_f$ to include the quark back-reaction to the gluonic sector. Following [75], I propose

$$T_0 = T_\tau \exp\left[\frac{-1}{\alpha_0 b(N_f, \mu_f)}\right] \text{ where } b(N_f, \mu_f) = \frac{11N_c - 2N_f}{6\pi} - \frac{16}{\pi}\sum_f\left(\frac{\mu_f}{T_\tau}\right)^2, \quad (5)$$

with $T_\tau = 1770$ MeV and $\alpha_0 = 0.304$. Also, $N_c$ is the number of colors, fixed to three in all this paper ($r$, $g$, $b$) and $N_f$ is the number of (approximately) massless flavors. In the expression of $b(N_f, \mu_f)$, the first term corresponds to the $N_f$ correction, and the second to the $\mu_f$ one. With this modification, the pure gauge value $T_0 \approx 270$ MeV is reached only when $N_f = 0$ and at null chemical potentials. In this document, the name "$\mu$PNJL" [81] will be used to designate the calculations that use (5), whereas the "PNJL" label will be reserved to the ones that consider a constant $T_0$.

In (3), $\Phi$ and $\bar{\Phi}$ are, respectively, the average of the Polyakov field and the average of its complex conjugate. In pure gauge calculations, $\Phi$ is used as an order parameter to describe the phase transition between the color confined regime (of the gluons) at low temperatures, and the deconfined one at high temperatures. The confined regime corresponds to $\Phi = 0$, and the deconfined one to $\Phi \neq 0$. An advantage of the form (3) is $\Phi \rightarrow 1$ at high temperatures with $\Phi < 1$, as expected [59, 60]. As visible in [60, 72], the behavior of the $\Phi$ that minimizes $\mathcal{U}$ describes a first order phase transition when $T = T_0$ in pure gauge calculations [55]. When the quarks are added in the modeling, one keeps the terminology of "confined" and "deconfined" regime but with quotation marks, in order to recall that the inclusion of the Polyakov loop only mimics a mechanism of quark confinement [63]. In addition, the presence of the quarks leads to modify the nature of the phase transition according to $T$, which becomes a crossover [55]. Furthermore, $\Phi$ and $\bar{\Phi}$ are expressed by

$$\Phi = \frac{1}{N_c}\operatorname{Tr}_c(L) \text{ and } \bar{\Phi} = \frac{1}{N_c}\operatorname{Tr}_c(L^\dagger), \tag{6}$$

where $\operatorname{Tr}_c$ is a trace over the colors, and

$$L(\vec{x}) = \mathcal{P} \exp\left(i\int_0^\beta A_4(\vec{x},\tau)\,d\tau\right) \tag{7}$$

is the Polyakov line and $L^\dagger$ is its conjugate one. In this equation, $\mathcal{P}$ is a path ordering operator, $\beta = 1/T$ is the inverse of the temperature. Also, $A_4$ corresponds to the temporal component of the Euclidian gauge field. In the mean field approximation, the Polyakov line and its conjugate one are rewritten as $L = \exp(i\beta A_4)$ and $L^\dagger = \exp(-i\beta A_4)$. As explained in [55], it is possible to work in a basis in which $A_4$ is a diagonal matrix,

$$A_4 = \operatorname{diag}\left(A_{4(11)}, A_{4(22)}, A_{4(33)}\right). \tag{8}$$

In this reference, $\beta A_4 \equiv \operatorname{diag}[\phi, \phi', -(\phi+\phi')]$. In [59], a different writing is proposed, i.e. $\beta A_4 = \phi_3 \lambda_3 + \phi_8 \lambda_8$, where $\phi_3, \phi_8$ are real numbers (there are "angles") and $\lambda_3, \lambda_8$ are Gell-Mann matrices, leading to

$$\beta A_{4(11)} = \phi_3 + \frac{1}{\sqrt{3}}\phi_8, \quad \beta A_{4(22)} = -\phi_3 + \frac{1}{\sqrt{3}}\phi_8 \text{ and } \beta A_{4(33)} = -\frac{2}{\sqrt{3}}\phi_8. \tag{9}$$

Obviously, these two choices are equivalent.

*2.3 Detail on the $\Omega_M$ term*

In the general case, the Fermi-gas term $\Omega_M$ [19] used in (2) is written as

$$\Omega_M = -\frac{T}{2}\int\frac{d^3p}{(2\pi)^3}\sum_n \operatorname{Tr}\left\{\ln\left[\frac{\tilde{S}^{-1}(i\omega_n,\vec{p})}{T}\right]\right\}, \tag{10}$$

in which $\tilde{S}^{-1}(i\omega_n,\vec{p})$ is the inverse propagator of the quarks and charge conjugate quarks (assimilated to antiquarks), where $\omega_n = (2n+1)\pi T$ are fermionic Matsubara frequencies, with $n \in \mathbb{Z}$. In the Nambu-Gorkov formalism [15], one considers the spinors

$$\Psi \equiv \begin{pmatrix} \psi \\ \psi^C \end{pmatrix}, \tag{11}$$

which define a basis where $\psi$ are quark spinors and $\psi^C$ are charge conjugate quark spinors. $\tilde{S}^{-1}$ is a matrix that is divided into four submatrices and written in the Nambu-Gorkov basis as [14-16, 95-97]

$$\tilde{S}^{-1} = \begin{bmatrix} \left(S_0^+\right)^{-1} & \Delta^- \\ \Delta^+ & \left(S_0^-\right)^{-1} \end{bmatrix}. \tag{12}$$

The two submatrices $\left(S_0^\pm\right)^{-1}$ gather the inverses of the propagators for each flavor $f$ quarks/antiquarks

$$\left(S_0^\pm\right)^{-1} = \begin{bmatrix} \left(S_u^\pm\right)^{-1} & 0 & 0 \\ 0 & \left(S_d^\pm\right)^{-1} & 0 \\ 0 & 0 & \left(S_s^\pm\right)^{-1} \end{bmatrix}, \tag{13}$$

where $\left(S_f^{\pm}\right)^{-1} = \not{p} \pm \gamma_0 \tilde{\mu}_f - m_f$, with $f = u, d, s$. In this relation, $\left(S_f^{+}\right)^{-1} \equiv \left(S_f\right)^{-1}$ concerns the quarks, and $\left(S_f^{-}\right)^{-1} \equiv \left(S_{\bar{f}}\right)^{-1}$ the antiquarks. Each of these flavor terms is divisible into $N_c$ parts, in order to explicit the $r, g, b$ color terms. More precisely, $\tilde{\mu}_f$ is a diagonal matrix, written as $\tilde{\mu}_f = \mathrm{diag}\left(\mu_{f,r}, \mu_{f,g}, \mu_{f,b}\right)$, and so

$$\left(S_f^{\pm}\right)^{-1} = \begin{bmatrix} \not{p} \pm \gamma_0 \mu_{f,r} - m_f & 0 & 0 \\ 0 & \not{p} \pm \gamma_0 \mu_{f,g} - m_f & 0 \\ 0 & 0 & \not{p} \pm \gamma_0 \mu_{f,b} - m_f \end{bmatrix}. \tag{14}$$

In the NJL description, $\mu_{f,r} = \mu_{f,g} = \mu_{f,b}$, so $\tilde{\mu}_f \equiv \mu_f 1_3$, in which $1_3$ is the $3 \times 3$ identity matrix (omitted in the writing of the equations). In the ($\mu$)PNJL models, the modification $\tilde{\mu}_f \to \tilde{\mu}_f - iA_4$ is performed, where $A_4$ was detailed upstream.

The submatrices $\Delta^{\mp}$ translate the pairings between the quarks. In this paper, I focus on scalar quark/quark interactions, associated with the scalar diquark condensates $\left\langle\!\left\langle \bar{\psi}_f^C \gamma_5 \psi_{f'} \right\rangle\!\right\rangle$. These ones are written for convenience via the shorthand notation $\left\langle\!\left\langle \psi_f \psi_{f'} \right\rangle\!\right\rangle$ in (2) and later in this work. The presence of the $\gamma_5$ matrix, to describe scalar diquark condensates, is imposed by the use of the charge conjugation operator $C = i\gamma_0\gamma_2$ [29]. The submatrix $\Delta^-$ is written in a condensed form as [21]

$$\Delta^- = \Delta_{ff',cc'} \gamma_5 \otimes \tau_{ff'} \otimes \lambda_{cc'}. \tag{15}$$

The $\tau_{ff'}$ and $\lambda_{cc'}$ are, respectively, the $SU(3)_{flavor}$ and $SU(3)_{color}$ generators without the factor ½, i.e. both correspond to Gell-Mann matrices. $\Delta_{ff',cc'}$ are the energy gaps of the concerned quark/quark pairs. They are related to the diquark condensates by the relation [16, 19]

$$\left|\Delta_{ff'}\right| = 2G_{DIQ} \left\langle\!\left\langle \psi_f \psi_{f'} \right\rangle\!\right\rangle. \tag{16}$$

In the scalar channel, three pairings, i.e. three scalar quark/quark pairs $ud$, $us$ and $ds$, are identifiable. They are antisymmetric in flavor and in color. These pairs are associated, respectively, with the $\tau_2 \otimes \lambda_2$, $\tau_5 \otimes \lambda_5$ and $\tau_7 \otimes \lambda_7$ Gell-Mann matrices, and with the energy gaps $\Delta_{ud}$, $\Delta_{us}$ and $\Delta_{ds}$. These gaps correspond to the binding energies of these Cooper pairs of the color superconductivity [98]. The $\tau_2 \otimes \lambda_2$ coupling only allows the $u,r$ quarks to interact with the $d,g$ quarks, and the $u,g$ quarks with the $d,r$ ones. In the same way, the $\tau_5 \otimes \lambda_5$ coupling involves the $u$ and $s$ flavors and the $r$ and $b$ colors, and the $\tau_7 \otimes \lambda_7$ coupling concerns the $d$ and $s$ flavors and the $g$ and $b$ colors. When the writing of $\Delta^-$ is developed according to the flavor terms, as done for $\left(S_0^{\pm}\right)^{-1}$ in the equation (13), it leads to [22]

$$\Delta^- = \begin{bmatrix} 0 & -i\gamma_5 \Delta_{ud} \lambda_2 & -i\gamma_5 \Delta_{us} \lambda_5 \\ i\gamma_5 \Delta_{ud} \lambda_2 & 0 & -i\gamma_5 \Delta_{ds} \lambda_7 \\ i\gamma_5 \Delta_{us} \lambda_5 & i\gamma_5 \Delta_{ds} \lambda_7 & 0 \end{bmatrix}, \tag{17}$$

and, in the same way [16],

$$\Delta^+ = \gamma_0 \left(\Delta^-\right)^{\dagger} \gamma_0 = \begin{bmatrix} 0 & i\gamma_5 \Delta_{ud}^* \lambda_2 & i\gamma_5 \Delta_{us}^* \lambda_5 \\ -i\gamma_5 \Delta_{ud}^* \lambda_2 & 0 & i\gamma_5 \Delta_{ds}^* \lambda_7 \\ -i\gamma_5 \Delta_{us}^* \lambda_5 & -i\gamma_5 \Delta_{ds}^* \lambda_7 & 0 \end{bmatrix}. \tag{18}$$

Thanks to the writings of $\left(S_0^{\pm}\right)^{-1}$ or $\Delta^{\mp}$, one observes that each of them is a $36 \times 36$ matrix in the framework of the $SU(3)_f \times SU(3)_c$ description. The three $\Delta_{ff'}$, or $\langle\langle\psi_f\psi_{f'}\rangle\rangle$, are used as order parameters to identify the various phases/subphases encountered in the description of the color superconductivity, and to study the phase transitions between them. We notably have [21, 22]

- Normal quark (NQ) phase:      $\Delta_{ud}=0$, $\Delta_{us}=0$ and $\Delta_{ds}=0$
- 2-flavor color-superconducting (2SC) phase:  $\Delta_{ud}\neq 0$, $\Delta_{us}=0$ and $\Delta_{ds}=0$
- 2SC*us* phase:           $\Delta_{ud}=0$, $\Delta_{us}\neq 0$ and $\Delta_{ds}=0$
- 2SC*ds* phase:           $\Delta_{ud}=0$, $\Delta_{us}=0$ and $\Delta_{ds}\neq 0$
- *u*SC phase:            $\Delta_{ud}\neq 0$, $\Delta_{us}\neq 0$ and $\Delta_{ds}=0$
- *d*SC phase:            $\Delta_{ud}\neq 0$, $\Delta_{us}=0$ and $\Delta_{ds}\neq 0$
- *s*SC phase:            $\Delta_{ud}=0$, $\Delta_{us}\neq 0$ and $\Delta_{ds}\neq 0$
- Color-flavor-locked (CFL) phase:     $\Delta_{ud}\neq 0$, $\Delta_{us}\neq 0$ and $\Delta_{ds}\neq 0$

The NQ phase is modeled by the NJL, PNJL and $\mu$PNJL approaches, section 3. At the opposite, I propose to use the names $\Delta$NJL, $\Delta$PNJL and $\Delta\mu$PNJL to designate the versions of these models that include non-null submatrices $\Delta^{\mp}$. In practice, the two color superconducting phases that are the most studied are the 2SC and the CFL ones [19], because of the frequent use of the isospin symmetry. However, this list is not exhaustive as regards the various phases imagined at the present state of the art: other phases not treated in this work should be mentioned [19, 23]. It concerns the gapless color superconducting phases when the neutral quark matter conditions and the $\beta$ equilibrium are considered, the crystalline phase, etc.

*2.4 Global description of the performed calculations*

2.4.1 Treatment of the $\Omega_M$ term

For each case, the first crucial stage is to calculate the $\Omega_M$ term described above. I give in this part the general ideas to perform this operation. The specific results obtained for each configuration are described in the devoted sections, i.e. sections 3 to 6. Firstly, as explained at several occasions in the literature [67, 99], the relation

$$\mathrm{Tr}\left[\ln\left(\tilde{S}^{-1}\right)\right]=\ln\left[\det\left(\tilde{S}^{-1}\right)\right] \quad (19)$$

allows replacing the trace of the logarithm of a matrix by the logarithm of the determinant of this matrix. To be able to perform this calculation analytically, I consider the method described in [85], whose objective is to calculate the determinant of block matrices. More precisely, for an $(nN)\times(nN)$ complex matrix $S$ partitioned into $n\times n$ blocks, like

$$S=\begin{bmatrix} S_{1,1} & S_{1,2} & \cdots & S_{1,N} \\ S_{2,1} & S_{2,2} & \cdots & S_{2,N} \\ \vdots & \vdots & \ddots & \vdots \\ S_{N,1} & S_{N,2} & \cdots & S_{N,N} \end{bmatrix}, \quad (20)$$

the quoted reference explains that

$$\det(S)=\prod_{k=1}^{N}\det\left(\alpha_{k,k}^{(N-k)}\right), \quad (21)$$

with the initialization $\alpha_{i,j}^{(0)}=S_{i,j}$ and the recursive formula, for $k\geq 1$,

$$\alpha_{i,j}^{(k+1)}=\alpha_{i,j}^{(k)}-\alpha_{i,N-k}^{(k)}\left(\alpha_{N-k,N-k}^{(k)}\right)^{-1}\alpha_{N-k,j}^{(k)}. \quad (22)$$

I performed the calculations with the partitioning visible in (13) for $\left(S_0^\pm\right)^{-1}$ and in (17) and (18) for $\Delta^\mp$. In other words, I do not explicit the Dirac matrices and the color terms. With $N_f = N_c = 3$, it leads to consider $\tilde{S}^{-1}$, equation (12), as a matrix composed by $6\times6$ blocks, where each block is a $12\times12$ matrix. Then, when the determinant is found, the next operation consists in performing the Matsubara summation of the fermionic frequencies $\omega_n = (2n+1)\pi T$, with the relation [99]

$$T \sum_{n=-\infty}^{+\infty} \ln\left(\frac{\omega_n^2 + \lambda_k^2}{T^2}\right) = \lambda_k + 2T \ln\left[1 + \exp(-\lambda_k/T)\right] + Cst. \tag{23}$$

The establishment of this formula is discussed in the appendix A. The simplification of the $\Omega_M$ writing makes it possible to estimate the quantities evoked upstream: the chiral condensate that appears in (1), the density term also present in this equation, the gaps $\Delta_{ud}$, $\Delta_{us}$ and $\Delta_{ds}$, and the Polyakov fields $\Phi, \bar{\Phi}$ or $\phi_3, \phi_8$. The associated calculations, detailed later in this work, include differentiations of $\Omega_M$ in order to minimize the grand potential $\Omega$ [1]. It implies that the constant term $Cst$ visible in (23) disappears. So, this constant can be omitted [99].

### 2.4.2 The chiral condensates

A first possibility to estimate the condensate $\langle\langle \bar{\psi}_f \psi_f \rangle\rangle$ is to consider the relation [19]

$$\frac{\partial \Omega}{\partial \langle\langle \bar{\psi}_f \psi_f \rangle\rangle} = 0. \tag{24}$$

However, if we calculate this derivative for all the treated flavors in (2) and if we use the quark gap equations, i.e. the first line of (1), we conclude that the condensate also satisfies the following equation, visible in an equivalent form in [19]

$$\langle\langle \bar{\psi}_f \psi_f \rangle\rangle = \frac{\partial \Omega_M}{\partial m_f}. \tag{25}$$

The advantage of these two relations is they stay valid for all the treated configurations, using or not the Polyakov loop or the diquark condensate gaps. Nevertheless, the expressions of the chiral condensates can also be found with another approach [16]. Its objective is to estimate the normal propagator $S^+$ of the quarks and the one of antiquarks $S^-$ starting directly from the definition of $\tilde{S}^{-1}$, equation (12). These ones are obtained with the relation [15, 16, 96, 97, 100, 101]

$$S^\pm = \left[\left(S_0^\pm\right)^{-1} - \Sigma^\pm\right]^{-1} \text{ with } \Sigma^\pm = \Delta^\mp S_0^\mp \Delta^\pm. \tag{26}$$

The normal propagators for each flavor/antiflavor are then extracted from $S^+$ and $S^-$. To make the link with the quoted literature, I note them

$$S_f \equiv S_f^+ \equiv G_f^+ \text{ and } S_{\bar{f}} \equiv S_f^- \equiv G_f^-. \tag{27}$$

By this method, the condensate of the flavor $f$ quark is obtained with [16, 19, 30, 35]

$$\langle\langle \bar{\psi}_f \psi_f \rangle\rangle = \frac{1}{\beta} \sum_n \int \frac{d^3p}{(2\pi)^3} \text{Tr}\left(S_f\right). \tag{28}$$

The trace is performed on the Dirac terms. As a consequence, one should understand $\text{Tr}(S_f)$ as $\text{Tr}(S_{f,r}) + \text{Tr}(S_{f,g}) + \text{Tr}(S_{f,b})$, i.e. one treats separately each color and then one sums these three traces. In practice, the equation (28) is usable to check the results found with (25). In some cases, it also allows proposing an alternative writing that does not use derivatives, which constitutes an advantage in numerical calculations.

---

[1] As argued in [19], since $G_V > 0$, Table 1, $\Omega$ should not be minimized with respect to $\mu_f$, but maximized. This remark is extended to the derivative of $\Omega$ with respect to $\langle\langle \psi_f^+ \psi_f \rangle\rangle$ [52], which appears in (29).

2.4.3 The density terms $\langle\langle \psi_f^+ \psi_f \rangle\rangle$

The condensates $\langle\langle \psi_f^+ \psi_f \rangle\rangle$ that are used to evaluate the densities satisfy the relation [19]

$$\frac{\partial \Omega}{\partial \langle\langle \psi_f^+ \psi_f \rangle\rangle} = 0. \tag{29}$$

As done for the chiral condensates, if one uses this equation in the expression of the grand potential $\Omega$, equation (2), with the equation [19, 29, 30, 72, 102]

$$\mu_f = \mu_{0f} - 4G_V \langle\langle \psi_f^+ \psi_f \rangle\rangle, \tag{30}$$

one finds [19, 52, 86, 103]

$$\langle\langle \psi_f^+ \psi_f \rangle\rangle = -\frac{\partial \Omega_M}{\partial \mu_f}. \tag{31}$$

The $\mu_f$ are solutions of the system (1). In fact, the $\mu_{0f}$ visible in (30) are the bare chemical potentials, and the $\mu_f$ are effectives ones. Nevertheless, with the values of $G_V$ used in Table 1, the differences between $\mu_{0f}$ and $\mu_f$ stay rather weak, except at very high densities [72]. As with (28), $\langle\langle \psi_f^+ \psi_f \rangle\rangle$ can be evaluated from the normal quark propagators $S_f$, with the relation [19, 30, 35]

$$\langle\langle \psi_f^+ \psi_f \rangle\rangle = \frac{1}{\beta} \sum_n \int \frac{d^3 p}{(2\pi)^3} \mathrm{Tr}(\gamma_0 S_f). \tag{32}$$

2.4.4 The energy gaps $\Delta_{ff'}$

The equation to be solved to find the values of the gaps $\Delta_{ff'}$ is [15, 16, 19, 104-106]

$$\frac{\partial \Omega}{\partial |\Delta_{ff'}|} = 0. \tag{33}$$

Thanks to the equation (16) that indicates the link between the gaps $\Delta_{ff'}$ and the diquark condensates $\langle\langle \psi_f \psi_{f'} \rangle\rangle$, one remarks that the $G_{DIQ}\left(\langle\langle \psi_u \psi_d \rangle\rangle^2 + \langle\langle \psi_u \psi_s \rangle\rangle^2 + \langle\langle \psi_d \psi_s \rangle\rangle^2\right)$ term present in (2) is fully equivalent to the $\frac{|\Delta_{ud}|^2 + |\Delta_{us}|^2 + |\Delta_{ds}|^2}{4G_{DIQ}}$ term found in the literature [19, 22, 24, 60, 84]. It also implies that (33) is also rewritable as

$$\frac{\partial \Omega}{\partial \langle\langle \psi_f \psi_{f'} \rangle\rangle} = 0. \tag{34}$$

As done for the chiral condensates and the density terms, the gaps $\Delta_{ff'}$ can be estimated with the alternative method. This one considers now the abnormal quarks/antiquarks propagators $\Xi^\pm$ [15, 16, 96, 97, 100, 101], estimated thanks to the relation

$$\Xi^\pm = -S^\mp \Delta^\pm S_0^\pm. \tag{35}$$

With this method, the relation proposed in [16] can be written as

$$\left|\Delta_{ff',cc'}\right| = 2G_{DIQ} \frac{1}{\beta} \sum_n \int \frac{d^3 p}{(2\pi)^3} \left|\mathrm{Tr}\left(\Xi^-_{\substack{ff' \\ cc'}} \gamma_5\right)\right|. \tag{36}$$

As with (28) and (32), the trace is performed on the Dirac terms. The gap $\Delta_{ff'}$ is found with the relation $|\Delta_{ff'}| = |\Delta_{ff',cc'}| + |\Delta_{f'f,cc'}| + |\Delta_{ff',c'c}| + |\Delta_{f'f,c'c}|$, i.e. $|\Delta_{ff'}| = 4|\Delta_{ff',cc'}|$.

2.4.5 The Polyakov fields

In the PNJL and $\mu$ PNJL models, the values of the average Polyakov field $\Phi$ and its conjugate one $\bar{\Phi}$ have to minimize the grand potential $\Omega$. It leads to use the relations [67, 94, 107, 108]

$$\frac{\partial \Omega}{\partial \Phi} = 0 \text{ and } \frac{\partial \Omega}{\partial \bar{\Phi}} = 0. \tag{37}$$

In this case, $\Phi$ and $\bar{\Phi}$ are treated as real and independent variables. In the writing of $\Omega$, equation (2), $\mathcal{U}$ and $\Omega_M$ depend on these fields. So, these ones are also real numbers. Furthermore, with the potential described in (3), an equivalent writing of $\partial \Omega / \partial \Phi = 0$ is

$$-IT^4 + \frac{\partial \Omega_M}{\partial \Phi} = 0 \text{ with } I = \frac{a(T)}{2}\bar{\Phi} + 6\frac{b(T)(\bar{\Phi} - 2\Phi^2 + \bar{\Phi}^2 \Phi)}{1 - 6\Phi\bar{\Phi} + 4(\Phi^3 + \bar{\Phi}^3) - 3(\bar{\Phi}\Phi)^2}, \tag{38}$$

and so on for $\bar{\Phi}$. However, in the $\Delta$ PNJL and $\Delta \mu$ PNJL descriptions, the terms forming the diagonal of the matrix $A_4$ cannot be factorized out to form expressions that depend on $\Phi$ and $\bar{\Phi}$ [60]. In the mean field approximation, thanks to (6) to (8), $\Phi$ and $\bar{\Phi}$ are written as

$$\Phi = \frac{1}{N_c} \sum_{j=1}^{3} \exp\left[i\beta A_{4(jj)}\right] \text{ and } \bar{\Phi} = \frac{1}{N_c} \sum_{j=1}^{3} \exp\left[-i\beta A_{4(jj)}\right], \tag{39}$$

with the $A_{4(jj)}$ defined equation (9). In this configuration, the fields $\phi_3$ and $\phi_8$ that compose $A_{4(jj)}$ become the variables to be taken into account in the resolution of the equations, and not $\Phi$ and $\bar{\Phi}$ that are now complex numbers. It implies that the grand potential $\Omega$ becomes complex, too. As explained in [60, 68], it constitutes in this context a (fermion) sign problem, because the minimization of a complex function has no sense. In practice, as in [59, 60, 93], the minimization is performed on the real part of the grand potential. Consequently, (37) is then replaced by

$$\frac{\partial \operatorname{Re}(\Omega)}{\partial \phi_3} = 0 \text{ and } \frac{\partial \operatorname{Re}(\Omega)}{\partial \phi_8} = 0, \tag{40}$$

and the derivatives of the potential with respect to these fields are expressed as

$$\begin{aligned}\frac{\partial \mathcal{U}}{\partial \phi_3} &= \frac{4T^4}{N_c} \sin(\phi_3) \operatorname{Re}\left[I \exp\left(i\frac{1}{\sqrt{3}}\phi_8\right)\right] \\ \frac{\partial \mathcal{U}}{\partial \phi_8} &= \frac{2T^4}{\sqrt{3}N_c} \operatorname{Re}\left(-iI\left\{\exp\left[i\beta A_{4(11)}\right] + \exp\left[i\beta A_{4(22)}\right] - 2\exp\left[i\beta A_{4(33)}\right]\right\}\right)\end{aligned}. \tag{41}$$

This approximation does not invalidate the equations described in the sub-subsections 2.4.2 and 2.4.3, but it imposes to take the real part of $\Omega$ or $\Omega_M$ in (24), (25), (29), (31), (33) and (34), and to take the real part of the results found with the method described equations (28), (32) and (36). In the mean field approximation, this solution is justified in [60]. Works that go beyond the mean field approximation [59] consider an approach to take into account the neglected imaginary part.

## 3. NJL, PNJL and $\mu$ PNJL descriptions without the inclusion of the superconductivity

*3.1 NJL equations*

When the two submatrices $\Delta^+$ and $\Delta^-$ are null, the inverse of the quarks/antiquarks propagator $\tilde{S}^{-1}$ is a diagonal block matrix. The calculation of its determinant is then straightforward: it corresponds to the product of the determinants of each block that composes its diagonal, i.e.

$$\det\left(\tilde{S}^{-1}\right) = \prod_{f=u,d,s} \det\left[\left(S_f^+\right)^{-1}\right]\det\left[\left(S_f^-\right)^{-1}\right]. \tag{42}$$

Thanks to (14), the determinant of $\tilde{S}^{-1}$ is written as

$$\det\left(\tilde{S}^{-1}\right) = \prod_{f=u,d,s}\left[\det\left(\not{p}+\gamma_0\mu_f-m_f\right)\det\left(\not{p}-\gamma_0\mu_f-m_f\right)\right]^{N_c}. \tag{43}$$

With the relation

$$\det\left(\not{p}\pm\gamma_0\mu_f-m_f\right) = \left[\left(i\omega_n\pm\mu_f\right)^2-E_f^2\right]^2, \tag{44}$$

where $E_f = \sqrt{\vec{p}^2+m_f^2}$ is the energy of a flavor $f$ quark, one obtains

$$\det\left(\tilde{S}^{-1}\right) = \prod_{f=u,d,s}\left[\left(i\omega_n+\mu_f\right)^2-E_f^2\right]^{2N_c}\left[\left(i\omega_n-\mu_f\right)^2-E_f^2\right]^{2N_c}. \tag{45}$$

One uses the equation $\left[(a+b)^2-c^2\right]\left[(a-b)^2-c^2\right] = \left[a^2-(b-c)^2\right]\left[a^2-(b+c)^2\right]$ and one finds

$$\begin{aligned}\sum_n \mathrm{Tr}\left[\ln\left(\frac{\tilde{S}^{-1}}{T}\right)\right] &= \sum_n \ln\left[\det\left(\frac{\tilde{S}^{-1}}{T}\right)\right]\\ &= 2N_c\sum_n\sum_{f=u,d,s}\ln\left[\frac{\omega_n^2+\left(E_f-\mu_f\right)^2}{T^2}\right]+\ln\left[\frac{\omega_n^2+\left(E_f+\mu_f\right)^2}{T^2}\right]\end{aligned}. \tag{46}$$

The summation of the fermionic Matsubara frequencies $\omega_n$ is performed thanks to (23), which allows finding again the well known expression of $\Omega_M$ in the framework of the NJL model,

$$\begin{aligned}\Omega_M &= -\frac{T}{2}\int\frac{d^3p}{(2\pi)^3}\sum_n \mathrm{Tr}\left[\ln\left(\frac{\tilde{S}^{-1}}{T}\right)\right]\\ &= -2N_c\int\frac{d^3p}{(2\pi)^3}\sum_{f=u,d,s}E_f+T\ln\left[1+\exp\left(-\frac{E_f-\mu_f}{T}\right)\right]+T\ln\left[1+\exp\left(-\frac{E_f+\mu_f}{T}\right)\right]\end{aligned}. \tag{47}$$

With (25), (31) and (47), or with (28), (32) and the propagators obtained via (14), one also recovers the expressions of $\langle\langle\bar{\psi}_f\psi_f\rangle\rangle$ and $\langle\langle\psi_f^+\psi_f\rangle\rangle$ visible in NJL publications [29, 30], i.e.

$$\langle\langle\bar{\psi}_f\psi_f\rangle\rangle = -2N_c\int\frac{d^3p}{(2\pi)^3}\frac{m_f}{E_f}\left[1-f\left(E_f-\mu_f\right)-f\left(E_f+\mu_f\right)\right]\bigg|_{f=u,d,s} \tag{48}$$

and

$$\langle\langle\psi_f^+\psi_f\rangle\rangle = 2N_c\int\frac{d^3p}{(2\pi)^3}\left[f\left(E_f-\mu_f\right)-f\left(E_f+\mu_f\right)\right]\bigg|_{f=u,d,s}, \tag{49}$$

in which the Fermi-Dirac distribution $f(x) = \left(e^{\beta x}+1\right)^{-1}$ is employed.

*3.2 PNJL and $\mu$PNJL equations*

The equation (42) is still valid in the PNJL and $\mu$PNJL models. But, with the replacement $\tilde{\mu}_f \to \tilde{\mu}_f - iA_4$ indicated in the subsection 2.3, (46) is rewritten as

$$\sum_n \ln\left[\det\left(\frac{\tilde{S}^{-1}}{T}\right)\right] = 2\sum_n \sum_{f=u,d,s} \sum_{j=1}^{N_c} \ln\left[\frac{\omega_n^2 + (E_f - \mu_{f,j})^2}{T^2}\right] + \ln\left[\frac{\omega_n^2 + (E_f + \mu_{f,j})^2}{T^2}\right], \quad (50)$$

where the shorthand notation $\mu_{f,j} = \mu_f - iA_{4(jj)}$ is used. As confirmed in the Appendix A, (23) is usable when $\lambda_k$ is complex, with the condition $\text{Im}(\lambda_k/T) \in ]-\pi;\pi[$. It was checked during the numerical calculations that this constraint was always verified in all the presented results, including or not the color superconductivity. It leads to the relation

$$\begin{aligned}
\Omega_M = &-2N_c \int \frac{d^3p}{(2\pi)^3} \sum_{f=u,d,s} E_f \\
&+ \frac{1}{N_c}\sum_{j=1}^{N_c} T\ln\left[1+\exp\left(-\frac{E_f - \mu_{f,j}}{T}\right)\right] + T\ln\left[1+\exp\left(-\frac{E_f + \mu_{f,j}}{T}\right)\right]
\end{aligned}, \quad (51)$$

or, with $L = \exp(i\beta A_4)$ and $L^\dagger = \exp(-i\beta A_4)$, it gives the form encountered in the PNJL literature

$$\begin{aligned}
\Omega_M = &-2N_c \int \frac{d^3p}{(2\pi)^3} \sum_{f=u,d,s} E_f \\
&+ \frac{1}{N_c}\text{Tr}_c\left\{T\ln\left[1+L^\dagger\exp\left(-\frac{E_f - \mu_f}{T}\right)\right] + T\ln\left[1+L\exp\left(-\frac{E_f + \mu_f}{T}\right)\right]\right\}
\end{aligned}, \quad (52)$$

where $\text{Tr}_c$ is the trace over the colors. Furthermore, the expressions of $\langle\langle \bar{\psi}_f \psi_f \rangle\rangle$ and $\langle\langle \psi_f^+ \psi_f \rangle\rangle$ in the ($\mu$)PNJL models can be estimated and then written as in the NJL approach, equations (48) and (49), if one replaces $f(E_f - \mu_f)$ by $f_\Phi^+(E_f - \mu_f)$ and $f(E_f + \mu_f)$ by $f_\Phi^-(E_f + \mu_f)$. In fact, $f_\Phi^+$ and $f_\Phi^-$ are the "modified" Fermi-Dirac distributions for, respectively, the quarks and antiquarks. They can be expressed, e.g., as

$$f_\Phi^+(E_f - \mu_f) = \frac{1}{N_c}\text{Tr}_c\left\{\frac{1}{1+L\exp[\beta(E_f - \mu_f)]}\right\} \quad (53)$$

and

$$f_\Phi^-(E_f + \mu_f) = \frac{1}{N_c}\text{Tr}_c\left\{\frac{1}{1+L^\dagger\exp[\beta(E_f + \mu_f)]}\right\}. \quad (54)$$

The calculations related to their writing in the ($\mu$)PNJL models are proposed in the Appendix B.

*3.3 NJL and PNJL results: link between $T, \rho_B$ and $T, \mu_q$ graphs*

Even if (P)NJL results were found outside of the framework of the isospin symmetry [69, 72], I focus in this part on the ones that consider this symmetry, with the P1 parameter set. The resolution of the system of equations (1), with (37) in the PNJL model, allows obtaining the masses of the dressed quarks, according to the temperature and the baryonic density $\rho_B = \frac{2}{3}\rho_q$ [29, 35], with $\rho_s$ fixed to zero, Figure 1. In this figure, the inclusion of the Polyakov loop leads to the shifting of the values towards higher temperatures, as mentioned in the introduction. These equations also enable to find the evolution of the chemical potential $\mu_q = \mu_B/3$ of the light quarks in the $T, \rho_B$ plane, Figure 2(a). This one underlines the non-trivial evolution of the chemical potential according to the temperature and the baryonic density, especially at low temperatures [62, 94].

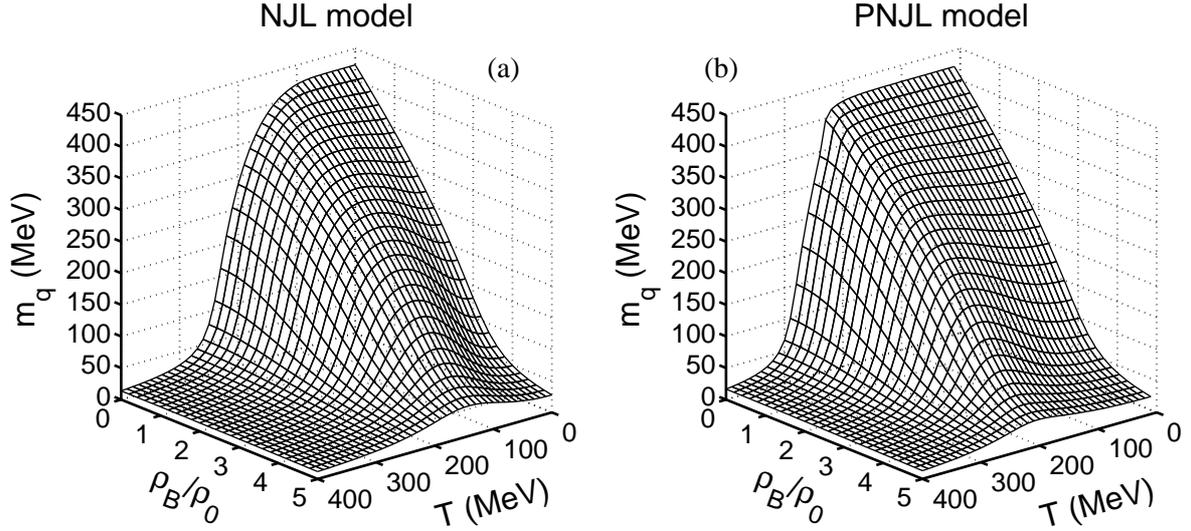

**Figure 1.** Mass of the $q$ quarks in the $T, \rho_B$ plane, in the (a) NJL and (b) PNJL models. $\rho_0 \approx 0.16$ fm$^{-3}$ designates the standard nuclear density.

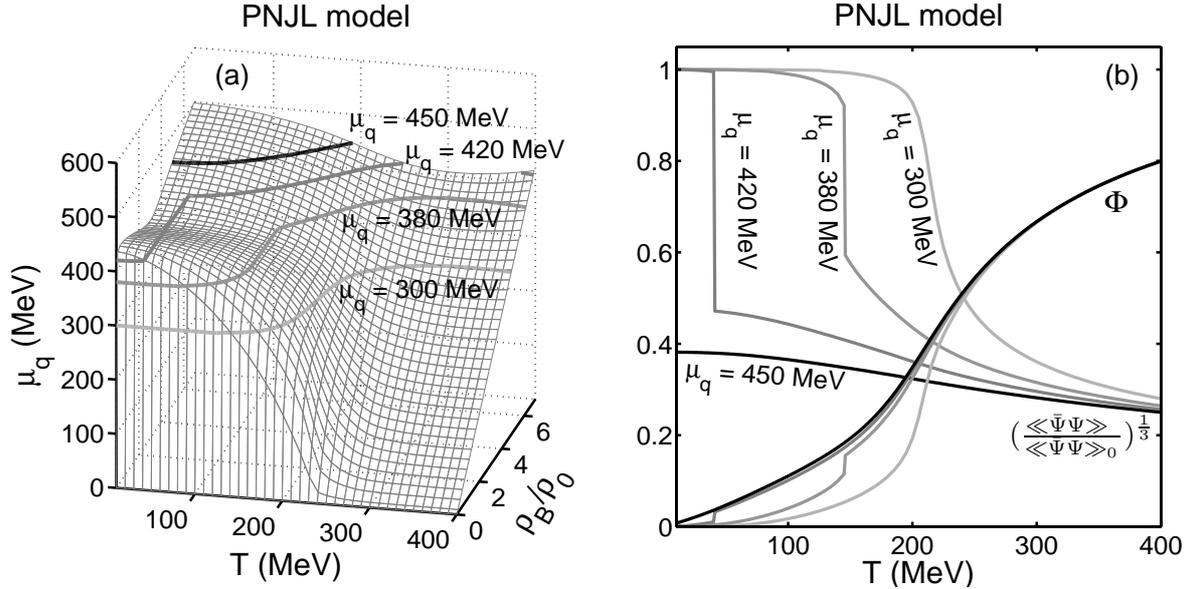

**Figure 2.** (a) Chemical potential $\mu_q$ in the $T, \rho_B$ plane, with four contour lines. (b) Polyakov field $\Phi$ and cubic root of the normalized chiral condensate $\langle\langle \bar{\psi}_q \psi_q \rangle\rangle$ according to the temperature $T$, for several $\mu_q$.

The Figure 2(b) exhibits the evolution of the Polyakov field $\Phi$ and the cubic root of the chiral condensate of the light quarks. This one is normalized by its value $\langle\langle \bar{\psi}_q \psi_q \rangle\rangle_0$ found at $T = 0$ and $\mu_q = 0$. Similar graphs are found in [55, 67, 107]. They display the well-known discontinuities of the chiral condensate according to $T$, which intervene for certain values of $\mu_q$. It also exhibit small discontinuities of the Polyakov field that occur at the same temperatures. They result from the coupling between $\langle\langle \bar{\psi}_q \psi_q \rangle\rangle$ and $\Phi$ [77]. At the opposite, such discontinuities of these order parameters are not observed in the $T, \rho_B$ plane, Figure 3. This difference is graphically explainable thanks to the Figure 2(a). Indeed, the contour lines represented in this figure allows finding a correspondence between $\mu_q, \rho_B, T$. This correspondence is then employable in the graphs presented in the Figure 3. If one observes these tridimensional graphs upon, respectively, the $T,\Phi$ and the $T,\langle\langle\bar{\psi}\psi\rangle\rangle^{1/3}$ planes, one recovers the curves of the Figure 2(b). At moderate chemical potentials, e.g. $\mu_q = 300$ MeV, the evolution of the contour line is smooth. At the opposite, at $\mu_q = 380$ MeV, a portion of the contour line in the Figure 2(a) indicates that several densities correspond to the same chemical potential for a temperature close to 150 MeV. It permits to understand the discontinuities visible in the Figure 2(b) for this chemical potential. Similar explanations can be done for $\mu_q = 420$ MeV, for which the discontinuity of the chiral condensate is stronger. This behavior is related to the non-monotonic variations of the chemical potential in the Figure 2(a). Above about $\mu_q \approx 450$ MeV, since the contour lines are then above this non-monotonic zone, they are smooth and the discontinuities disappear.

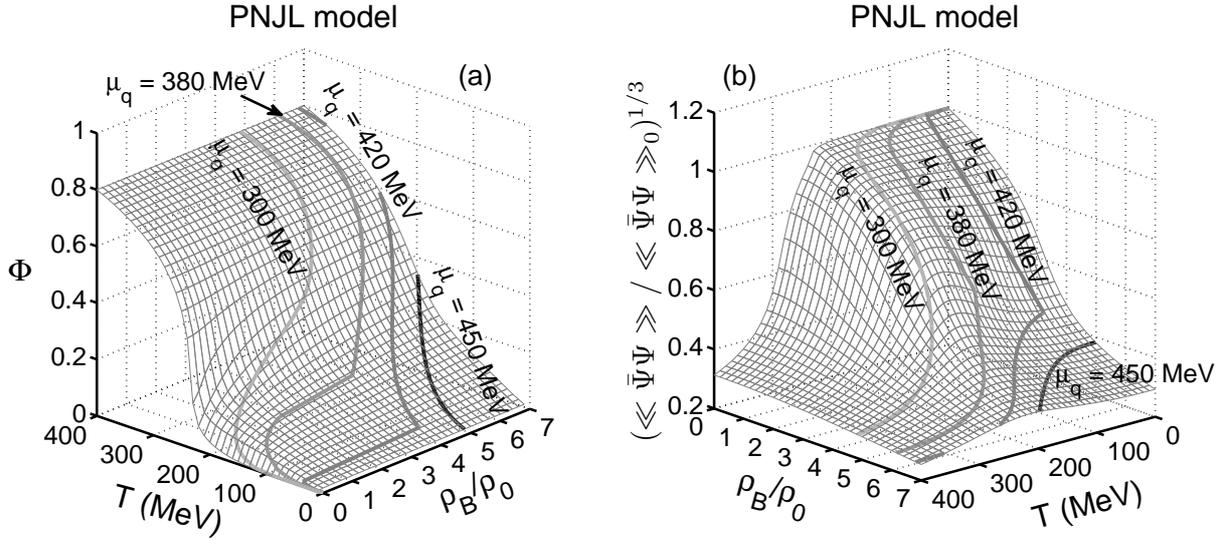

**Figure 3.** Contour lines of four fixed chemical potentials $\mu_q$ in graphs displaying the (a) Polyakov field $\Phi$ and the (b) normalized chiral condensate $\langle\langle\bar{\psi}_q\psi_q\rangle\rangle$, in the $T,\rho_B$ plane.

However, for high chemical potentials, a limitation of the employed numerical method is reached. Indeed, since the (P)NJL models are not renormalizable, a cutoff $\Lambda$ is used[2], Table 1. It corresponds to the upper bound of the integrals performed numerically. When the chemical potentials become too close to $\Lambda$, the results become unreliable. This aspect can be illustrated, e.g., in the calculations of the densities at null temperature with (49), in the NJL model. In this example, if one neglects the mass of the particle in the reasoning, it can be observed that one "misses" in the integration domain a part of the rectangular function, formed by the two Fermi-Dirac statistics at $T = 0$, when $\mu_f > \Lambda$.

*3.4 The problem of matter stability*

From a physical point of view, the mentioned discontinuities of the chiral condensate are interpreted as a first order phase transition in the $T,\mu_q$ plane [109]. However, one neglects the fact that the value of the chiral condensate does not fall to zero after the discontinuities in the Figure 2(b), i.e. when the chiral symmetry is restored. This behavior is due to the non-null masses $m_{0f}$ of the naked quarks. Furthermore, in a first order phase transition, metastable/unstable states should exist near the discontinuity. In this subsection, a simple method is proposed to observe them.

---

[2] Some approaches avoid to use a cutoff, for example with a form factor [77].

To work in the $T, \mu_f$ plane, the chemical potentials are considered as input parameters. So, the set of equations described upstream is solved without the second line of (1). Far from the discontinuity, for given values of $T, \mu_f$, it exists one unique solution: the stable one. However, near the discontinuity, several solutions can be found. Indeed, the minimization of the grand potential is performed by setting its derivatives equal to zero, subsection 2.4. Consequently, not only the global minimum (stable state) can be obtained: a local minimum (metastable state) or a maximum (unstable state) are also possible solutions. In this case, the implemented algorithms very frequently chose the stable one. This remark is particularly verified when the stability constraints $\partial^2 \Omega / \partial \zeta^2 > 0$ are applied to discard unstable solutions [20], where $\zeta$ corresponds to the chiral condensates, the Polyakov fields or the gaps $\Delta_{ff'}$.

At the opposite, in $T, \rho_B$ calculations, the second line of (1) constitutes an additional constraint that restraints the number of solutions. For given values of $T, \rho_B$, it exists one unique solution. However, this one can correspond to a stable, metastable or unstable state. This found solution notably gives the masses of the dressed quarks and the associated chemical potentials. In the Figure 4 and Figure 5, such solutions are plotted according to $T$ and the obtained $\mu_q$. So, these graphs include metastable and unstable states, and not only stable ones. The Figure 4 shows the evolution of $m_q$. Compared with the Figure 9(a), the discontinuity is replaced at low temperatures by the "inverted S" structure associated with a first order phase transition. For increasing values of $\mu_q$, the curve firstly goes beyond the discontinuity (metastable states), then it goes backward (unstable states), and finally it goes forward again (metastable states before reaching the discontinuity). This behavior was described in [62], at null temperature. In the Figure 5, the order parameters of the PNJL model are studied. This figure present strong similarities with results published in [110]. However, as in the Figure 4, the discontinuities visible in the graphs of this reference are replaced by the "inverted S" structures. In the Figure 5(a), this phenomenon is characterized by a strong local deformation of the surface: its aspect is comparable to a folded sheet of paper. In the graphs of the Figure 5, the curves of the Figure 2(b) were added. Their evolution is consistent with the plotted surfaces, notably at the level of the described structures.

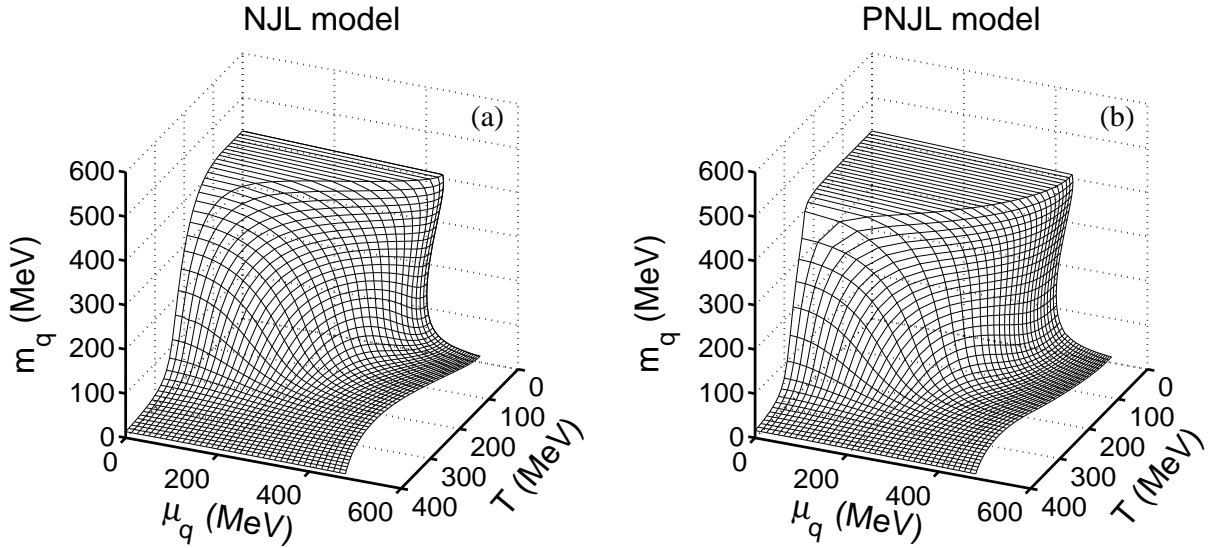

**Figure 4.** Mass of the light quarks $m_q$ in the $T, \mu_q$ plane, in the (a) NJL and (b) PNJL models. The represented surfaces are formed by lines that correspond to constant temperatures or constant baryonic densities.

Consequently, the Figure 4 and Figure 5 allow confirming the first order phase transition of the chiral condensate according to $\mu_q$, at low $T$. With the P1 parameter set, it intervenes when $\mu_q \approx 400$ MeV. Furthermore, they underline that states found according to $T, \rho_B$ are metastable or unstable. This aspect was identified in [46, 111], and mentioned in papers like [62, 84, 112, 113]. It is known as the *problem of matter stability*, notably because the standard nuclear density $\rho_0$ is found to be unstable at low $T$. Here, at null temperature, baryonic densities between $0.42\,\rho_0$ and $3.0\rho_0$ correspond to unstable states. This unstable zone is surrounded by metastable states for $0 < \rho_B < 0.42\,\rho_0$ and $3.0\,\rho_0 < \rho_B < 3.7\,\rho_0$. For the other $\rho_B$, including $\rho_B = 0$, the found solutions are stable. These values are comparable to the ones mentioned in [46, 109, 111]. At $T \neq 0$, the widths of these metastable and unstable zones decrease when $T$ is growing, until their disappearance at the level of the critical endpoint (CEP), when $T \approx 79$ MeV in the NJL model, versus $T \approx 158$ MeV in the PNJL one, both at $\rho_B \approx 2.8\rho_0$. These temperatures are in agreement, respectively, with [19, 114] and [62].

The problem of matter stability constitutes a feature of the chiral models [46, 111][3]. In [46], it was imagined that in the metastable/unstable conditions, the quarks could evolve to form droplets surrounded by the vacuum. Inside of these droplets, a density above the unstable ones is reached, and the chiral symmetry is restored. It leads to consider a mixture of two stable phases: the droplets and the vacuum. Consequently, the metastable/unstable phases become the *mixed phase* [86]. In these conditions, one proposes to see $\rho_B$ as an average value, weighted by the volume of the vacuum and the one of the droplets. So, the volume of the droplets phase should grow continuously when $\rho_B$ is increasing, and conversely. In the same manner, one assimilates the dressed quark mass to a weighted average value, performed with the mass of a quark plunged into the vacuum and the one of a quark added into a droplet. The reasoning is extendable to other quantities, like the order parameters. This vision corresponds well to the apparent crossover observed in the $T, \rho_B$ plane, Figure 1 and Figure 3. Nevertheless, as indicated in [19], the order of a phase transition should be determined via a study according to the $\mu_f$ and/or with the grand potential, and not only via the resolution of gap equations.

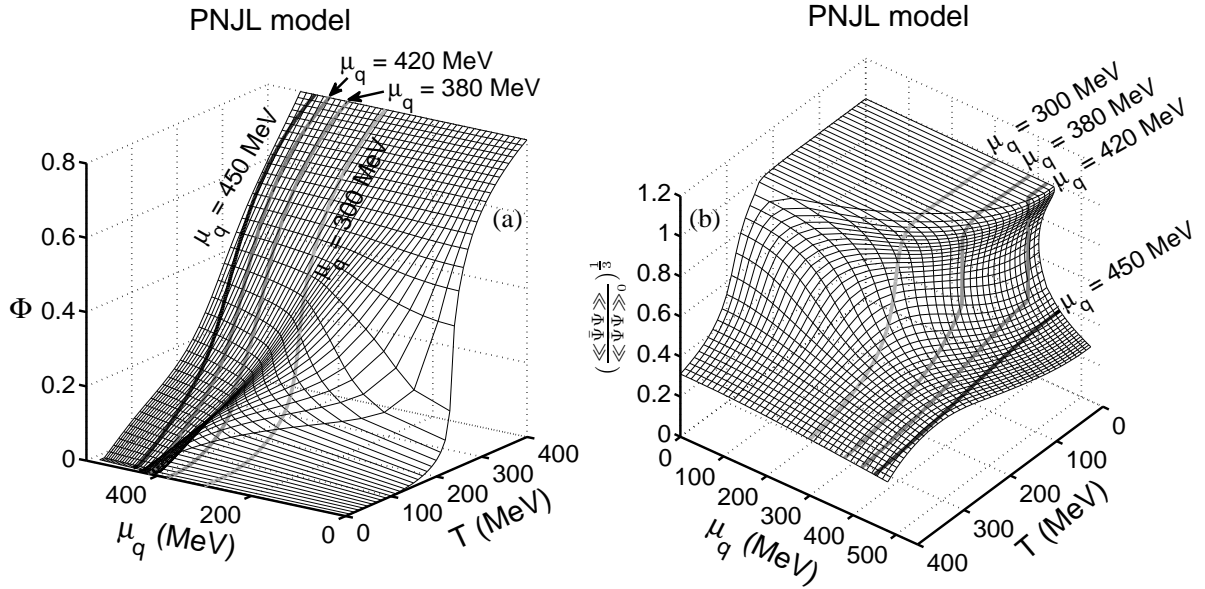

**Figure 5.** (a) Polyakov field $\Phi$ and (b) normalized cubic root of $\left\langle\left\langle \bar{\psi}_q \psi_q \right\rangle\right\rangle$, in the $T, \mu_q$ plane.

*3.5 Results related to the $\mu$PNJL model*

In the $\mu$PNJL model, since $T_0$ becomes a variable in (5), it appears relevant to study its variations, Figure 6. The $N_f$ correction leads to lower its value compared to the pure gauge one. In this work,

---
[3] As recalled in the first of these two references, this phenomenon can disappear for certain parameter sets.

$N_f = 2+1$. It corresponds to $T_0 \approx 187$ MeV [75] at $\mu_f = 0$ (so at $\rho_f = 0$), with the effective value $N_f = 2.7$. When $\rho_B$ becomes non-null, $\mu_q$ presents a discontinuity, Figure 2(a). Since $T_0$ directly depends on the chemical potentials, this discontinuity also appears in its evolution in the $\rho_B, T$ plane. However, it does not affect the other observables, as $\Phi$ or $m_q$, Figure 7 and Figure 8. In the Figure 6(a), at moderate $\rho_B$ and low $T$, the non-monotonic variations of $T_0$ according to $\rho_B$ comes from the variations of $\mu_q$ in the $\rho_B, T$ plane. These ones are reversed according to $\rho_B$ in the unstable phase, Figure 2(a). As a consequence, the contour line $T = T_0$ is associated with an increasing function with respect to $\rho_B$ only in the unstable zone, Figure 6(a) and Figure 16(b). At the opposite, this behavior is not visible when $K = 0$, Figure 6(b). Indeed, with the used parameter set, the first order phase transition of the chiral condensate becomes a crossover when the 't Hooft term $K$ is removed, which leads to the disappearance of the unstable zone.

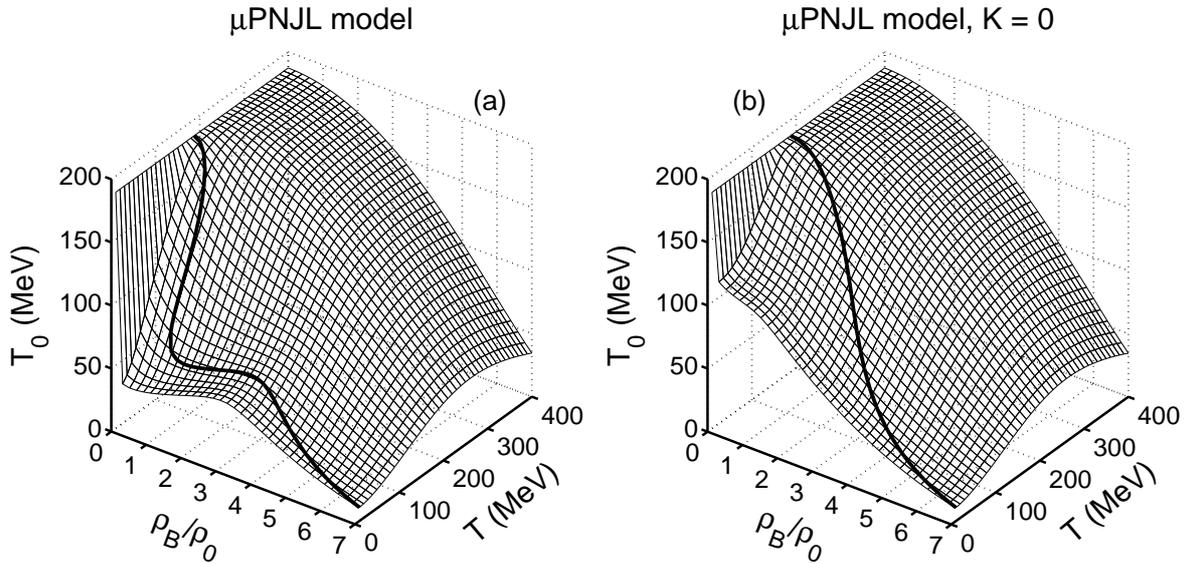

**Figure 6.** Evolution of the transition temperature $T_0$ in the $\rho_B, T$ plane, with (a) the P1 parameter set ($K \neq 0$) and (b) $K = 0$. The thick curve indicates when $T = T_0$.

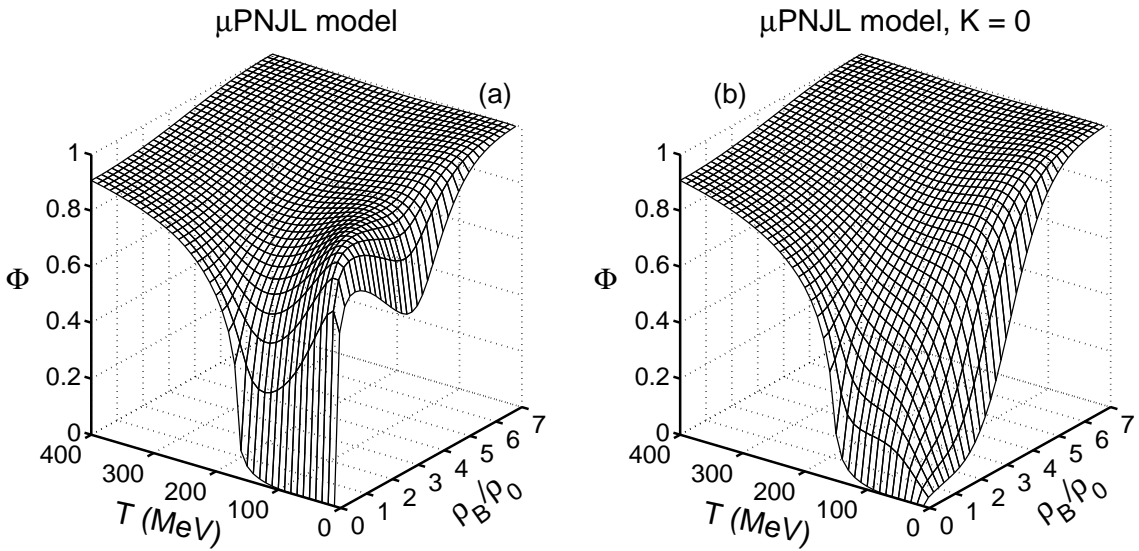

**Figure 7.** Polyakov field $\Phi$ according to $\rho_B, T$, in the $\mu$PNJL model, with (a) $K \neq 0$ and (b) $K = 0$.

The evolution of $T_0$ has a strong influence on the one of the Polyakov loop $\Phi$ (and $\bar{\Phi}$). It explains the variations of $\Phi$ in the mixed phase, Figure 7(a). Moreover, two differences with the PNJL model should be mentioned, Figure 3(a). Firstly, at null/low densities and high temperatures, $\Phi$ reaches stronger values in the $\mu$PNJL model because of the $N_f$ correction. Then, with the $\mu$-dependence, $\Phi$ also increases according to $\rho_B$, except in the unstable phase, instead of the observed stagnation in the PNJL model. As a consequence, $\Phi \approx 1$ for the highest temperatures and densities studied in the graphs. Thanks to the Polyakov loop potential described in (3), one keeps $\Phi < 1$. Compared to other $\mu$PNJL results found in the literature, the Figure 7(b) is in agreement with [81], which uses $N_f = 2$ and no vector interaction.

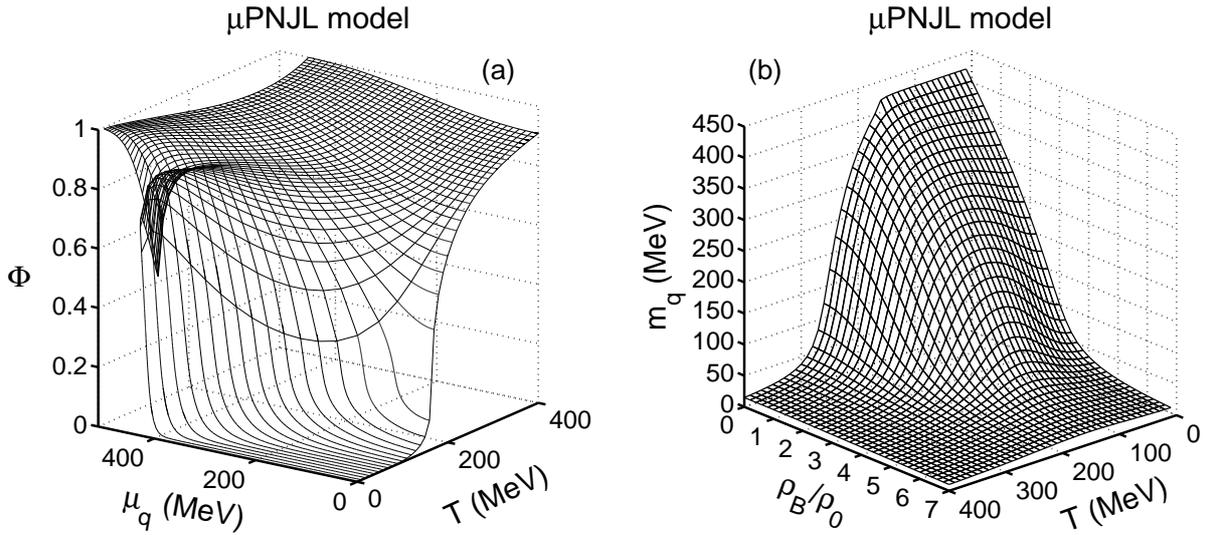

**Figure 8.** (a) Polyakov field $\Phi$ according to the chemical potential $\mu_q$ and the temperature $T$. (b) Masses of the light quarks $m_q$ in the $\rho_B, T$ plane, in the $\mu$PNJL description.

Concerning the evolution of $\Phi$ in the $T, \mu_q$ plane, Figure 8(a), two distinct regions are identifiable: a zone where $\Phi \approx 0$ ("confined" phase) at low $T$ and $\mu_q$, and another one where $\Phi \approx 1$ ("deconfined" phase) at high $T$ and/or at high $\mu_q$. At low $\mu_q$, the transition between these two regimes is still performed by a crossover, as in the PNJL model. However, a brutal transition is now observed according to $\mu_q$, at low $T$. The amplitude of the structure identified Figure 5(a) strongly increases [76]. In the $\mu$PNJL model, this one is interpreted as a first order phase transition of $\Phi$ according to $\mu_q$. The literature names it the "deconfinement" transition. It seems to coincide with the first order phase transition of $\langle\langle \bar{\psi}_q \psi_q \rangle\rangle$ [78, 80], whose position at low $T$ is approximately unchanged compared to NJL results. Indeed, the behavior of the other observables, as $m_q$ or $\langle\langle \bar{\psi}_q \psi_q \rangle\rangle$, is very comparable to the one observed in the NJL model, Figure 8(b). The position of the CEP in the $\mu$PNJL model is close to the NJL one: same $\rho_B$ and $T \approx 76.5$ MeV with the P1 parameter set. From a physical point of view, the PNJL results suggest the existence of a *quarkyonic* phase [110, 115, 116], in which the quark matter is still confined ($\Phi \approx 0$) and the chiral symmetry is restored. The $\mu$PNJL model seems to contradict this theory, since it indicates the disappearance or a strong diminution of this phase [77, 80].

**4. Study of the 2SC phase in the framework of the isospin symmetry**

## 4.1 ΔNJL equations

In the 2SC phase, the energy gap $\Delta_{ud}$ is non-null. The matrix $\tilde{S}^{-1}$ is no longer block diagonal, so (42) is not applicable. But, with the method described in [85], I obtain

$$\det(\tilde{S}^{-1}) = \left\{\det\left[\slashed{p} + \gamma_0 \mu_q - m_q + \delta(\slashed{p} - \gamma_0 \mu_q - m_q)\right]\right\}^4$$
$$\times \left[\det(\slashed{p} + \gamma_0 \mu_q - m_q)\right]^2 \left[\det(\slashed{p} - \gamma_0 \mu_q - m_q)\right]^6 \quad , \tag{55}$$
$$\times \left[\det(\slashed{p} + \gamma_0 \mu_s - m_s)\right]^3 \left[\det(\slashed{p} - \gamma_0 \mu_s - m_s)\right]^3$$

where the $\delta$ term is defined as

$$\delta = \frac{-|\Delta_{ud}|^2}{(i\omega_n - \mu_q)^2 - E_q^2}. \tag{56}$$

The evaluation of the first determinant in (55) is done thanks to the relation [85]

$$\det\left[\slashed{p} + \gamma_0 \mu_q - m_q + \delta(\slashed{p} - \gamma_0 \mu_q - m_q)\right] = \left\{\frac{\left[(i\omega_n)^2 - (E_q^-)^2\right]\left[(i\omega_n)^2 - (E_q^+)^2\right]}{(i\omega_n - \mu_q - E_q)(i\omega_n - \mu_q + E_q)}\right\}^2, \tag{57}$$

where $E_q^{\pm} = \sqrt{(E_q \pm \mu_q)^2 + |\Delta_{ud}|^2}$. The other determinants are estimated with (44). As a consequence, $\ln\left[\det(\tilde{S}^{-1}/T)\right]$ is then writable as a sum of terms whose structure is $(\omega_n)^2 + \lambda_k^2$. It implies that (23) is usable to perform the summation over $n$. After few manipulations, I obtain the result [16, 19]

$$\Omega_M = -2 \int \frac{d^3 p}{(2\pi)^3} \left\{ 2E_q^- + 2E_q^+ + 4T\ln\left[1 + \exp\left(-\frac{E_q^-}{T}\right)\right] + 4T\ln\left[1 + \exp\left(-\frac{E_q^+}{T}\right)\right] \right.$$
$$+ 2E_q + 2T\ln\left[1 + \exp\left(-\frac{E_f - \mu_f}{T}\right)\right] + 2T\ln\left[1 + \exp\left(-\frac{E_f + \mu_f}{T}\right)\right] \tag{58}$$
$$\left. + 3E_s + 3T\ln\left[1 + \exp\left(-\frac{E_s - \mu_s}{T}\right)\right] + 3T\ln\left[1 + \exp\left(-\frac{E_s + \mu_s}{T}\right)\right] \right\}$$

In this relation, the second line corresponds to the blue $q$ quarks. These ones are unpaired as indicated by (17) if one takes $\Delta_{qs} = 0$. Indeed, I do not include quark/quark pairs symmetric in color in this work, because of the expected weakness of this spin-1 pairing [19, 20, 46]. The strange quarks associated with the third line of (58) are also unpaired. Since these $s$ quarks are included in this modeling, one could speak about "2SC+s" phase instead of 2SC. As with the other expressions of $\Omega_M$ presented in this paper, a possible test to verify the calculation is to check if the total sum of the coefficients placed in front of the zero-point energy terms [99] (including the $-2$ in front of the integral) is equal to $-2N_f N_c$. In the same way, the sum obtained with the coefficients in front of the $T\ln(\ )$ terms should correspond to $-4N_f N_c$. With the same methods as in the section 2, I find

$$\langle\langle\bar{\psi}_q \psi_q\rangle\rangle = \frac{1}{2}(\langle\langle\bar{\psi}_u \psi_u\rangle\rangle + \langle\langle\bar{\psi}_d \psi_d\rangle\rangle)$$
$$= -2 \int \frac{d^3 p}{(2\pi)^3} \frac{m_q}{E_q} \left\{ \frac{E_q - \mu_q}{E_q^-}\left[1 - 2f(E_q^-)\right] + \frac{E_q + \mu_q}{E_q^+}\left[1 - 2f(E_q^+)\right] \right. \tag{59}$$
$$\left. + 1 - f(E_q - \mu_q) - f(E_q + \mu_q) \right\}$$

and

$$\langle\langle\psi_q^+ \psi_q \rangle\rangle = 2\int \frac{d^3p}{(2\pi)^3}\left\{\frac{E_q - \mu_q}{E_q^-}\left[2f(E_q^-)-1\right] + f(E_q - \mu_q) \right.$$
$$\left. - \frac{E_q + \mu_q}{E_q^+}\left[2f(E_q^+)-1\right] - f(E_q + \mu_q)\right\}. \tag{60}$$

The associated expressions for the strange quarks correspond to (48) and (49). Furthermore, the application of (33) or (36) allows establishing a gap equation for $\Delta_{ud}$, written in the form [16]

$$1 = 4G_{DIQ}\left|\int \frac{d^3p}{(2\pi)^3}\frac{1}{E_q^-}\left[1-2f(E_q^-)\right] + \frac{1}{E_q^+}\left[1-2f(E_q^+)\right]\right|. \tag{61}$$

So, in the framework of the $\Delta$ NJL model devoted to describe the 2SC phase, it is required to solve the system of equations formed by (1) and (61).

*4.2 $\Delta$ PNJL and $\Delta\mu$ PNJL equations*

When the Polyakov loop is added in the modeling performed in the subsection 4.1, (55) is rewritten as

$$\det(\tilde{S}^{-1}) = \left\{\det\left[\slashed{p} + \gamma_0\mu_{q,1} - m_q + \delta_2\left(\slashed{p} - \gamma_0\mu_{q,2} - m_q\right)\right]\right\}^2$$
$$\times\left\{\det\left[\slashed{p} + \gamma_0\mu_{q,2} - m_q + \delta_1\left(\slashed{p} - \gamma_0\mu_{q,1} - m_q\right)\right]\right\}^2$$
$$\times\left[\det\left(\slashed{p} + \gamma_0\mu_{q,3} - m_q\right)\right]^2 \prod_{j=1}^{3}\left[\det\left(\slashed{p} - \gamma_0\mu_{q,j} - m_q\right)\right]^2 \tag{62}$$
$$\times\prod_{j=1}^{3}\left[\det\left(\slashed{p} + \gamma_0\mu_{s,j} - m_s\right)\det\left(\slashed{p} - \gamma_0\mu_{s,j} - m_s\right)\right]$$

with $\mu_{f,j} = \mu_f - iA_{4(jj)}$. The $\delta$ term is updated to take into account the components of the matrix $A_4$,

$$\delta_j = \frac{-|\Delta_{ud}|^2}{(i\omega_n - \mu_{q,j})^2 - E_q^2}. \tag{63}$$

The evaluation of the two first determinants in (62) is performed with the formula

$$\det\left[\slashed{p} + \gamma_0\mu_{q,1} - m_q + \delta_2\left(\slashed{p} - \gamma_0\mu_{q,2} - m_q\right)\right]\det\left[\slashed{p} + \gamma_0\mu_{q,2} - m_q + \delta_1\left(\slashed{p} - \gamma_0\mu_{q,1} - m_q\right)\right]$$
$$= \left\{\frac{\left[(i\omega_n)^2 - (E_q^{--})^2\right]\left[(i\omega_n)^2 - (E_q^{-+})^2\right]\left[(i\omega_n)^2 - (E_q^{+-})^2\right]\left[(i\omega_n)^2 - (E_q^{++})^2\right]}{(i\omega_n - \mu_{q,1} - E_q)(i\omega_n - \mu_{q,1} + E_q)(i\omega_n - \mu_{q,2} - E_q)(i\omega_n - \mu_{q,2} + E_q)}\right\}^2, \tag{64}$$

in which

$$E_q^{+\pm} = \bar{E}_q^+ \pm \frac{iA_{4(11)} - iA_{4(22)}}{2}$$
$$E_q^{-\pm} = \bar{E}_q^- \pm \frac{iA_{4(11)} - iA_{4(22)}}{2}, \quad \text{with } \bar{E}_q^{\pm} = \sqrt{\left(E_q \pm \frac{\mu_{q,1} + \mu_{q,2}}{2}\right)^2 + |\Delta_{ud}|^2}. \tag{65}$$

After the summation of the Matsubara frequencies with (23), I find

$$\Omega_M = -2\int \frac{d^3p}{(2\pi)^3}\left\{2\bar{E}_q^- + 2T\ln\left[1+\exp\left(-\frac{E_q^{-+}}{T}\right)\right] + 2T\ln\left[1+\exp\left(-\frac{E_q^{--}}{T}\right)\right]\right.$$

$$+2\bar{E}_q^+ + 2T\ln\left[1+\exp\left(-\frac{E_q^{++}}{T}\right)\right] + 2T\ln\left[1+\exp\left(-\frac{E_q^{+-}}{T}\right)\right]$$

$$+2E_q + 2T\ln\left[1+\exp\left(-\frac{E_f - \mu_{f,3}}{T}\right)\right] + 2T\ln\left[1+\exp\left(-\frac{E_f + \mu_{f,3}}{T}\right)\right]$$

$$\left.+\sum_{j=1}^{N_c} E_s + T\ln\left[1+\exp\left(-\frac{E_s - \mu_{s,j}}{T}\right)\right] + T\ln\left[1+\exp\left(-\frac{E_s + \mu_{s,j}}{T}\right)\right]\right\}$$ (66)

If (9) is employed in (66), the writing of $\Omega_M$ proposed in [59, 93] is obtained. As explained in these references and upstream, we are unable in the $\Delta(\mu)$PNJL models to form expressions that depend only on $\Phi$ and $\bar{\Phi}$ (and not on $\phi_3$ and $\phi_8$). So, the "modified" Fermi-Dirac distributions $f_\Phi^\pm$, (53) and (54), are usable in the $\Delta(\mu)$PNJL approaches to describe the $N_c$ colors of the strange quarks/antiquarks, but not the $q$ quarks. Also, $\Phi$, $\bar{\Phi}$, $f_\Phi^\pm$ and $\Omega_M$ become complex. One saw that it imposes to take the real part of the found results, including the ones of the $s$ quarks. I propose

$$\langle\langle\bar{\psi}_q\psi_q\rangle\rangle = -2\,\text{Re}\left(\int \frac{d^3p}{(2\pi)^3}\frac{m_q}{E_q}\left\{\frac{E_q - 1/2(\mu_{q,1}+\mu_{q,2})}{\bar{E}_q^-}\left[1 - f\left(E_q^{-+}\right) - f\left(E_q^{--}\right)\right]\right.\right.$$
$$\left.\left.+\frac{E_q + 1/2(\mu_{q,1}+\mu_{q,2})}{\bar{E}_q^+}\left[1 - f\left(E_q^{++}\right) - f\left(E_q^{+-}\right)\right] + 1 - f\left(E_q - \mu_{q,3}\right) - f\left(E_q + \mu_{q,3}\right)\right\}\right)$$ (67)

and

$$\langle\langle\psi_q^+\psi_q\rangle\rangle = 2\,\text{Re}\left(\int \frac{d^3p}{(2\pi)^3}\left\{\frac{E_q - 1/2(\mu_{q,1}+\mu_{q,2})}{\bar{E}_q^-}\left[f\left(E_q^{-+}\right) + f\left(E_q^{--}\right) - 1\right] + f\left(E_q - \mu_{q,3}\right)\right.\right.$$
$$\left.\left.-\frac{E_q + 1/2(\mu_{q,1}+\mu_{q,2})}{\bar{E}_q^+}\left[f\left(E_q^{++}\right) + f\left(E_q^{+-}\right) - 1\right] - f\left(E_q + \mu_{q,3}\right)\right\}\right)$$ (68)

The equations related to the strange quarks correspond to (48) and (49) with the replacements $f(E_f - \mu_f) \to f_\Phi^+(E_f - \mu_f)$ and $f(E_f + \mu_f) \to f_\Phi^-(E_f + \mu_f)$ proposed in the subsection 3.2. Also, the gap equation for $\Delta_{ud}$ becomes

$$1 = 4G_{DIQ}\left|\text{Re}\left\{\int \frac{d^3p}{(2\pi)^3}\frac{1}{\bar{E}_q^-}\left[1 - f\left(E_q^{-+}\right) - f\left(E_q^{--}\right)\right] + \frac{1}{\bar{E}_q^+}\left[1 - f\left(E_q^{++}\right) - f\left(E_q^{+-}\right)\right]\right\}\right|.$$ (69)

*4.3 Results*

The results that describe the 2SC phase are presented in the Figure 9 to Figure 16. These ones are shared between this subsection and the next one, which focuses more specifically on the behavior of $\Phi$. The calculations were performed with the P1 parameter set. The strangeness density is fixed to zero. In the Figure 9(a), the discontinuity of $m_q$ corresponds to the restoration of the chiral symmetry, via the first order phase transition of $\langle\langle\bar{\psi}_q\psi_q\rangle\rangle$, at $\mu_q \approx 394$ MeV in the $\Delta$NJL model. Moreover, $\Delta_{ud}$ is an order parameter of the transition between the normal quark ($\Delta_{ud}=0$) and 2SC ($\Delta_{ud}\neq 0$) phases. At low $T$, its discontinuity indicates that the NQ/2SC transition is a first order one according to

$\mu_q$, Figure 9(b). This phase transition and the chiral one seem to intervene at the same $\mu_q$, as interpreted, e.g., in [19, 46, 84, 86]. Since the chiral and the diquark condensates involve the same quarks, there is competition between them. Furthermore, the quark-quark pairing is relatively weak in comparison with the chiral condensate. Consequently, a scenario imagines that $\Delta_{ud}$ can be non-null only if $\langle\langle \bar{\psi}_q \psi_q \rangle\rangle$ is enough reduced. So, at low $T$, the 2SC phase could be present if the chiral symmetry is restored. However, a limitation of this scenario was shown in [52]. This reference indicates the presence of a *coexisting* phase, in which $\Delta_{ud}$ presents low (but non-null) values at very low temperatures, just before the restoration of the chiral symmetry. This behavior is not visible in the Figure 9 because of the used parameter set, which leads to consider rather high $m_q$ values. If these ones are lowered, the coexisting phase appears, as explained in [52]. I verified this affirmation with the replacement $G\Lambda^2 = 1.835$ in the P1 parameter set. When the Polyakov loop is added, all the descriptions performed above in this paragraph are still valid at low $T$, Figure 16(a). For higher $T$, the effect of the temperature is to weaken the quark-quark pairing: $\Delta_{ud}$ continuously decreases to zero. It reveals a second order phase transition [86], which intervenes when $T > 40$ MeV in the $\Delta$ NJL model.

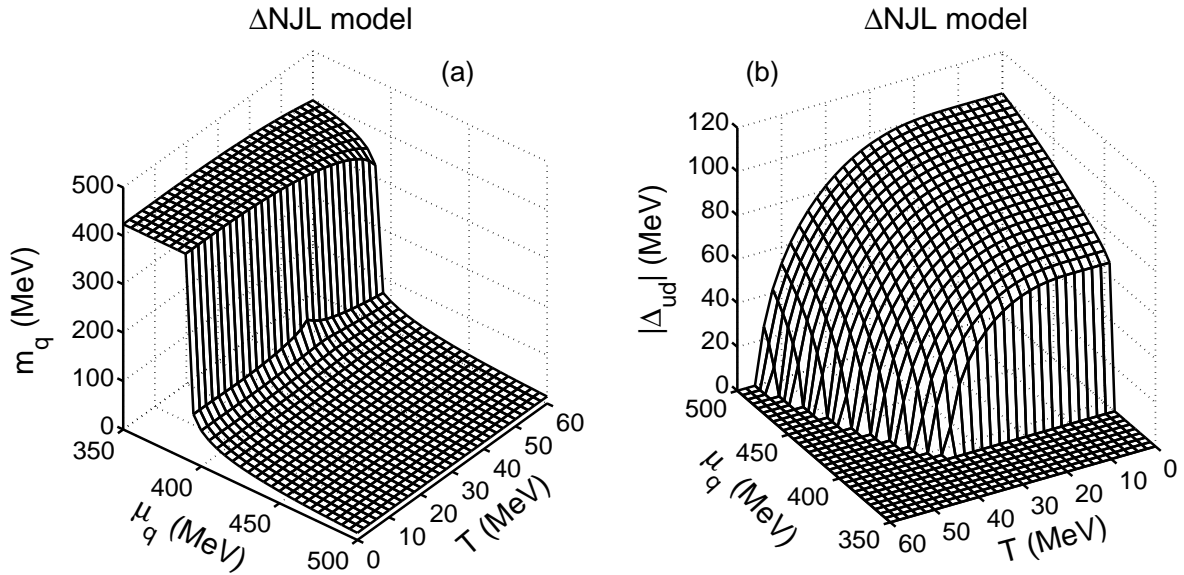

**Figure 9.** (a) Mass of the light quarks and (b) gap $\Delta_{ud}$ according to $\mu_q$ and $T$.

In the Figure 10(a), $\Delta_{ud}$ is studied in the $T, \rho_B$ plane. It reveals that non-null values of this gap are found at very low density, even lower than the standard nuclear density, in agreement with [52]. At low $T$, a wide range of low densities corresponds to metastable/unstable states, subsection 3.4. In the interpretation of [46], the system evolves towards the formation of droplets. In these ones, the chiral symmetry is restored. Consequently, the chiral condensate cannot prevent the quark-quark pairing in the droplets. It explains this result, which finally appears as a manifestation of the lack of confinement in the $\Delta$ NJL model[4]. Indeed, as recalled in [47], the confinement should strongly favor ordinary matter over quark-quark pairs. Concretely, its effect could be, e.g., to impose the coupling of the scalar quark-quark pairs with the unpaired blue quarks, in order to form baryons...

---

[4] The $\Delta$ PNJL and $\Delta\mu$ PNJL results described hereafter confirm that the mechanism of confinement simulated by the inclusion of the Polyakov loop does not modify this phenomenon. So, when [25] asks if the confinement preludes the color superconductivity, it should concern the *real* confinement, not this confinement mechanism.

The data obtained according to $T$ and $\rho_B$ are used to study the metastable/unstable states in the $T, \mu_q$ plane, Figure 10(b). In this one, the surface of the Figure 9(b) is superimposed. When $\partial |\Delta_{ud}|/\partial \mu_q > 0$, i.e. here for stable states, the two surfaces strictly coincide. It notably includes the second order phase transition. On the other hand, instead of the discontinuity, a deformation associated with the metastable/unstable states is visible. In the $T, \mu_q$ plane, the metastable zone involving non-null gap values, located at $\mu_q \approx 394$ MeV, appears to be reduced. At the opposite, the unstable zone, for which $\partial |\Delta_{ud}|/\partial \mu_q < 0$, is more extended. In the other figures, the stability criterion $\partial^2 \Omega / \partial \langle\langle \bar{\psi}_q \psi_q \rangle\rangle^2 > 0$ is employed to draw this unstable zone in the $T, \rho_B$ plane, as in [62, 86, 114]. In order to not weight down the graphs, the metastable zones are not represented. At low $T$, the density range of the unstable zone is comparable to the one found subsection 3.4. It reveals the modest influence of $\Delta_{ud}$, in agreement with [46]. For example in Figure 11, it leads to a slight deformation of the limit of the unstable zone when $\rho_B \approx 0.5 \rho_0$, and it shifts the highest unstable densities from $3.0 \rho_0$ to $3.1 \rho_0$, except in the $\Delta \mu$ PNJL model. Stable states are found at $T = 0$ when $\rho_B > 3.5 \rho_0$.

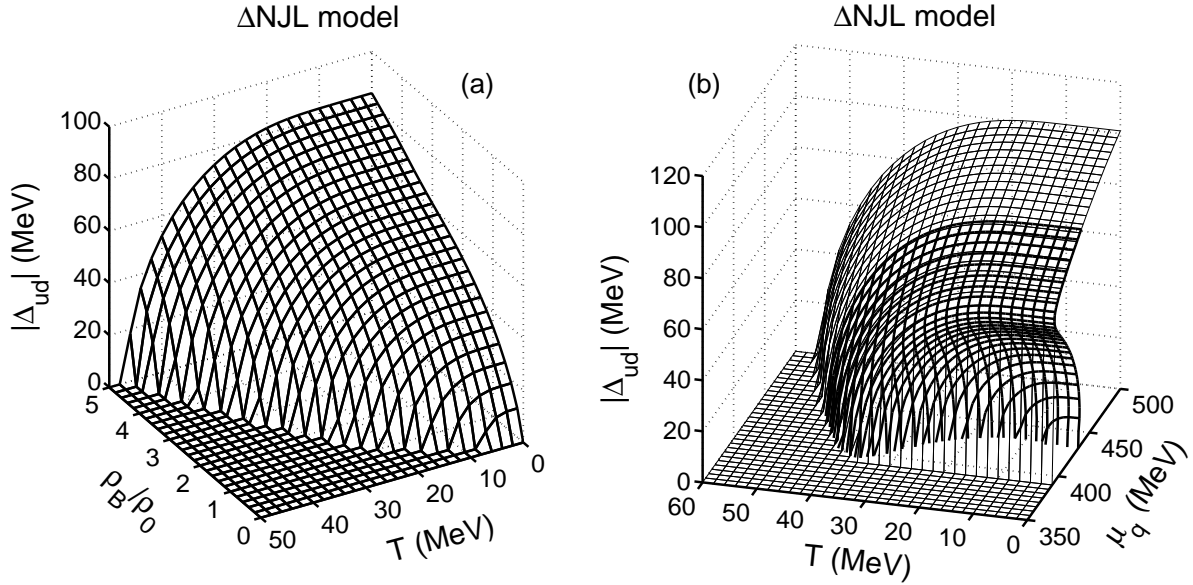

**Figure 10.** Gap $\Delta_{ud}$ in the (a) $\rho_B$, $T$ plane and in the (b) $\mu_q$, $T$ plane. In the graph (b), the surface plotted with the thick black lines was found with a calculation that uses the temperature and the densities as input parameters. The other surface is the one of the Figure 9(b).

The Figure 11 focuses on the values of $\Delta_{ud}$ in the $T, \rho_B$ plane, with or without the Polyakov loop. $\Delta_{ud}$ continuously increases when $\rho_B$ is growing and/or when $T$ is decreasing. In the $\Delta$ NJL, $\Delta$ PNJL and $\Delta \mu$ PNJL models, $\Delta_{ud} \approx 84$ MeV at $T = 0$ and $\rho_B \approx 5\rho_0$. As with $m_q$ in the Figure 1, the inclusion of the Polyakov loop in the $\Delta$ PNJL description leads to a non-negligible shifting of the $\Delta_{ud}$ graph towards higher temperatures. At the opposite, the $\Delta \mu$ PNJL results are rather close to the $\Delta$ NJL ones. In both cases, the values of $\Delta_{ud}$ are not modified by the loop, but shifted. At $\rho_B = 5\rho_0$, the critical temperature $T_\Delta$ of the 2SC phase is about 47 MeV in the $\Delta$ NJL and $\Delta \mu$ PNJL models, versus 63 MeV with $\Delta$ PNJL. Consequently, the BCS relation $T_\Delta = 0.57 \Delta_{T=0}$ [14, 19, 20, 86] is well verified in the $\Delta$ NJL and $\Delta \mu$ PNJL descriptions. In the graph, this remark is particularly true when $\rho_B \geq 3\rho_0$. This agreement was reported in [14] for the $\Delta$ NJL model. On the other hand, the Polyakov loop potential (3) without the corrections (5) seems to affect the behavior of the *ud* pair according to the temperature, via the observed shifting of $T_\Delta$ [25, 60].

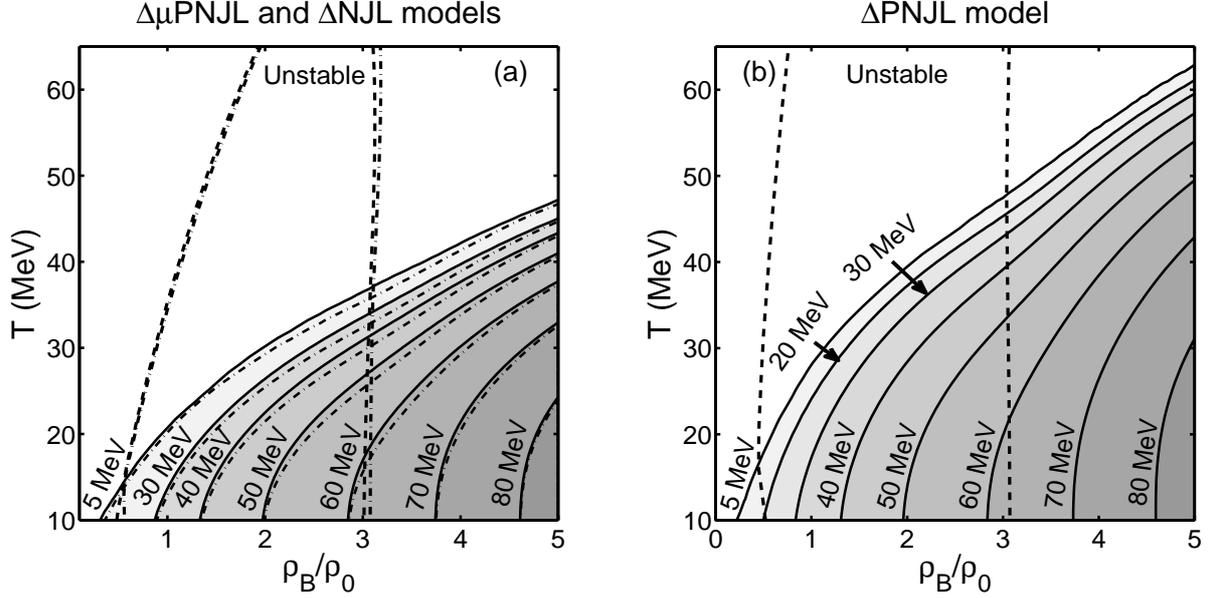

**Figure 11.** Gap $\Delta_{ud}$ in the $\rho_B, T$ plane, in the (a) $\Delta\mu$ PNJL (solid curves), $\Delta$ NJL (dash-dotted curves) and in the (b) $\Delta$ PNJL models. The unstable zone is delimited by dashed ($\Delta(\mu)$PNJL) or dash-dotted ($\Delta$ NJL) curves

We now investigate the effects of the color superconductivity on some observables. Firstly, it leads to an increase in the mass $m_q$ of the light quarks, in the domain visible in the Figure 12. The maximum mass gap is slightly upper than 12 MeV, at low $T$ and $\rho_B \approx 3.1\rho_0$, i.e. the upper limit of the unstable zone. In these conditions, $m_q$ exceeds 200 MeV. At higher $\rho_B$, because of the restoration of chiral symmetry, $m_q$ presents a strong diminution. It leads to the observed decrease in the mass gap. As a whole, the effect of the color superconductivity on the quark masses stay rather modest, even at very low $T$ and high $\rho_B$. It does not seem to contradict (P)NJL works that did not include the color superconductivity: the Figure 1 would be very slightly modified by the effect of the 2SC phase. Moreover, because of the strong link between $m_q$ and $\langle\langle\bar\psi_q\psi_q\rangle\rangle$, first line of (1), this result also indicates that $\Delta_{ud}$ has a limited influence on this chiral condensate, equations (59) and (67).

In the same manner, the Figure 13 shows the influence of the 2SC phase on the chemical potential $\mu_q$, which is found when one solves the system of equations that includes (1). The color superconductivity tends to decrease the values of $\mu_q$. This decrease is maximum at low $T$ and at densities located in the stable zone, i.e. when $\rho_B > 3.5\rho_0$. However, as with $m_q$, the effect appears to be rather reduced. With (30), it allows saying that $\Delta_{ud}$ has a weak influence on $\langle\langle\psi_q^+\psi_q\rangle\rangle$ in (60) and (68). The corrections to be done to the Figure 2(a) seem to be modest. This conclusion is also applicable to the graphs performed in the $T, \mu_q$ plane. For example, at low $T$, the critical $\mu_q$ of the chiral phase transition is only reduced by a few MeV, Figure 4 and Figure 9(a). This remark is confirmed by the Figure 16(a).

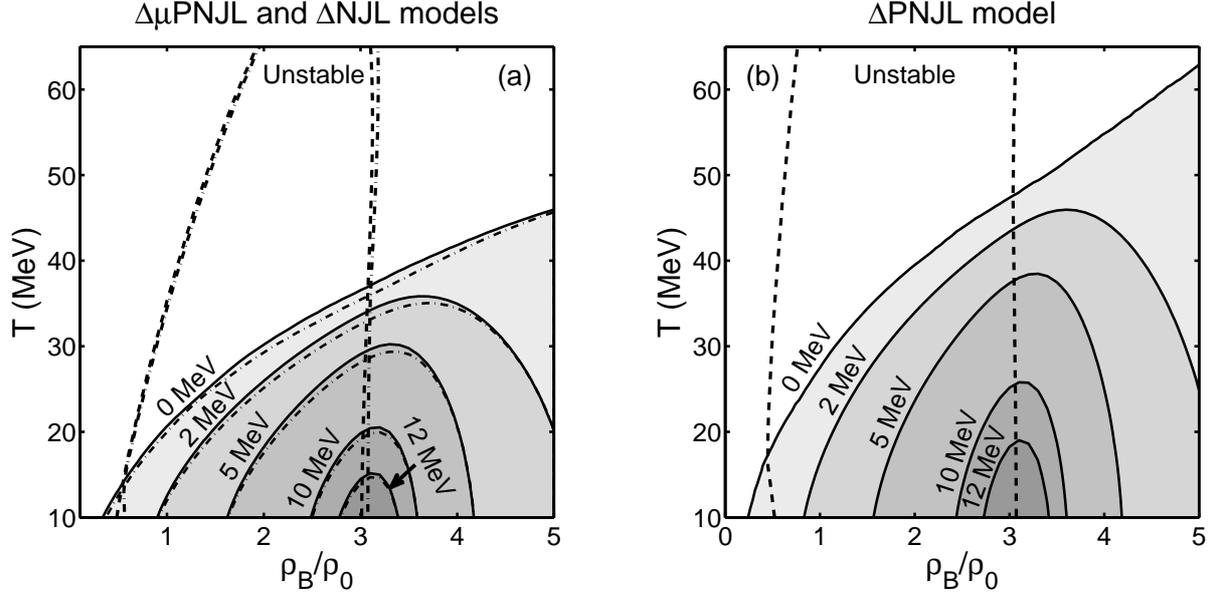

**Figure 12.** Increase in the mass $m_q$ due to the inclusion of the color superconductivity in the (a) $\Delta\mu$ PNJL (solid), $\Delta$ NJL (dash-dotted) and (b) $\Delta$ PNJL models.

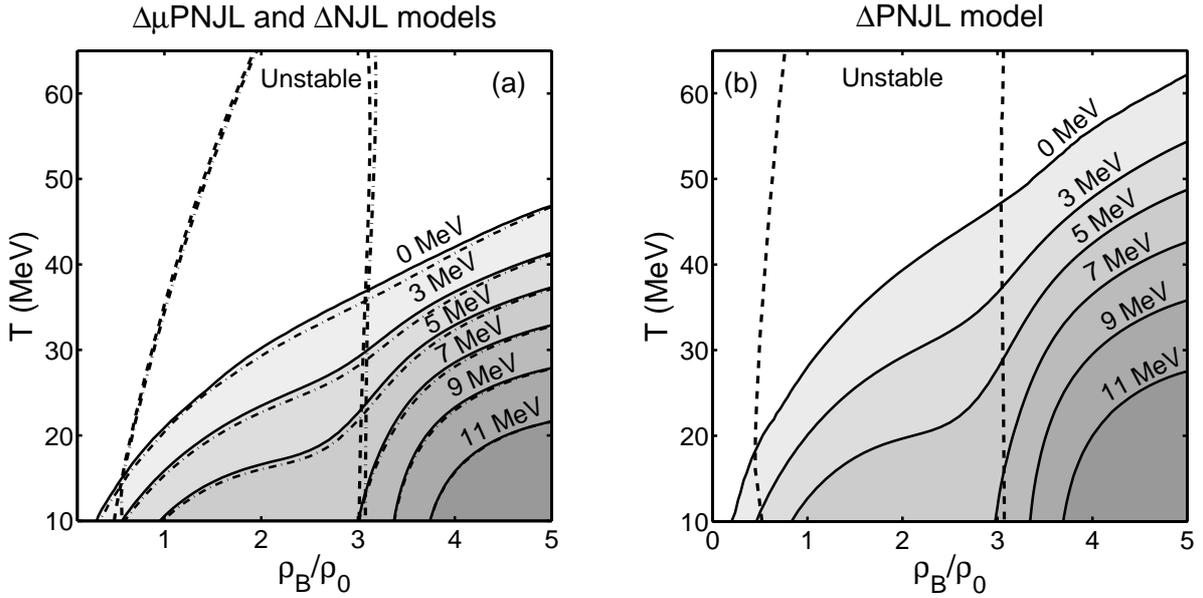

**Figure 13.** Decrease in the chemical potential $\mu_q$ caused by the inclusion of the color superconductivity in the (a) $\Delta\mu$ PNJL (solid), $\Delta$ NJL (dash-dotted) and (b) $\Delta$ PNJL models.

*4.4 Evolution of $\Phi$ in the 2SC phase*

As indicated in [60], the method that is used to estimate $\Phi$ and its conjugate $\bar{\Phi}$ differs in the standard PNJL model [61, 63] and in the $\Delta$ PNJL version that includes color superconductivity [59, 60]. In the first case (37) is employed, versus (40) for the second approach. These two methods are compared hereafter, for the PNJL / $\Delta$ PNJL models and for the $\mu$ PNJL / $\Delta\mu$ PNJL ones. The calculations were performed with (37) and with (40) while imposing $\Delta_{ud} = 0$. The differences are plotted in the Figure 14. In (37), $\Phi$ and $\bar{\Phi}$ are treated as real and independent variables. This approximation leads to an overestimation of $\Phi$ and an overestimation of the difference between $\Phi$ and $\bar{\Phi}$ [60]. The consequences of this simplification appear to be limited, at least in the studied zone. For example, at $T = 60$ MeV, $\Phi \approx 0.04$ in the Figure 3(a), whereas the differences visible in the Figure 14(b) are close to 0.006 or 0.007. In the $\mu$PNJL/$\Delta\mu$PNJL versions, since higher values of $\Phi$ are obtained, the differences are highly negligible, except at low densities, Figure 7(a) and Figure 14(a).

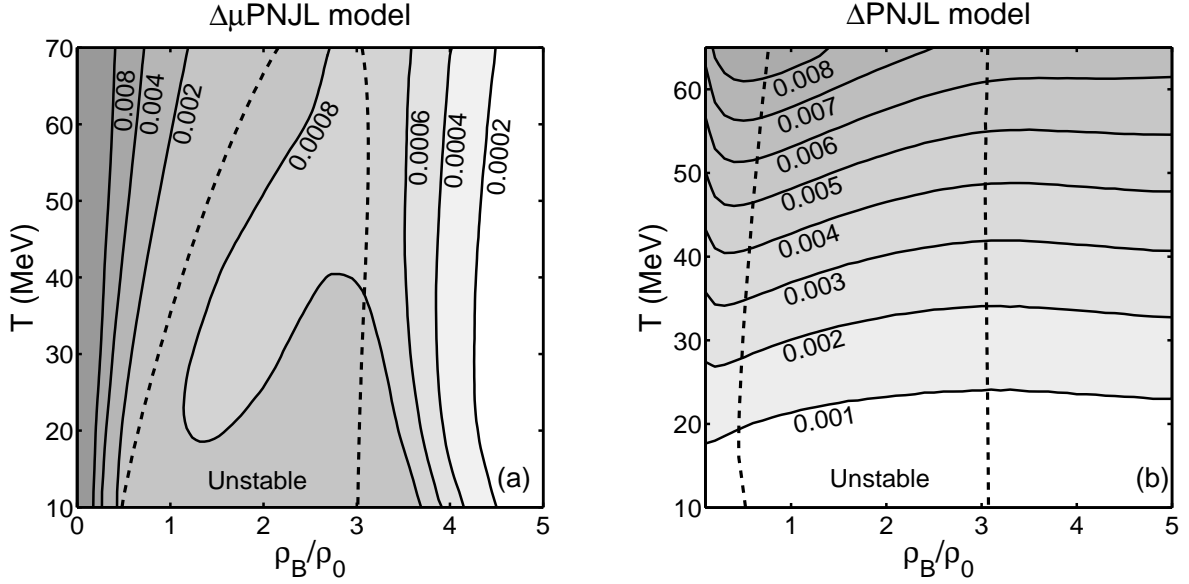

**Figure 14.** (a) Difference between the Polyakov field $\Phi$ found in the $\mu$PNJL model and $\Phi$ obtained in the $\Delta\mu$PNJL model when $\Delta_{ud} = 0$ is imposed. (b) Idem when the $\mu$-dependence in the Polyakov loop potential is not used (PNJL and $\Delta$PNJL models when $\Delta_{ud} = 0$ is imposed).

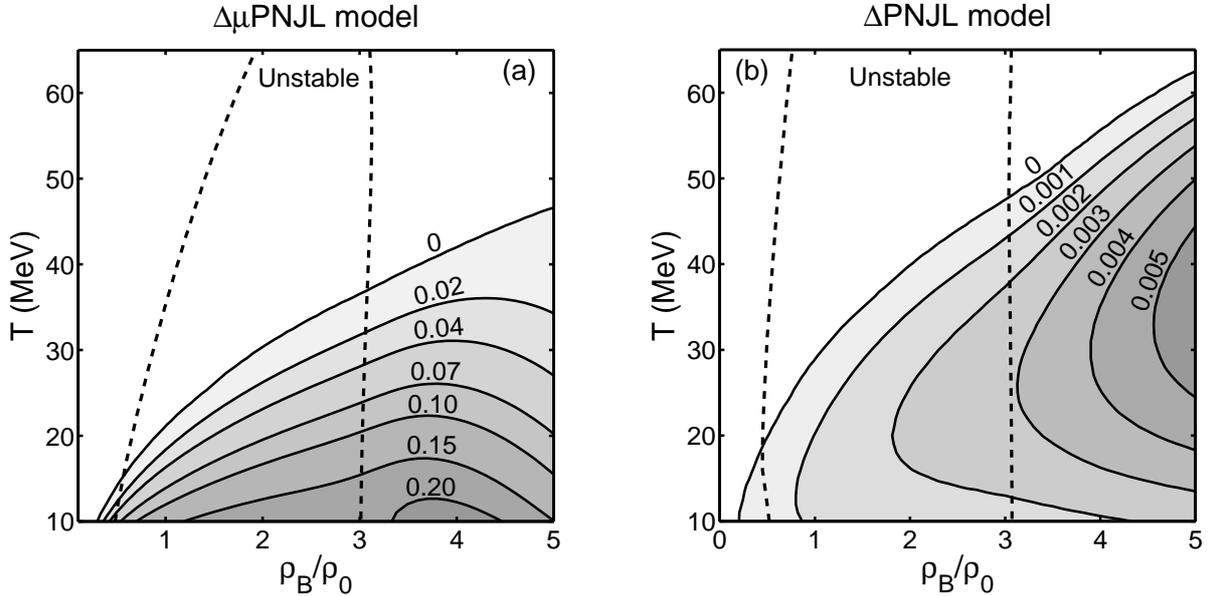

**Figure 15.** Decrease in $\text{Re}(\Phi)$ caused by the 2SC phase, in the (a) $\Delta\mu$PNJL and (b) $\Delta$PNJL models.

Then, one investigates the effect of the 2SC phase on $\Phi$. The calculations were done with (40), allowing or not $\Delta_{ud}$ to be non-null. With this equation, $\Phi$ is real only when $\Delta_{ud} = 0$. So, $\Phi$ is complex in the 2SC phase, Figure 16(a). In the $\Delta$PNJL and $\Delta\mu$PNJL models, the 2SC phase leads to a decrease in the real part of $\Phi$, Figure 15. However, the amplitude of this decrease is not comparable for these two approaches. In the $\Delta$ PNJL model, the effect of the color superconductivity on $\Phi$ is weak, like the one of the approximation described upstream, Figure 14(b). Its effect increases when $\rho_B$ is growing, when $T \approx 20-40$ MeV. In this description, its influence is more limited at low $T$, because $\Phi \approx 0$ in this zone. At the opposite, in the $\Delta\mu$ PNJL model, the differences are more important, especially at very low $T$ and at densities around $4\rho_B$, i.e. after the chiral restoration. In these conditions, the $\text{Re}(\Phi)$ gap exceeds 0.2, whereas $\text{Re}(\Phi) \approx 0.35-0.65$ in this region of the 2SC phase. Similar results are found with $|\Phi|$ because $\text{Re}(\Phi) \gg \text{Im}(\Phi)$ in all the $\Delta_{ud} \neq 0$ regime.

In the Figure 16(a), the chiral condensate $\langle\langle\bar{\psi}_q\psi_q\rangle\rangle$, the polyakov field $\Phi$ and the energy gap $\Delta_{ud}$ are studied according to $\mu_q$, at a fixed temperature, in the $\Delta\mu$ PNJL model. This figure confirms that the 2SC phase leads to shift the critical $\mu_q$ of the first order phase transitions that rules the chiral restoration and the "deconfinement" transition from 400 MeV to 394 MeV with the used parameter set. Furthermore, the first order NQ/2SC phase transition coincides with them. In other words, these three transitions occur simultaneously at $\mu_q \approx 394$ MeV. This result is extendable to the low $T$ regime. Consequently, the color superconductivity does not seem to preclude the disappearance of the quarkyonic phase in the $\Delta\mu$ PNJL description. This aspect contradicts scenarios based on a $\mu$-independent Polyakov loop potential [25], i.e. $\Delta$ PNJL descriptions.

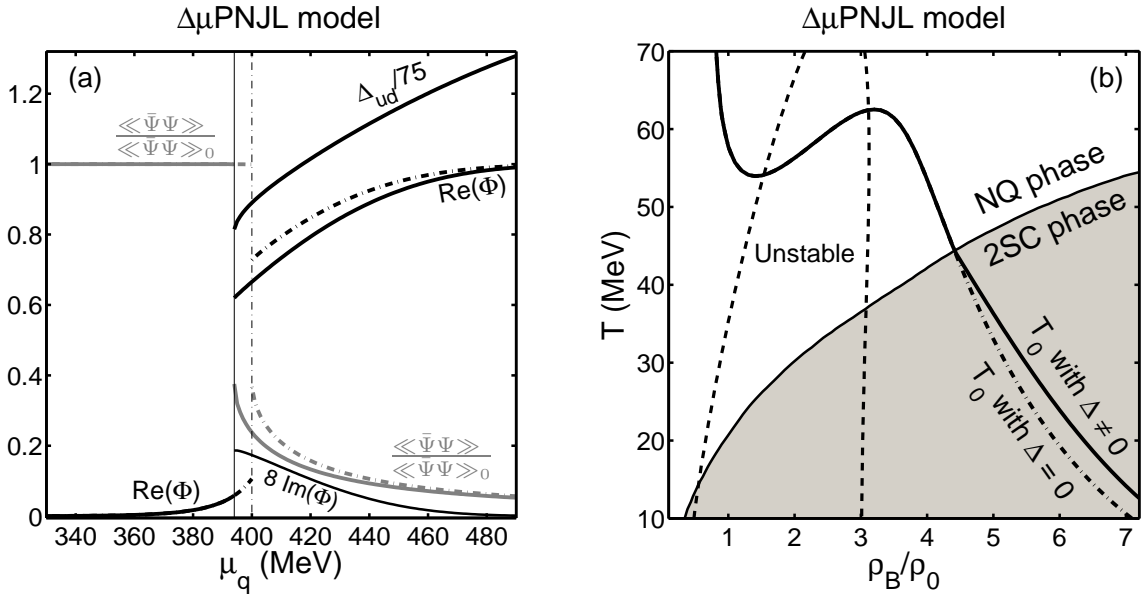

**Figure 16.** (a) Evolution of the 3 order parameters according to $\mu_q$, at $T = 20$ MeV. The dash-dotted curves correspond to the results found with the constraint $\Delta_{ud} = 0$. (b) Effect of the 2SC phase on the temperature $T_0$ in the $\rho_B, T$ plane: the solid ($\Delta_{uq} \neq 0$) and dash-dotted ($\Delta_{uq} = 0$) curves indicate when $T = T_0$.

Moreover, the decrease in $\Phi$ due to $\Delta_{ud}$ is strong just after the chiral restoration, Figure 16(a). With the results of the Figure 15(a), it allows indicating that this effect is growing when $T$ decreases. At the opposite, the increase in $\Delta_{ud}$ at high $\mu_q$ does not prevent $\Phi$ to reach its limit value: the two $\text{Re}(\Phi)$ curves (with or without the 2SC phase) converge towards 1. These results are complemented by a study of the transition temperature $T_0$ in the $\rho_B, T$ plane, Figure 16(b). The curves $T = T_0$ show that $T_0$ is shifted towards higher $\rho_B$ and higher $T$ when the 2SC phase is included in the modeling. This result is in agreement with the behavior of $\Phi$ observed Figure 15(a), cf. subsection 3.5.

## 5. The 2SC phase beyond the isospin symmetry

*5.1 Equations*

To study the 2SC phase in the $\Delta$ NJL approach without the isospin symmetry, i.e. $m_u \neq m_d$, I use again the method of [85], and I find

$$\det(\tilde{S}^{-1}) = \left\{\det\left[\not{p} + \gamma_0 \mu_u - m_u + \delta_d\left(\not{p} - \gamma_0 \mu_d - m_d\right)\right]\right\}^2$$
$$\times \left\{\det\left[\not{p} + \gamma_0 \mu_d - m_d + \delta_u\left(\not{p} - \gamma_0 \mu_u - m_u\right)\right]\right\}^2$$
$$\times \det\left(\not{p} + \gamma_0 \mu_u - m_u\right)\left[\det\left(\not{p} - \gamma_0 \mu_u - m_u\right)\right]^3 \det\left(\not{p} + \gamma_0 \mu_d - m_d\right)\left[\det\left(\not{p} - \gamma_0 \mu_d - m_d\right)\right]^3 \quad , \quad (70)$$
$$\times \left[\det\left(\not{p} + \gamma_0 \mu_s - m_s\right)\right]^3 \left[\det\left(\not{p} - \gamma_0 \mu_s - m_s\right)\right]^3$$

where I define

$$\delta_f = \frac{-|\Delta_{ud}|^2}{\left(i\omega_n - \mu_f\right)^2 - E_f^2} \ . \tag{71}$$

Thanks to the calculations presented in the appendix C, the first and the second determinants can be estimated together, with the formula

$$\det\left[\not{p} + \gamma_0 \mu_u - m_u + \delta_d\left(\not{p} - \gamma_0 \mu_d - m_d\right)\right]\det\left[\not{p} + \gamma_0 \mu_d - m_d + \delta_u\left(\not{p} - \gamma_0 \mu_u - m_u\right)\right]$$
$$= \frac{\left[\prod_{k=1}^{4}(i\omega_n)^2 - \lambda_k^2\right]^2}{\left[(i\omega_n - \mu_u)^2 - E_u^2\right]^2 \left[(i\omega_n - \mu_d)^2 - E_d^2\right]^2} \ , \tag{72}$$

where the four roots $\lambda_k$ are defined in this appendix. This writing allows using again (23) to perform the summation over $n$ analytically, which leads to the expression

$$\Omega_M = -2\int \frac{d^3p}{(2\pi)^3}\left\{\sum_{k=1}^{4}\lambda_k + 2T\ln\left[1 + \exp\left(-\frac{\lambda_k}{T}\right)\right]\right.$$
$$+ E_u + T\ln\left[1 + \exp\left(-\frac{E_u - \mu_u}{T}\right)\right] + T\ln\left[1 + \exp\left(-\frac{E_u + \mu_u}{T}\right)\right]$$
$$+ E_d + T\ln\left[1 + \exp\left(-\frac{E_d - \mu_d}{T}\right)\right] + T\ln\left[1 + \exp\left(-\frac{E_d + \mu_d}{T}\right)\right] \quad . \tag{73}$$
$$\left. + 3E_s + 3T\ln\left[1 + \exp\left(-\frac{E_s - \mu_s}{T}\right)\right] + 3T\ln\left[1 + \exp\left(-\frac{E_s + \mu_s}{T}\right)\right]\right\}$$

In the first line of this relation, it is no longer possible to extract or to separate the partition functions of the quarks and antiquarks, as done in the appendix B for the ($\mu$)PNJL models. Then, with (25) and thanks to the property underlined in the equation (C4), it can be found, e.g., for the $u$ quarks,

$$\langle\langle\bar{\psi}_u\psi_u\rangle\rangle = -2\int\frac{d^3p}{(2\pi)^3}\left\{-2\sum_{k=1}^{4}\frac{\partial\lambda_k}{\partial m_u}f(\lambda_k) + \frac{m_u}{E_u}\left[1 - f(E_u - \mu_u) - f(E_u + \mu_u)\right]\right\} \ . \tag{74}$$

The application of (28) allows expressing the derivative of $\lambda_k$ with respect to $m_u$ as

$$\frac{\partial\lambda_k}{\partial m_u} = 2\frac{m_u\left[(\lambda_k - \mu_d)^2 - E_d^2\right] - m_d|\Delta_{ud}|^2}{\prod_{\substack{j=1 \\ j\neq k}}^{4}(\lambda_k - \lambda_j)} \ , \tag{75}$$

using the propagators $S_{u,r}$ and $S_{u,g}$ of, respectively, the red and the green $u$ quarks

$$S_{u,r} = S_{u,g} = \frac{\left[(i\omega_n - \mu_d)^2 - E_d^2\right](\not{p} + \gamma_0 \mu_u + m_u) - |\Delta_{ud}|^2 (\not{p} - \gamma_0 \mu_d + m_d)}{\prod_{j=1}^{4}(i\omega_n - \lambda_j)}. \quad (76)$$

In the same way, one obtains

$$\langle\langle \psi_u^+ \psi_u \rangle\rangle = 2\int \frac{d^3p}{(2\pi)^3} \left\{ \sum_{k=1}^{4} \frac{\partial \lambda_k}{\partial \mu_u}\left[1 - 2f(\lambda_k)\right] + f(E_u - \mu_u) - f(E_u + \mu_u) \right\}, \quad (77)$$

and the gap equation for $\Delta_{ud}$ is, in a condensed form,

$$|\Delta_{ud}| = 8 G_{DIQ} \left| \int \frac{d^3p}{(2\pi)^3} \sum_{k=1}^{4} \frac{\partial \lambda_k}{\partial |\Delta_{ud}|} f(\lambda_k) \right|. \quad (78)$$

Concerning the $\Delta(\mu)$PNJL versions, the modifications to be done are comparable to what I did in the subsection 4.2. The $\tilde{\mu}_f \to \tilde{\mu}_f - iA_4$ replacement leads to obtain 18 different determinants. The equation (72) must be applied to the determinants that gather red $u$ and green $d$ quarks on one side, and to the determinants that consider green $u$ and red $d$ quarks on the other side. It leads to the apparition of two sets of different roots $\lambda_k$ and $\lambda'_k$, which are now complex numbers. So, in (73), $\sum_{k=1}^{4} \lambda_k + 2T\ln\left(1 + e^{-\beta\lambda_k}\right)$ is replaced by $\frac{1}{2}\sum_{k=1}^{4} \lambda_k + 2T\ln\left(1 + e^{-\beta\lambda_k}\right) + \frac{1}{2}\sum_{k=1}^{4} \lambda'_k + 2T\ln\left(1 + e^{-\beta\lambda'_k}\right)$. Also, the unpaired quarks, i.e. the blue $u$ and $d$ quarks and the strange quarks, are written in a similar manner as in (66), and so on for the chiral condensates, density terms and the gap equation for $\Delta_{ud}$. As done in the subsection 4.2, one has to take the real part of the found relations.

*5.2 Results*

The results obtained with the EB parameter set are presented in the Figure 17 to Figure 19 for the $\Delta$ NJL model. The Figure 20 focuses on the influence of the Polyakov loop. As previously, $\mu_s$ and $\rho_s$ are fixed to zero. The Figure 17 shows the behavior of $\Delta_{ud}$ according to $\mu_u, \mu_d$ and $\rho_u, \rho_d$. For both, the graphs appear to be symmetric according, respectively, to the $\mu_u = \mu_d$ and $\rho_u = \rho_d$ planes. Along these planes, one recovers the results presented in the subsection 4.3. This symmetry is due to the fact that the naked masses of the $u$ and $d$ quarks are very close in the EB parameter set. When it is not the case, this symmetry is lost, as with the $q$ and $s$ quarks. This affirmation is confirmed by the Figure 21(b) and Figure 24(b), which are associated with the CFL phase. In the $\mu_u, \mu_d$ plane, Figure 17(a), the visible frontier between the NQ and the 2SC phases is materialized by a discontinuity. It reveals a first order phase transition between these two phases, in agreement with the literature. The appearance of the graph is comparable to the ones of [102] or [21], even if this second reference also included pion/kaon condensation. In the part of the 2SC phase studied in this graph, the increase in the values of $\Delta_{ud}$ appears to be rather regular, especially along the $\mu_u = \mu_d$ plane.

In the $\rho_u, \rho_d$ plane, Figure 17(b), $\Delta_{ud} \neq 0$ only if $\rho_u \approx \rho_d$. It confirms the fragility of the 2SC phase with respect to the asymmetric matter in flavor [117]. Also, the evolution of $\Delta_{ud}$ recalls the Figure 10(a), i.e. smooth variations without discontinuities. So, the first order phase transition according to $\mu_{u,d}$ is invisible in the $\rho_u, \rho_d$ plane. The part of the 2SC phase located at low $\rho_{u,d}$ is formed by metastable/unstable states. Consequently, as in the subsection 4.3, the model of quark droplets [46] is usable to interpret the presence of the 2SC phase at such densities. The description of the unstable zone in the $\rho_u, \rho_d$ plane allows observing that its frontiers are slightly deformed by the 2SC phase. Around $\rho_{u,d} \approx 0{,}12$ fm$^{-3}$ and $\rho_{u,d} \approx 0{,}74$ fm$^{-3}$, its effect is to shift the unstable zone towards higher densities. As done Figure 10(b), the data found according to $\rho_u, \rho_d$ are used to represent the metastable/unstable states in the $\mu_u, \mu_d$ plane, Figure 18. To facilitate its reading, I only used the data for which $\rho_{u,d} < 0.8 \text{ fm}^{-3}$ and $\Delta_{ud} \neq 0$ to plot the opaque gray surface. Its slope along the $\mu_u = \mu_d$ direction is reversed compared with the one of the surface associated with stable states, except near the discontinuity. So, the great majority of the states represented by the gray surface are unstable. Its dots located at the highest chemical potentials correspond to the lowest densities, and conversely. In agreement with the Figure 10(b), the metastable zone involving $\Delta_{ud} \neq 0$, located between the unstable and the stable zones, appears to be rather reduced.

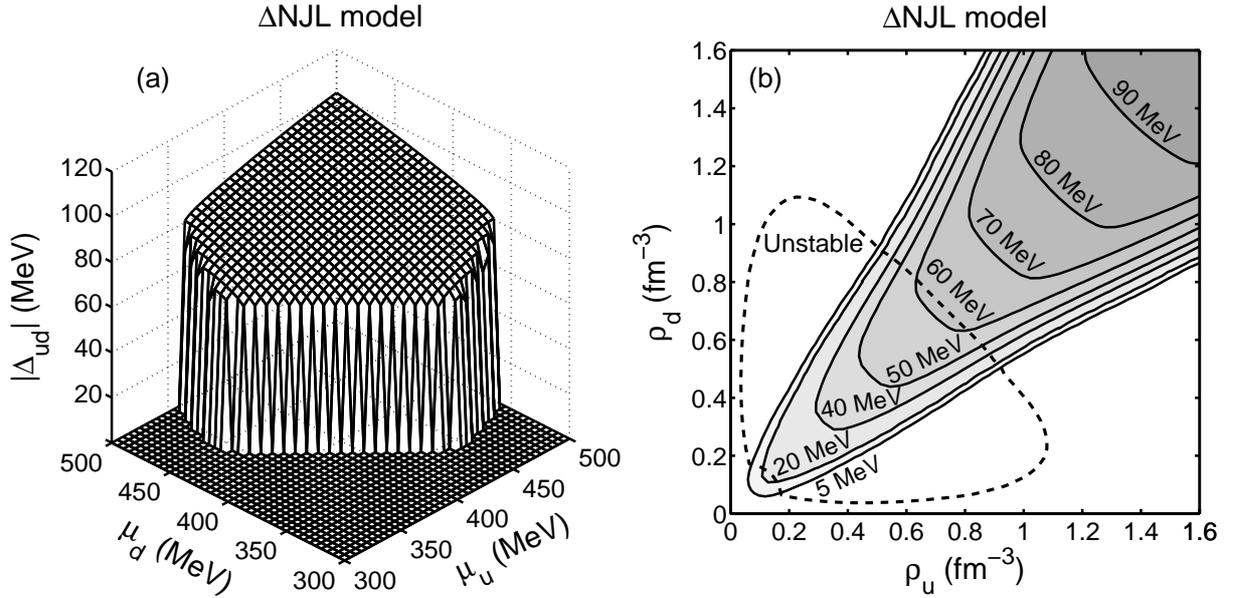

**Figure 17.** Gap $\Delta_{ud}$ (a) according to $\mu_{u,d}$ and (b) according to the densities $\rho_{u,d}$, at $T = 5$ MeV.

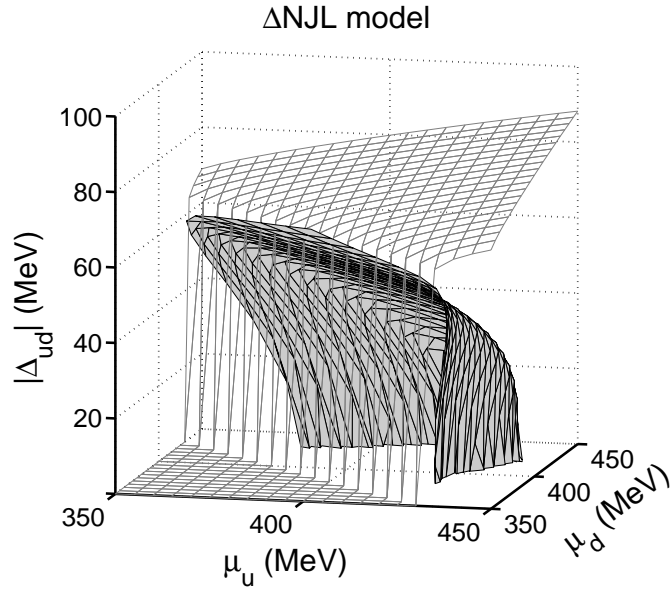

**Figure 18.** $\Delta_{ud}$ according to $\mu_u$ and $\mu_d$, at $T = 5$ MeV. The opaque gray surface was found using the temperature and the densities as input parameters. The other surface is the one of the Figure 17(a).

The calculations at finite densities are extended to several temperatures, Figure 19. It includes a representation of the unstable zone, Figure 19(a), and the theoretical limits of the 2SC phase, Figure 19(b). The width of this phase, measured from the $\rho_u = \rho_d$ plane, tends to grow when the densities increase. Also, it slightly increases with $T$, but only when $T \approx 5-10$ MeV in the $\Delta$ NJL model and at high densities. However, this width decreases rapidly for higher temperatures, until the disappearance of the 2SC phase when $T$ is strong enough. The minimum densities for which the 2SC phase is expected to be found strongly increase with $T$. In parallel, the unstable zone evolves more slowly according to $T$. Consequently, when $T > 35$ MeV, the 2SC phase is entirely outside the unstable zone.

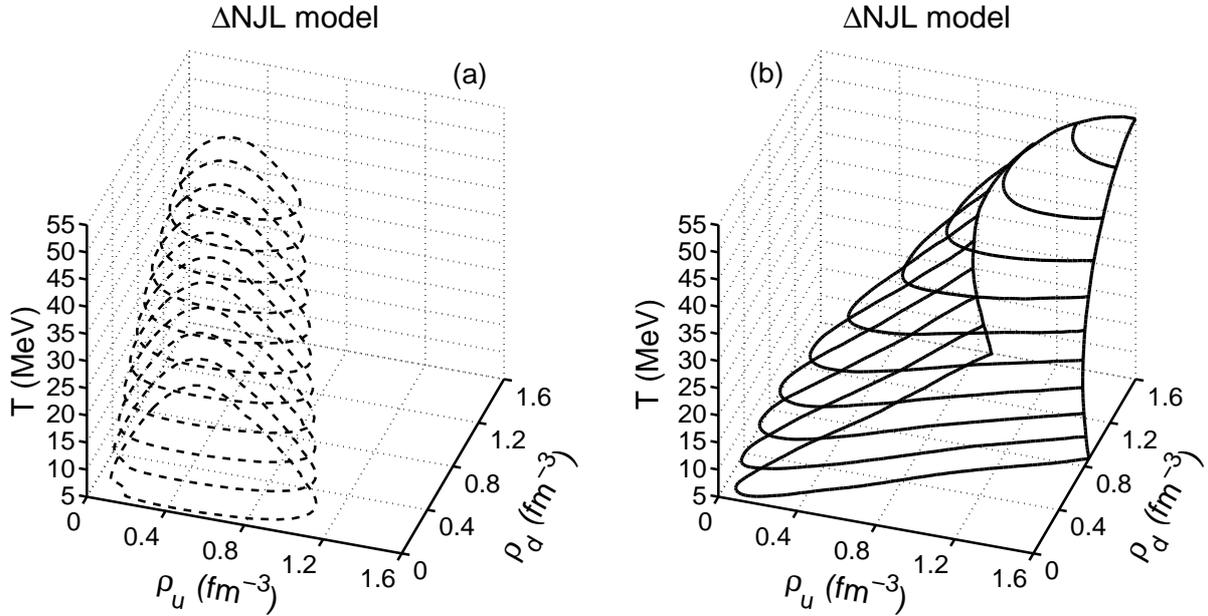

**Figure 19.** (a) Unstable zone, delimited by the dashed curves, and (b) maximum theoretical extension of the 2SC phase in the $\rho_u, \rho_d, T$ space, in the $\Delta$ NJL description. The 2SC phase is located inside the plotted surface.

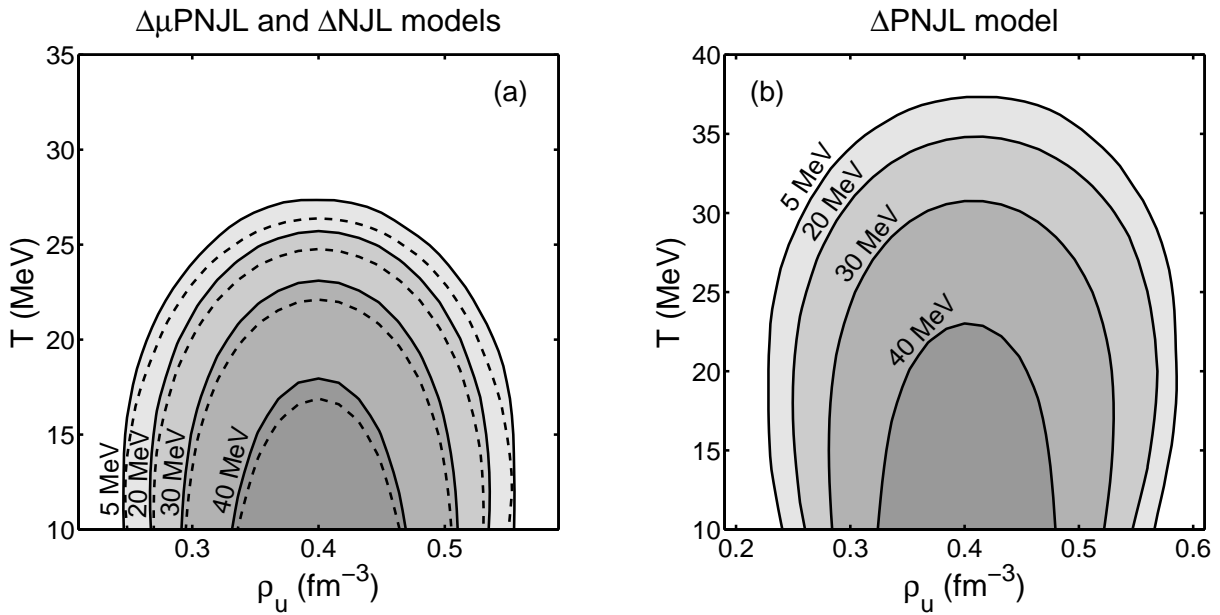

**Figure 20.** Gap $\Delta_{ud}$ according to the temperature and the densities $\rho_{u,d}$, with the constraint $\rho_u + \rho_d = 0.8$ fm$^{-3}$, in the (a) $\Delta\mu$ PNJL (solid curves), $\Delta$ NJL (dash-dotted curves) and (b) $\Delta$ PNJL models.

This description concerns the $\Delta$ NJL model, but it is easily transposable to the $\Delta$ PNJL and $\Delta\mu$ PNJL ones. Indeed, the graphs found with these approaches are qualitatively comparable. Firstly, the $\Delta$ NJL and $\Delta$ PNJL results coincide at very low $T$. So, the Figure 17 and Figure 18 also describe $\Delta$ PNJL results. Also, the shifting of the $\Delta_{ud}$ values towards higher $T$, observed subsection 4.3, is confirmed in the Figure 20. In this figure, it leads to an increase in the 2SC critical temperature from 27 MeV in the $\Delta$ NJL model to 38 MeV in the $\Delta$ PNJL one. Consequently, the aspect of the 2SC phase in the $\rho_u, \rho_d, T$ space can be found in the $\Delta$ PNJL model while applying the shifting along the $T$ axis in the Figure 19. Secondly, about the $\Delta\mu$ PNJL model, the Figure 20(a) indicates that the $\Delta$ NJL and $\Delta\mu$ PNJL results are rather similar, as previously observed in this paper for the masses and $\Delta_{ud}$. So, the data presented in the Figure 17 to Figure 19 should be slightly modified in the $\Delta\mu$ PNJL model.

## 6. Study of the CFL phase

*6.1 Equations*

To describe the CFL phase, the three gaps $\Delta_{ff'}$ must be non-null. With the method [85] used in this paper, the calculation of the determinant $\det(\tilde{S}^{-1})$ is still possible. However, the factorizations and simplifications performed in the previous sections cannot be applied in this configuration. The equations found with the $\Delta(\mu)$PNJL models or without the isospin symmetry appear in an intractable analytical form. So, I do not use them in this paper to describe the CFL phase. I focus on the $\Delta$ NJL model with $m_u = m_d \equiv m_q \neq m_s$ and $\mu_u = \mu_d \equiv \mu_q \neq \mu_s$. In this case, I obtain the following form

$$\begin{aligned}
\det(\tilde{S}^{-1}) = \det&\left\{\left[\not{p} - \gamma_0\mu_q + m_q + 2\bar{\delta}_{s(q)}(\not{p} + \gamma_0\mu_s + m_s)\right]\right.\\
&\left.\times\left[\not{p} + \gamma_0\mu_q - m_q + 2\delta_{s(q)}(\not{p} - \gamma_0\mu_s - m_s)\right] - |\Delta_{ud}|^2\right\}\\
&\times\left\{\det\left[\not{p} + \gamma_0\mu_s - m_s + \delta_{q(s)}(\not{p} - \gamma_0\mu_q - m_q)\right]\right\}^2\\
&\times\left\{\det\left[\not{p} + \gamma_0\mu_q - m_q + \delta_{s(q)}(\not{p} - \gamma_0\mu_s - m_s)\right]\right\}^2\\
&\times\left\{\det\left[\not{p} + \gamma_0\mu_q - m_q + \delta_{q(q)}(\not{p} - \gamma_0\mu_q - m_q)\right]\right\}^3\\
&\times\left[\det(\not{p} - \gamma_0\mu_q - m_q)\right]^5 \det(\not{p} + \gamma_0\mu_s - m_s)\left[\det(\not{p} - \gamma_0\mu_s - m_s)\right]^3
\end{aligned} \quad (79)$$

with

$$\delta_{f(f')} = \frac{-|\Delta_{ff'}|^2}{(i\omega_n - \mu_f)^2 - E_f^2} \quad \text{and} \quad \bar{\delta}_{f(f')} = \frac{-|\Delta_{ff'}|^2}{(i\omega_n + \mu_f)^2 - E_f^2}. \quad (80)$$

The second and the third determinants are estimated together with (72), the fourth determinant with (57) and the three last ones with (44). However, the first determinant must be evaluated differently. In a similar manner as in the appendix C, it can be shown that this determinant leads to study a polynomial function of degree twelve. This one is only constituted by even monomials, which permits to affirm that its twelve roots are writable as $\pm L_k$, with $k = [\![1;6]\!]$. One comes back to the study of a sextic polynomial. As a consequence, I managed to express this determinant as

$$\det\left\{\left[\not{p}-\gamma_0\mu_q+m_q+2\bar{\delta}_{s(q)}\left(\not{p}+\gamma_0\mu_s+m_s\right)\right]\right.$$
$$\left.\times\left[\not{p}+\gamma_0\mu_q-m_q+2\delta_{s(q)}\left(\not{p}-\gamma_0\mu_s-m_s\right)\right]-\left|\Delta_{ud}\right|^2\right\}\quad. \tag{81}$$

$$=\frac{\left[\prod_{k=1}^{6}(i\omega_n)^2-L_k^2\right]^2}{\left[(i\omega_n+\mu_s)^2-E_s^2\right]^4\left[(i\omega_n-\mu_s)^2-E_s^2\right]^4}$$

Strategies were developed to treat a sextic polynomial analytically [118-122]. Nevertheless, the roots of such polynomials can also be estimated via numerical calculations, as I did in this work. Thanks to (81), the summation of the Matsubara frequencies is still feasible with (23), as with the other cases described upstream. It leads to

$$\Omega_M = -2\int\frac{d^3p}{(2\pi)^3}\left\{\sum_{k=1}^{4}\lambda_k + 2T\ln\left[1+\exp\left(-\frac{\lambda_k}{T}\right)\right] + \frac{1}{2}\sum_{k=1}^{6}L_k + 2T\ln\left[1+\exp\left(-\frac{L_k}{T}\right)\right]\right.$$
$$+\frac{3}{2}\left(E_q^- + E_q^+\right) + 3T\ln\left[1+\exp\left(-\frac{E_q^-}{T}\right)\right] + 3T\ln\left[1+\exp\left(-\frac{E_q^+}{T}\right)\right] \tag{82}$$
$$\left. - E_s - T\ln\left[1+\exp\left(-\frac{E_s-\mu_s}{T}\right)\right] - T\ln\left[1+\exp\left(-\frac{E_s+\mu_s}{T}\right)\right]\right\}$$

with $E_q^\pm = \sqrt{\left(E_q\pm\mu_q\right)^2 + \left|\Delta_{ud}\right|^2}$. In this configuration, the normal and abnormal propagators cannot easily be employed to obtain the expressions, e.g., of the chiral condensates. It leads to exclusively consider the derivatives of $\Omega_M$. It gives results as

$$\langle\langle\bar{\psi}_q\psi_q\rangle\rangle = -2\int\frac{d^3p}{(2\pi)^3}\left\{\frac{1}{2}\sum_{k=1}^{4}\frac{\partial\lambda_k}{\partial m_q}\left[1-2f(\lambda_k)\right] + \frac{1}{4}\sum_{k=1}^{6}\frac{\partial L_k}{\partial m_q}\left[1-2f(L_k)\right]\right.$$
$$\left. + \frac{3}{4}\frac{E_q-\mu_q}{E_q^-}\frac{m_q}{E_q}\left[1-2f(E_q^-)\right] + \frac{3}{4}\frac{E_q+\mu_q}{E_q^+}\frac{m_q}{E_q}\left[1-2f(E_q^+)\right]\right\}, \tag{83}$$

$$\langle\langle\bar{\psi}_s\psi_s\rangle\rangle = -2\int\frac{d^3p}{(2\pi)^3}\left\{\sum_{k=1}^{4}\frac{\partial\lambda_k}{\partial m_s}\left[1-2f(\lambda_k)\right] + \frac{1}{2}\sum_{k=1}^{6}\frac{\partial L_k}{\partial m_s}\left[1-2f(L_k)\right]\right.$$
$$\left. - \frac{m_s}{E_s}\left[1-f(E_s-\mu_s) - f(E_s+\mu_s)\right]\right\}. \tag{84}$$

The equations for $\langle\langle\psi_f^+\psi_f\rangle\rangle$ are writable in a similar form, and

$$\left|\Delta_{ud}\right| = 2G_{DIQ}\left|\int\frac{d^3p}{(2\pi)^3}\frac{3\left|\Delta_{ud}\right|}{E_q^-}\left[1-2f(E_q^-)\right] + \frac{3\left|\Delta_{ud}\right|}{E_q^+}\left[1-2f(E_q^+)\right] + \sum_{k=1}^{6}\frac{\partial L_k}{\partial\left|\Delta_{ud}\right|}\left[1-2f(L_k)\right]\right|, \tag{85}$$

$$\left|\Delta_{qs}\right| = 2G_{DIQ}\left|\int\frac{d^3p}{(2\pi)^3}\sum_{k=1}^{4}\frac{\partial\lambda_k}{\partial\left|\Delta_{qs}\right|}\left[1-2f(\lambda_k)\right] + \frac{1}{2}\sum_{k=1}^{6}\frac{\partial L_k}{\partial\left|\Delta_{qs}\right|}\left[1-2f(L_k)\right]\right|. \tag{86}$$

In a description of the sSC phase, (79) can be used and then it can be simplified thanks to the constraint $\left|\Delta_{ud}\right|=0$ associated with this phase. In this case, the roots $L_k$ presented in (81) are $E_s\pm\mu_s$ and $\hat{\lambda}_j$ with $j=[\![1;4]\!]$. $\hat{\lambda}_j$ are the roots of the numerator of the expression that is obtained when one uses a formula like (72) with

$$\det\left[\not{p}+\gamma_0\mu_s-m_s+2\delta_{q(s)}\left(\not{p}-\gamma_0\mu_q-m_q\right)\right]$$
$$\times\det\left[\not{p}+\gamma_0\mu_q-m_q+2\delta_{s(q)}\left(\not{p}-\gamma_0\mu_s-m_s\right)\right] \quad , \tag{87}$$

such terms being found thanks to the property

$$\det\left[\not{p}-\gamma_0\mu_q+m_q+2\bar{\delta}_{s(q)}\left(\not{p}+\gamma_0\mu_s+m_s\right)\right]$$
$$=\left[\frac{(i\omega_n-\mu_q)^2-E_q^2}{(i\omega_n+\mu_s)^2-E_s^2}\right]^2 \det\left[\not{p}+\gamma_0\mu_s-m_s+2\delta_{q(s)}\left(\not{p}-\gamma_0\mu_q-m_q\right)\right] \tag{88}$$

applied to the equation (81).

*6.2 Results at finite chemical potentials*

The Figure 21 studies the gaps $\Delta_{ud}$ and $\Delta_{qs}$ according to the chemical potentials $\mu_q,\mu_s$. Comparable graphs are found, e.g., in [84]. Nevertheless, some differences are visible, notably at high $\mu_s$. They are explainable by the approximation employed in this reference and the different parameter sets used. It includes the value of the cutoff $\Lambda$, which is larger in my work. It allows calculations with upper $\mu_s$. My results are also close to the ones of [21], in which pseudoscalar meson condensation is included but no 't Hooft interaction. In the Figure 21(a), the zone located at reduced $\mu_s$ is in agreement with the Figure 9(b). Indeed, $\Delta_{ud}=0$ until $\mu_q\approx 394$ MeV, it becomes non-null via a marked discontinuity, and then it increases for higher $\mu_q$. In parallel, $\Delta_{qs}=0$ in this zone. When $\mu_s$ increases, this behavior is not modified until $\mu_s\approx 550$ MeV. After this value, $\Delta_{qs}$ becomes non-null for some values of $\mu_q$ and $\mu_s$, Figure 21(b). The zone for which $\Delta_{qs}\neq 0$ is visible in the two graphs of the Figure 21, because $\Delta_{qs}\neq 0$ leads in the Figure 21(a) to a brutal diminution of $\Delta_{ud}$. It forms a structure comparable to a "canyon" in this graph. It illustrates the competition between the *ud* and *qs* pairs, since both involve at least one light quark [84]. Consequently, when the conditions are favorable to the formation of *qs* pairs, it is to the detriment of the *ud* ones. For the highest $\mu_s$ treated in the figure and for $\mu_q$ around 400 MeV, $\Delta_{qs}$ is null again but not $\Delta_{ud}$, except for the lowest values of $\mu_q$ visible in the graph. This zone of the Figure 21(a) is the only one where $\Delta_{ud}$ continuously decreases to zero. It reveals there a second order phase transition. In the rest of these two graphs, all the phase transitions involve discontinuities, so they are of first order. It includes when $\Delta_{qs}$ becomes non-null in the Figure 21(b).

The analysis of these first order phase transitions is complemented by the Figure 22, which uses the data that permitted to build Figure 24. In the Figure 22(a), the phase transition that intervenes when $\mu_s<550$ MeV seems to have a reduced zone for which it can exist metastable states with $\Delta_{ud}\neq 0$. The zone of the unstable states is more extended, in agreement with the Figure 10(b). The phase transition 2SC/CFL that intervenes when $\mu_s\approx 550$ MeV is also easily observable in the two graphs of the Figure 22, with a quite similar behavior. However, about the transition 2SC/CFL that occurs when $\mu_s\geq 580$ MeV, the deformations of the plotted surfaces draw more complex structures, notably in the zone for which $\mu_q\approx 380$ MeV and $\mu_s\approx 600$ MeV.

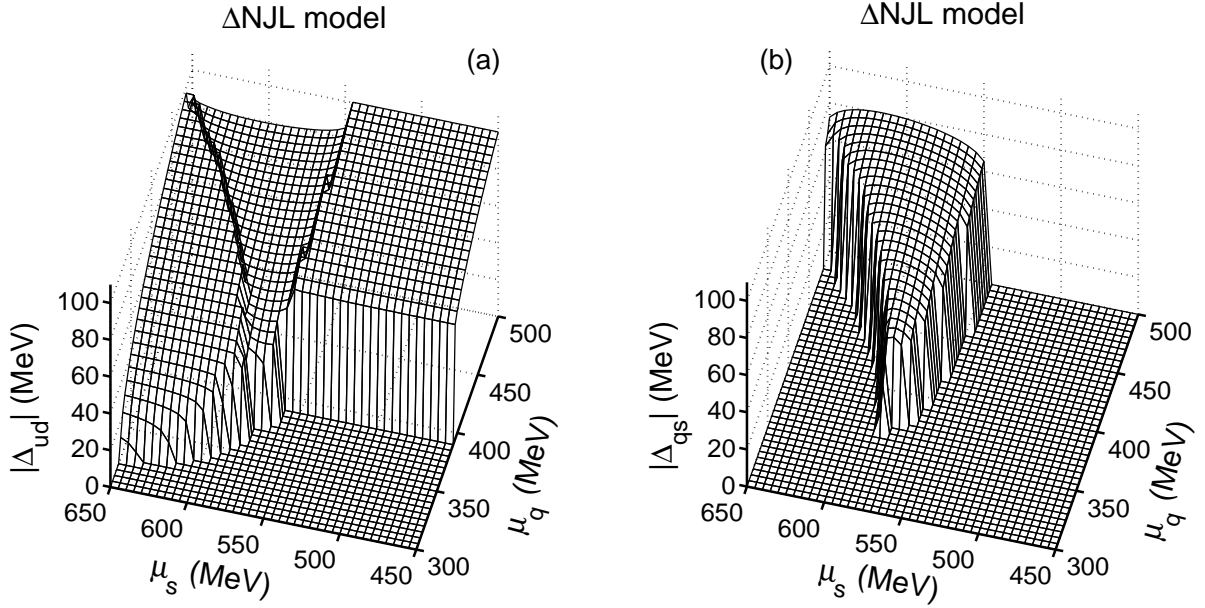

**Figure 21.** Gaps (a) $\Delta_{ud}$ and (b) $\Delta_{qs}$ according to the chemical potentials $\mu_q, \mu_s$, at $T = 20$ MeV.

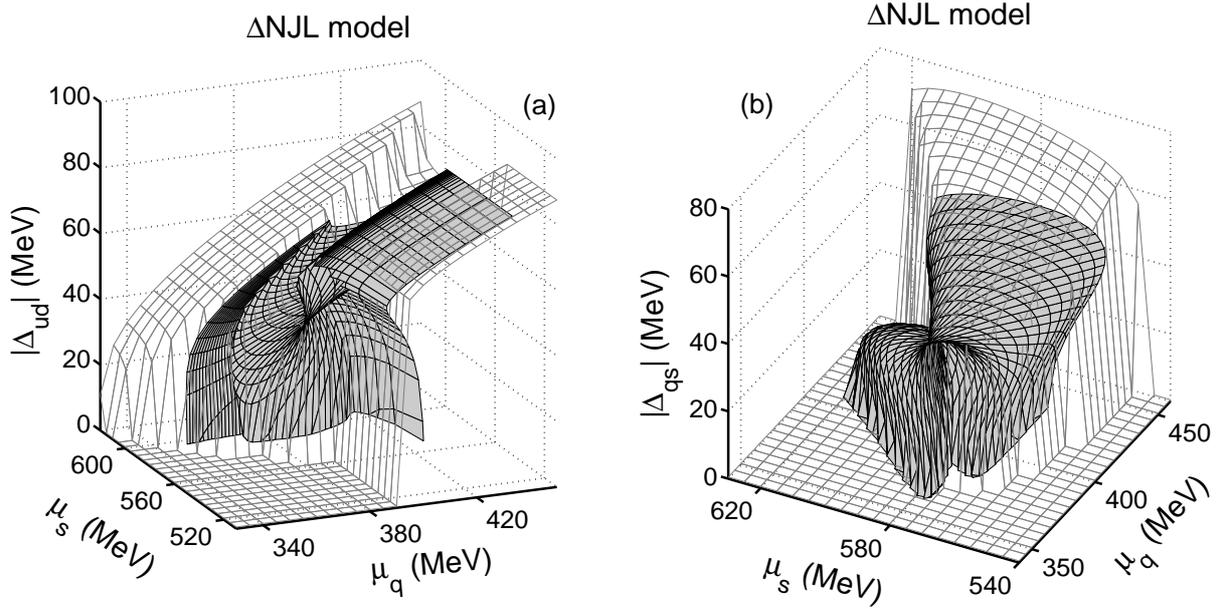

**Figure 22.** Gaps (a) $\Delta_{ud}$ and (b) $\Delta_{qs}$ according to $\mu_q, \mu_s$, at $T = 20$ MeV. The opaque gray surfaces were found using the temperature and the densities as input parameters. The other surfaces are the ones of Figure 21.

I propose now to focus on the particular case $\mu_q = \mu_s \equiv \mu$, which is frequently encountered in the literature. The P1 parameter set is used, as with the other results of this subsection. The Figure 23 shows the evolution of the quark masses $m_q, m_s$ and the gaps $\Delta_{ud}, \Delta_{qs}$ according to $\mu$. Only the stable states are represented. The temperature is fixed at $T = 20$ MeV, as in the Figure 21. A good agreement is qualitatively observed between the Figure 23 and certain results of [19]. It especially concerns the figure 5.4 of this reference, which includes the 't Hooft interaction, as I did in this work. However, numerical differences are observable: the masses of the quarks at low $\mu$, the chemical potentials for which the phase transitions occur, and the highest values reached by the gaps. They are explainable by the fact that I used a different parameter set. Furthermore, the curves exhibit much more variations than in the quoted figure of [19], because this one considers a lower temperature, i.e. $T = 0$ MeV. From a physical point of view, the Figure 23 is another illustration of the competition between the chiral and the diquark condensates. As with the Figure 9, the strong decrease in $m_q$ reveals the discontinuity of $\langle\langle \bar{\psi}_q \psi_q \rangle\rangle$ when $\mu \approx 394$ MeV. Because of the 't Hooft term, first line of (1), it affects $m_s$, which slightly decreases. Furthermore, the decrease in $\langle\langle \bar{\psi}_q \psi_q \rangle\rangle$ allows $\Delta_{ud}$ to be non-null: it permits the formation of the 2SC phase. Then, when $\mu \approx 550$ MeV, $m_s$ presents a strong discontinuity, interpretable as a fall of $\langle\langle \bar{\psi}_s \psi_s \rangle\rangle$ due to the restoration of the chiral symmetry for the strange quarks [19]. Consequently, it authorizes the apparition of the CFL phase, since $\Delta_{qs}$ is now non-null, like $\Delta_{ud}$. At this $\mu$, a slight increase in $m_q$ of about 5 MeV is visible, via a discontinuity. It cannot be explained by the 't Hooft term, because this discontinuity is also present when this term is absent, figure 5.1 of [19]. So, it translates the weak influence of $\Delta_{qs}$ on $m_q$, which is comparable to the effect of $\Delta_{ud}$ on this mass detected Figure 12.

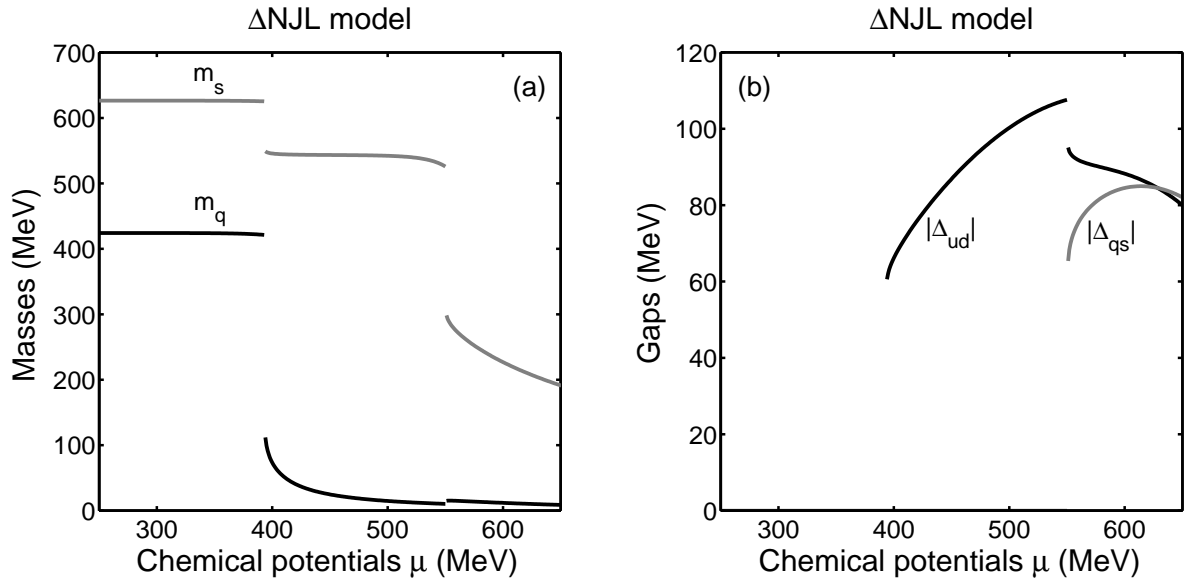

**Figure 23.** (a) Masses of the light and strange quarks and (b) gaps $\Delta_{ud}$ and $\Delta_{qs}$ according to the chemical potentials, in the configuration $\mu = \mu_q = \mu_s$, at $T = 20$ MeV.

*6.3 Results at finite densities*

The equations described in this section were also used to describe the CFL phase according to the densities. In order to facilitate the comparison with the results found at $\rho_s = 0$, subsection 4.3, the notation $\rho = \tfrac{2}{3}\rho_q$ is used to designate the density of the *non-strange* matter. In the Figure 24, the values of $\Delta_{ud}$ and $\Delta_{qs}$ are represented in the $\rho, \rho_s$ plane, at $T = 20$ MeV. These graphs present differences with the ones performed according to $\mu_q, \mu_s$. For example, the two gaps decrease to zero continuously. However, a common point is when $\Delta_{qs}$ becomes non-null, it leads to a decrease in $\Delta_{ud}$. This behavior is visible in the Figure 24(a), in which the gray curve corresponds to the $\Delta_{qs} = 0$ contour line. At the opposite, in the zone for which $\Delta_{qs} = 0$, the increase in $\Delta_{ud}$ according to $\rho$ is rather regular when this density is growing. It appears to be almost independent of the strange density $\rho_s$. This increase in $\Delta_{ud}$ at $\rho_s = 0$ is fully in agreement with the Figure 11(a). Furthermore, with the stability criteria $\partial^2\Omega/\partial\langle\langle\bar{\psi}_q\psi_q\rangle\rangle^2 > 0$ and $\partial^2\Omega/\partial\langle\langle\bar{\psi}_s\psi_s\rangle\rangle^2 > 0$, I included the limits of the unstable zones associated with the chiral condensates of, respectively, the light and strange quarks. A non-negligible part of the 2SC and CFL phases located at low and moderate $\rho$ corresponds to unstable states. Also, if one considers these unstable zones or the evolution of the gaps, one observes that the graphs are highly asymmetric according to $\rho, \rho_s$, as in the Figure 21 with $\mu_q, \mu_s$.

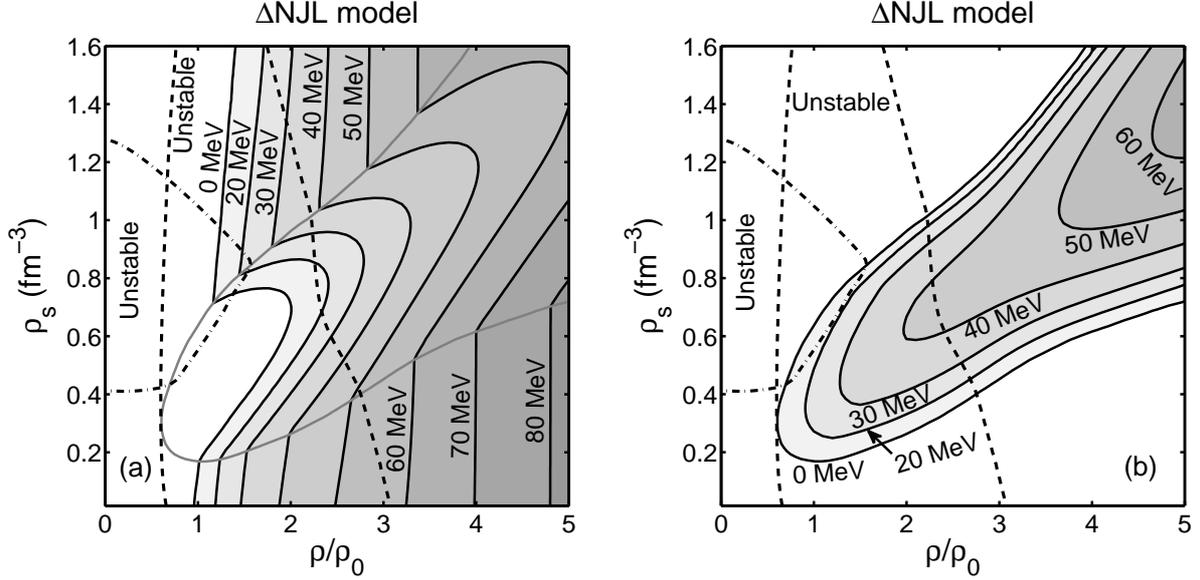

**Figure 24.** Gaps (a) $\Delta_{ud}$ and (b) $\Delta_{qs}$ in the $\rho, \rho_s$ plane, with $T = 20$ MeV. The dashed and the dash-dotted curves delimit the unstable zones associated, respectively, with the light and strange chiral condensates.

This asymmetry also appears in the Figure 25, in which $\Delta_{ud}$ and $\Delta_{qs}$ are used to identify the various phases observable in the $\rho, \rho_s$ plane, for several $T$. Four distinct phases are visible: the normal quark phase (NQ), the 2SC, the CFL and the sSC phases. At $T \approx 20$ MeV, the sSC phase is located in a zone where $\rho \approx \rho_0$ or slightly upper, and $\rho_s$ around 0.5 fm$^{-3}$. Its presence is rather unexpected, because it is absent in the Figure 21. However, this apparent disagreement is explainable by the fact that this sSC phase is entirely located in an unstable zone at this temperature, whereas the Figure 21 only displays stable states. In the Figure 22, it is also possible to check that when $\mu_q \approx 380$ MeV and $\mu_s \approx 580$ MeV, the criteria of existence of the sSC phase, i.e. $\Delta_{ud} = 0$ and $\Delta_{qs} \neq 0$, can be satisfied but not for stable states.

Furthermore, in the Figure 25, one observes how the various phases evolve according to $T$. In agreement with the Figure 10(a), the frontier between the NQ and 2SC phases is shifted towards higher $\rho$ when $T$ increases. For the studied temperatures, if the CFL and sSC phases are neglected, the strange density has a weak influence on the position of this frontier. At the opposite, the behavior of the CFL phase is strongly related to the two densities $\rho$ and $\rho_s$. When $T$ increases, its frontier is moved towards higher $\rho$ and $\rho_s$ simultaneously. It underlines that the formation of a $qs$ pair requires specific conditions with respect to $\rho, \rho_s$ (or $\mu_q, \mu_s$). Moreover, the values of the two gaps, and by extension the positions of the 2SC and CFL phases, do not evolve at the same rhythm according to $T$. It implies that the extension of the sSC phase in the $\rho, \rho_s$ plane strongly depends on the temperature. This area is reduced at low $T$, it seems to reach a maximum when $T \approx 20$ MeV, then it progressively decreases until the complete disappearance of the sSC phase from $T \approx 35$ MeV onwards. About the unstable zones, some deformations of their frontiers are observable, notably caused by the sSC and

CFL phases. These deformations are more marked at low temperatures, and they disappear for higher $T$. The superficies of the unstable zones evolve rather slowly according to $T$, as also observed in the Figure 19(a). As a consequence, the sSC phase can be partially located outside of these unstable zones for certain temperatures. At $T = 10$ MeV and $T = 30$ MeV, a part of this phase is in metastable zones.

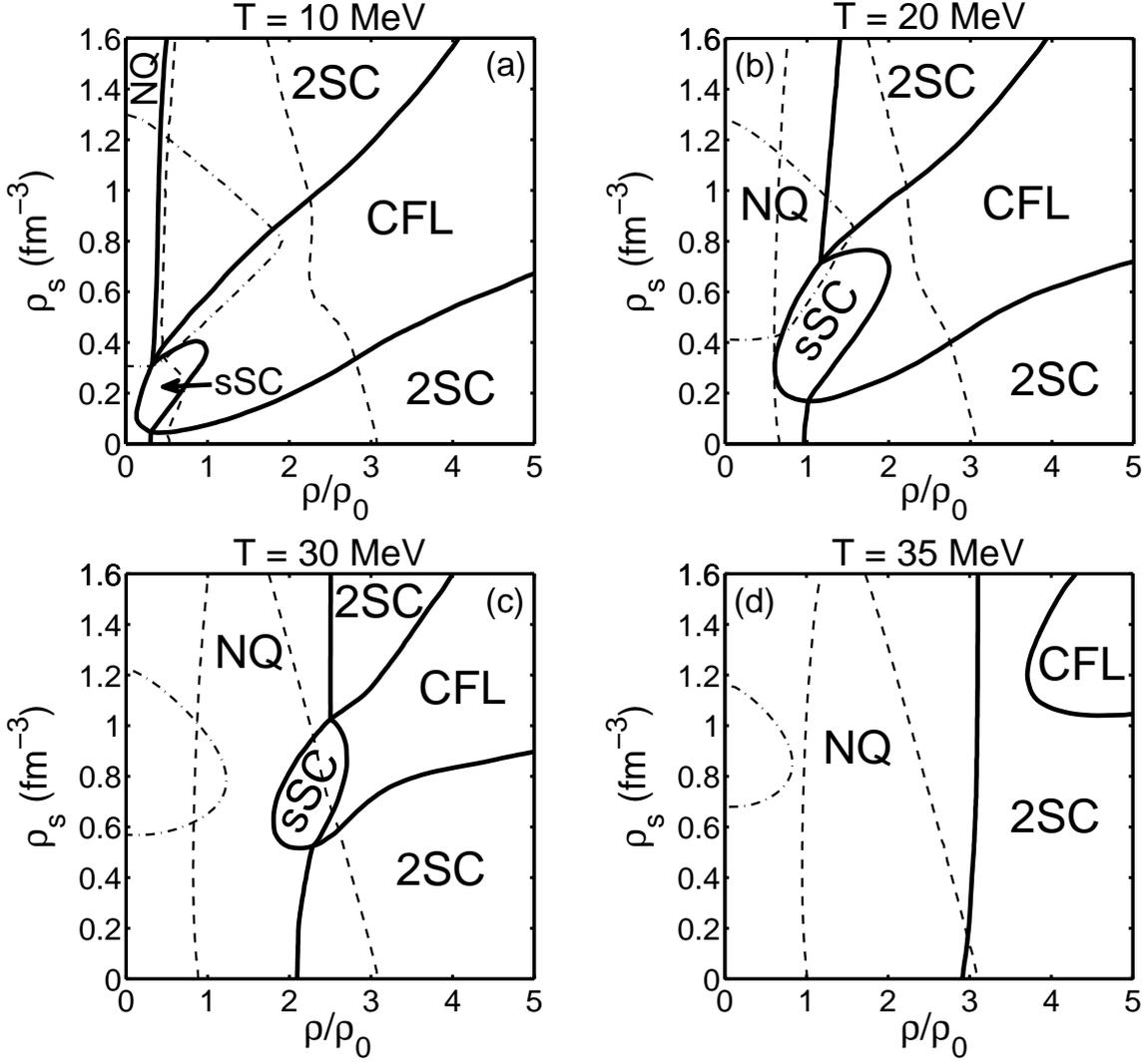

**Figure 25.** Phase diagrams in the $\rho, \rho_s$ plane, in the $\Delta$ NJL model, at (a) $T = 10$ MeV, (b) $T = 20$ MeV, (c) $T = 30$ MeV and (d) $T = 35$ MeV. The unstable states are located inside the zones delimited by the dashed and the dash-dotted curves, as in the Figure 24.

## 7. Conclusions

In this paper, the color superconductivity phenomenon was studied at finite temperatures, finite densities and finite chemical potentials, with adaptations of the NJL models. These adaptations concerned the coupling to a Polyakov loop (PNJL model) with a $\mu$-dependent Polyakov loop potential ($\mu$PNJL model) and the required modifications to describe the color superconductivity ($\Delta$ NJL, $\Delta$ PNJL and $\Delta\mu$ PNJL models). Firstly, the general relations were presented. Then, it was shown how they were used to obtain the equations to be solved, for each of the phases treated in this paper. They were the normal quark phase, the 2SC, the CFL and the sSC phases. Furthermore, it was underlined some mathematical aspects associated with this work. It notably concerned the application of the method presented in [85], which allowed calculating the determinant of the inverse propagator $\tilde{S}^{-1}$ analytically. Compared with a numerical estimation of this determinant, it presents some advantages. It increases the precision of the results and it speeds up the numerical calculations, since it avoids treating a 72×72 matrix numerically. One saw that the method permitted to find the equations related to the 2SC phase in the $\Delta$NJL, $\Delta$PNJL and $\Delta\mu$PNJL models, with or without the isospin symmetry. It also allowed establishing the equations associated with the CFL phase in the $\Delta$NJL model, with the isospin symmetry. Consequently, a possible extension of the presented works may concern the inclusion of the Polyakov loop in the description of the CFL phase [25, 123], and the treatment of these calculations beyond the isospin symmetry, in order to draw a complete phase diagram in the $T, \rho_{u,d,s}$ and $T, \mu_{u,d,s}$ spaces. For example, it could allow observing in the $T, \rho_{u,d,s}$ space the various "exotic" color superconducting phases found in [21] when $\mu_u = \mu_s \neq \mu_d$.

Concerning the found results, it was firstly studied the relation between the $T, \rho_B$ graphs and the $T, \mu_q$ ones, in the (P)NJL models. It led to see how the first order phase transition of the chiral condensate according to $\mu_q$ becomes an apparent crossover according to $\rho_B$. Then, it was recalled the problem of matter stability, characterized by the fact that a non-negligible part of the results found at low temperatures and moderate densities are metastable or unstable. This feature was interpreted with the model of quark droplets [46], which considers the metastable/unstable phases as a mixed phase formed by the droplets and the vacuum. This vision permitted to give a physical sense to the results found at finite densities in this regime. Then, it allowed explaining that the presence of the 2SC phase at very low densities is due to the lack of real confinement in the proposed descriptions. Moreover, one focused on the data produced by the version of the Polyakov loop potential that includes an $N_f$ correction and a $\mu$-dependence. In this $\mu$PNJL model, it was observed that the Polyakov field $\Phi$ describes a first order phase transition according to the chemical potential, i.e. a "deconfinement" transition. It was found that this one coincides with the chiral phase transition [78, 80]. It constitutes a major difference with the PNJL model, in which $\Phi$ only shows a crossover according to the temperature.

Then, it was found that the phase transitions associated with the gaps $\Delta_{ff'}$ are of first or second order in the $T, \mu_q$ or $\mu_q, \mu_s$ planes. The calculations performed at finite densities permitted to study the metastable/unstable states associated with these first order phase transitions, for the 2SC and the CFL phases. Also, it was studied the influence of the 2SC phase on some observables, in the $T, \rho_B$ plane. This phase tends to increase the quark masses and it tends to decrease the chemical potentials and the Polyakov field $\Phi$. In the $\Delta$PNJL model, this decrease in $\Phi$ is negligible, whereas it can be strong in the $\Delta\mu$PNJL model. In this description, it was observed that the 2SC phase leads to shift the critical chemical potential for which the chiral restoration and the "deconfinement" transitions occur, at low temperatures. Furthermore, a particularly interesting result is the observed triple coincidence between these two transitions and the normal quark/2SC phase transition according to $\mu$. Consequently, if the ($\Delta$)PNJL results indicated the presence of a possible quarkyonic phase [25], the ($\Delta$)$\mu$PNJL graphs were not in agreement with this hypothesis, with or without the color superconductivity. Another developed topic was the effect of the Polyakov loop on the results. In the ($\Delta$)PNJL models, these ones confirmed the behavior observed in previous PNJL publications, i.e. a shifting of the values towards higher temperatures. As an example, it leads to an extension of the zone in which the 2SC phase is expected to be present. At the opposite in the ($\Delta$)$\mu$PNJL results, the values of the quark masses, chemical potentials and $\Delta_{ud}$ are rather close to the ones found with the ($\Delta$)NJL descriptions.

Also, the fact to work beyond the isosopin symmetry for the 2SC phase allowed interesting developments. From a mathematical point of view, this study permitted to develop analytical methods, which were used later in the framework of the CFL phase. Also, from a physical point of view, this work allows modeling systems in which $\mu_u$ and $\mu_d$ are very different. As recalled in [19], it concerns heavy nuclei and neutron stars. This remark is extendable to quark stars and strange stars [19, 20, 117,

124-128]. For these examples, since the method permits to have $m_u \neq m_d$, it could be relevant to investigate the influence of this mass difference on the results. Moreover, concerning the CFL phase, it was shown that the calculations allowed recovering results available in the literature. Nevertheless, they also permitted to present new ones. It notably concerned the graphs established according to the densities. These graphs exhibited a rather unexpected result, i.e. the presence of an sSC phase. It was found that this one is located in the mixed phase.

In this paper, the color superconductivity was described with the scalar quark/quark interactions, as in the majority of the works found in the literature. So, a possible extension may concern the inclusion, e.g., of the pseudoscalar diquark condensates [23], or the spin-1 pairs symmetric in color (axial vector channel) [19, 20, 46]. It could be interesting to see how the method described in [85] could be employed in such a development. This work can also be complemented by the treatment of the pseudoscalar meson condensation [21], or by the inclusion of the crystalline color superconducting phase, also named LOFF (Larkin, Ovchinnikov, Fulde and Ferrell) phase [129]. Obviously, for a complete description of the color superconductivity, it appears indispensable to consider the neutrality conditions and the $\beta$ equilibrium, notably to be able to model the interior of the compact stars. As reported in the literature, these conditions lead to several modifications of the found results. One of them concerns the apparition of gapless phases: the g2SC [106, 130] and the gCFL [131]. It could lead to a description of these ones according to the densities, and not only according to the chemical potentials. Such a work could also take into account the influence on these phases of the extreme magnetic fields present in some compact stars (magnetars) [132-138]. Moreover, another possible continuation of the presented studies may be the update of the description of the mesons and baryons in the color superconducting regime, in order to complete the comparison between the NJL/PNJL/$\mu$PNJL descriptions and the $\Delta$NJL/$\Delta$PNJL/$\Delta\mu$PNJL ones [90, 101, 103, 139, 140].

**Appendix A. Establishment of the equation** (23)

*A.1. Preliminary calculations*

In order to study the establishment of (23) found in [99], and to extend it to the complex case in the ($\mu$)PNJL models, one has to evaluate $\sum_{n=-\infty}^{+\infty} \frac{1}{(2n+1)^2 \pi^2 + z^2}$, where $z$ is a complex number. Firstly, one proposes a method that uses the Psi (Digamma) function $\psi$ [141], which satisfies the relations

$$\psi(1+z) = -\gamma - \sum_{n=1}^{+\infty}\left(\frac{1}{n+z} - \frac{1}{n}\right) \text{ and } \psi(1+z) = \psi(z) + 1/z, \quad (A1)$$

where $\gamma$ is the Euler-Mascheroni constant. For two complexes $z_1 \neq z_2$, they permit to find

$$\sum_{n=0}^{+\infty} \frac{1}{(n+z_1)(n+z_2)} = \frac{\psi(z_1) - \psi(z_2)}{z_1 - z_2}, \quad (A2)$$

which can be found in [142]. Then, one writes

$$\sum_{n=0}^{+\infty} \frac{1}{(2n+1)^2 \pi^2 + z^2} = \sum_{n=0}^{+\infty} \frac{1}{4\pi^2} \frac{1}{\left(n + \frac{1}{2} + \frac{iz}{2\pi}\right)\left(n + \frac{1}{2} - \frac{iz}{2\pi}\right)}. \quad (A3)$$

The equation (A2) allows rewriting this sum as

$$\sum_{n=0}^{+\infty} \frac{1}{(2n+1)^2 \pi^2 + z^2} = \frac{1}{4i\pi z}\left[\psi\left(\frac{1}{2} + \frac{iz}{2\pi}\right) - \psi\left(\frac{1}{2} - \frac{iz}{2\pi}\right)\right]. \quad (A4)$$

If one uses the relation $\psi(1-z) = \psi(z) + \pi\cot(\pi z)$ [141] in the second $\psi$ of (A4), one obtains

$$\sum_{n=0}^{+\infty}\frac{1}{(2n+1)^2\pi^2+z^2}=-\frac{1}{4i\pi z}\cot\left[\pi\left(\frac{1}{2}+\frac{iz}{2\pi}\right)\right]=\frac{1}{4iz}\tan\left(\frac{iz}{2}\right)=\frac{1}{4z}\tanh\left(\frac{z}{2}\right). \tag{A5}$$

One checks that $\sum_{n=0}^{+\infty}\frac{1}{(2n+1)^2\pi^2+z^2}=\sum_{n=-1}^{-\infty}\frac{1}{(2n+1)^2\pi^2+z^2}$, which allows obtaining the wanted sum

$$\sum_{n=-\infty}^{+\infty}\frac{1}{(2n+1)^2\pi^2+z^2}=\frac{1}{2z}\tanh\left(\frac{z}{2}\right). \tag{A6}$$

The complex function $\tanh(z/2)/2z$ is defined for $z\in\mathbb{C}^*\setminus\{i(2n+1)\pi\}$, but this domain can be extended to $z\in\mathbb{C}\setminus\{i(2n+1)\pi\}$ if one takes the value $1/4$ at $z=0$.

*A.2. Matsubara summations by the application of the residue theorem*

One proposes to evoke the method that is used in practice to evaluate Matsubara summations $\frac{1}{\beta}\sum_n\phi(i\omega_n)$, where $\phi$ is a complex function. One restricts this description to the cases for which all its poles $z_e$ are simple. Also, $\omega_n=(2n+1)\pi T$ are the fermionic Matsubara frequencies, with $n\in\mathbb{Z}$. One considers a complex weighting function $w$, whose poles are $i\omega_n$. One can have, e.g., $w(z)=\frac{1}{2}\tanh\left(\frac{\beta z}{2}\right)$ [143] or $w(z)=f(\pm z)$, where

$$f(z)=\frac{1}{e^{\beta z}+1} \tag{A7}$$

is the Fermi-Dirac distribution extended to the complex plane. The function $\phi w$ is then integrated in the complex plane, over a circular contour $C$ of radius $R\to\infty$ centered at the origin. Consequently, the contour includes all the poles $z_e$ and $i\omega_n$. When $|z|\to\infty$, if $\phi(z)$ converges towards 0 faster than $|z|^{-1}$, this integral is null according to Jordan's first lemma. Furthermore, if the poles $z_e$ and $i\omega_n$ are always different, the residue theorem gives [72]

$$\underbrace{\oint_C\frac{dz}{2i\pi}\phi(z)w(z)}_{0}=\sum_n\phi(i\omega_n)\,res(w,i\omega_n)+\sum_{z_e}w(z_e)\,res(\phi,z_e). \tag{A8}$$

One has $res[f(z),i\omega_n]=\frac{-1}{\beta}$ and $res[f(-z),i\omega_n]=res\left[\frac{1}{2}\tanh\left(\frac{\beta z}{2}\right),i\omega_n\right]=\frac{1}{\beta}$, so that one finds

$$\frac{1}{\beta}\sum_n\phi(i\omega_n)=\sum_{z_e}\left[f(z_e)-k\right]res(\phi,z_e), \tag{A9}$$

where $k=0$ when $w(z)=f(z)$, $k=1/2$ when $w(z)=\frac{1}{2}\tanh\left(\frac{\beta z}{2}\right)$ and $k=1$ when $w(z)=f(-z)$.

In practice, if $\phi$ satisfies the convergence criterion, the choice of the weighting function, among the three proposed ones, is without consequence on the result. In addition, since the $z_e$ are simple poles, the residues $res(\phi,z_e)$ can be estimated thanks to the property

$$\text{if }\phi(z)\equiv\frac{P(z)}{Q(z)},\text{ then }res(\phi,z_e)=\frac{P(z_e)}{\left.\frac{\partial Q}{\partial z}\right|_{z=z_e}}. \tag{A10}$$

The sum treated in the subsection A.1 can be written in the form

$$\sum_{n=-\infty}^{+\infty}\frac{1}{(2n+1)^2\pi^2+(\lambda_k/T)^2}=\frac{1}{\beta^2}\sum_{n=-\infty}^{+\infty}\frac{1}{(\lambda_k-i\omega_n)(\lambda_k+i\omega_n)}\equiv\frac{1}{\beta^2}\sum_{n=-\infty}^{+\infty}\phi(i\omega_n),\quad\text{(A11)}$$

where this complex function $\phi$ and its two poles satisfy all the criteria listed upstream. Consequently, the equations (A9) and (A10) are usable, leading to

$$\sum_{n=-\infty}^{+\infty}\frac{1}{(2n+1)^2\pi^2+(\lambda_k/T)^2}=\frac{T}{2\lambda_k}\left[1-2f(\lambda_k)\right].\quad\text{(A12)}$$

Thanks to the property $\tanh(\beta\lambda_k/2)=1-2f(\lambda_k)$, one finds again the equation (A6), with $\lambda_k\in\mathbb{C}\setminus\{i\omega_n\}$ if one uses the same continuous extension at $\lambda_k=0$ by taking the value $1/4$. Furthermore, (A9) is also usable to perform the Matsubara summations in (28), (31) and (36). However, the sums found with (32) do not satisfy the convergence criterion. It imposes some precautions during the calculations. It notably concerns the inclusion of factors $\exp(i\omega_n\eta)$, visible e.g. in [30, 144], to allow the convergence.

*A.3. Equation (23) in the real case*

As in [99], one checks that if $\lambda_k$ is real,

$$\ln\left(\frac{\omega_n^2+\lambda_k^2}{T^2}\right)=\int_1^{\lambda_k^2/T^2}\frac{dx}{\omega_n^2/T^2+x}+\ln\left(\frac{\omega_n^2}{T^2}+1\right).\quad\text{(A13)}$$

Then, if one writes $\omega_n=(2n+1)\pi T$, with $n\in\mathbb{Z}$, and if one sums over $n$, (A13) is written as

$$\sum_{n=-\infty}^{+\infty}\ln\left[(2n+1)^2\pi^2+\frac{\lambda_k^2}{T^2}\right]=\int_1^{\lambda_k^2/T^2}\sum_{n=-\infty}^{+\infty}\frac{dx}{(2n+1)^2\pi^2+x}+\sum_{n=-\infty}^{+\infty}\ln\left[(2n+1)^2\pi^2+1\right].\quad\text{(A14)}$$

The $\sum_{n=-\infty}^{+\infty}\ln\left[(2n+1)^2\pi^2+1\right]$ term found in this relation is divergent; I note it as $\kappa$ hereafter. However, as argued in [99], only the non-constant terms can be conserved in the calculations, in order to focus on the terms that describe the involved physics. Thanks to (A6) adapted to the real case, one rewrites the integral in (A14) as

$$\int_1^{\lambda_k^2/T^2}\sum_{n=-\infty}^{+\infty}\frac{dx}{(2n+1)^2\pi^2+x}=\int_1^{\lambda_k^2/T^2}dx\,\frac{1}{2\sqrt{x}}\tanh\left(\frac{\sqrt{x}}{2}\right).\quad\text{(A15)}$$

One can have $x\geq 0$. So, the domain of definition of the function $\tanh(\sqrt{x}/2)/2\sqrt{x}$ is extended to $x=0$, as done with (A6). Then, one uses (A15) in (A14), and one has

$$\sum_{n=-\infty}^{+\infty}\ln\left[(2n+1)^2\pi^2+\frac{\lambda_k^2}{T^2}\right]=2\ln\left[\cosh\left(\frac{|\lambda_k|}{2T}\right)\right]-2\ln\left[\cosh(1/2)\right]+\kappa.\quad\text{(A16)}$$

After some manipulations, one obtains the well-known relation

$$T\sum_{n=-\infty}^{+\infty}\ln\left(\frac{\omega_n^2+\lambda_k^2}{T^2}\right)=\lambda_k+2T\ln\left[1+\exp(-\lambda_k/T)\right]+Cst,\quad\text{(A17)}$$

with $\lambda_k\in\mathbb{R}$ and $Cst=-2T\ln(2)-2T\ln\left[\cosh(1/2)\right]+T\kappa$ gathers all the constant terms. Since cosh is an even function, I used the property $\cosh(|x|)=\cosh(x)$ to the equation (A16). So, the equation (A17) does not exhibit absolute values as in [22, 23]. However, this equation and the ones of these references are strictly equivalent, even if (A17) is more tractable in the calculations.

## A.4. Equation (23) in the complex case

When $\lambda_k \in \mathbb{C}$, the method of the subsection A.3 is still relevant. However, specificities of the complex analysis [145] must be taken into account. Firstly, the complex logarithmic function $\mathrm{Ln}$, defined as one antiderivative of the complex function $1/z$ in the simply connected domain $\mathbb{C} \setminus \mathbb{R}^-$, is written as

$$\mathrm{Ln}(z) = \ln(|z|) + i \arg(z). \tag{A18}$$

For convenience, this complex logarithm is noted in this paper like the real one, i.e. $\ln$. With the definition (A18), its branch cut corresponds to $\mathbb{R}^-$. The complex function $\left(\omega_n^2/T^2 + z\right)^{-1}$ is analytic in the simply connected domain $\mathbb{C} \setminus \left]-\infty; -(\omega_n/T)^2\right[$. Its integration is path independent if the bounds of the integral are taken from this domain, according to Cauchy's theorem. So, (A13) is extendable to the complex case if

$$\lambda_k/T \notin \left\{ z \in i\mathbb{R} \text{ and } |\mathrm{Im}(z)| \geq |2n+1|\pi \right\}. \tag{A19}$$

The found equation is then summed over $n$, i.e.

$$\sum_{n=-\infty}^{+\infty} \ln\left(\frac{\omega_n^2 + \lambda_k^2}{T^2}\right) = \kappa + \sum_{n=-\infty}^{+\infty} \int_1^{\frac{\lambda_k^2}{T^2}} \frac{dz}{(2n+1)^2 \pi^2 + z}. \tag{A20}$$

Then, as in the real case, the positions of the summation and integration symbols are interchanged. One uses the equation (A6) to write

$$\int_1^{\frac{\lambda_k^2}{T^2}} \sum_{n=-\infty}^{+\infty} \frac{dz}{(2n+1)^2 \pi^2 + z} = \int_1^{\frac{\lambda_k^2}{T^2}} g(z) \, dz, \tag{A21}$$

where the complex function $g(z) = \frac{1}{2\sqrt{z}} \tanh(\sqrt{z}/2)$ is defined when $\sqrt{z} \neq i(2n+1)\pi$ and one poses $g(0) = 1/4$. With the Cauchy-Riemann condition $\partial g/\partial \bar{z} = 0$, it can be proved that $g$ is holomorphic and so analytic in its domain of definition. Consequently, the integration of this function is path independent in the simply connected domain $\mathbb{C} \setminus \left]-\infty; -\pi^2\right[$. It leads to

$$\int_1^{\frac{\lambda_k^2}{T^2}} \frac{1}{2\sqrt{z}} \tanh(\sqrt{z}/2) \, dz = 2\ln\left[\cosh\left(\sqrt{\lambda_k^2}/2T\right)\right] - 2\ln\left[\cosh(1/2)\right], \tag{A22}$$

in which $\cosh\left(\sqrt{\lambda_k^2}/2T\right)$ can be replaced by $\cosh(\lambda_k/2T)$, thanks to the property

$$\begin{cases} \sqrt{z^2} = |\mathrm{Re}(z)| + i\,\mathrm{sign}[\mathrm{Re}(z)]\,\mathrm{Im}(z) & \text{if } \mathrm{Re}(z) \neq 0 \\ \sqrt{z^2} = i\,|\mathrm{Im}(z)| & \text{otherwise} \end{cases}, \tag{A23}$$

where the sign function gives $+1$ in the case of a positive number, $-1$ in the case of a negative one. The equation (A22) is valid if the constraint (A19) is verified $\forall n$, which corresponds to branch cuts of $\ln\left[\cosh(\lambda_k/2T)\right]$ when $\lambda_k/T \in \left\{ z \in i\mathbb{R} \text{ and } |\mathrm{Im}(z)| \geq \pi \right\}$ and simple poles when $\lambda_k/T = i(2n+1)\pi$. Because of the definition of the complex logarithm, equation (A18), branch cuts are also observable when $\mathrm{Im}(\lambda_k/T) = (2n+1)2\pi$. Then, $\ln\left[\cosh(\lambda_k/2T)\right]$ is rewritten as $\ln\left[\exp(\lambda_k/2T) \frac{1 + \exp(-\lambda_k/T)}{2}\right]$. One uses $\ln(z_1 z_2) = \ln(z_1) + \ln(z_2)$, but this formula is only valid if $\arg(z_1) + \arg(z_2) \in ]-\pi; \pi[$, otherwise $\pm 2\pi i$ terms have to be added. This condition is realized when $|\text{Im}(\lambda_k/T)| < \pi$. Consequently, (A17) is extendable to the complex case, i.e.

$$T \sum_{n=-\infty}^{+\infty} \ln\left(\frac{\omega_n^2 + \lambda_k^2}{T^2}\right) = \lambda_k + 2T \ln\left[1 + \exp(-\lambda_k/T)\right] + Cst \text{ with } \text{Im}(\lambda_k/T) \in ]-\pi; \pi[. \quad (A24)$$

The mentioned condition allows using this relation in a domain for which the function $\lambda_k + 2T \ln\left[1 + \exp(-\lambda_k/T)\right]$ does not present poles or branch cuts. Furthermore, since this function is analytic in this domain, it can be manipulated like (A17), as expected.

*A.5. Alternative method*

It is also interesting to mention the method proposed in [16, 146] in the real case. It consists in differentiating $T \sum_{n=-\infty}^{+\infty} \ln\left[(2n+1)^2 \pi^2 + (\lambda_k/T)^2\right]$ with respect to $\lambda_k$. To be able to write $\frac{\partial}{\partial x} \sum_n f_n(x) = \sum_n \frac{\partial f_n(x)}{\partial x}$, the functions $f_n(x)$ must be differentiable $\forall n$, $\sum_n \frac{\partial f_n(x)}{\partial x}$ must converge uniformly and it must exist at least one number $x_0$ such that the series $\sum_n f_n(x_0)$ converges. Since $\sum_n \ln\left[(2n+1)^2 \pi^2 + \lambda_k^2/T^2\right]$ diverges $\forall \lambda_k$, I propose to include the divergent $\kappa$ term found in the equation (A14), which permits to write

$$\frac{\partial}{\partial \lambda_k}\left\{T \sum_{n=-\infty}^{+\infty} \ln\left[(2n+1)^2 \pi^2 + \frac{\lambda_k^2}{T^2}\right] - \ln\left[(2n+1)^2 \pi^2 + 1\right]\right\} = \frac{2\lambda_k}{T} \sum_{n=-\infty}^{+\infty} \frac{1}{(2n+1)^2 \pi^2 + \lambda_k^2/T^2}. \quad (A25)$$

Indeed, the convergence of the sum in the left hand side of (A25) is verified when $\lambda_k^2/T^2 = 1$. The other criteria mentioned upstream are also satisfied. The sum on the right hand side is then performed thanks to the relation (A12). One obtains

$$\frac{\partial}{\partial \lambda_k}\left\{T \sum_{n=-\infty}^{+\infty} \ln\left[(2n+1)^2 \pi^2 + \frac{\lambda_k^2}{T^2}\right] - \ln\left[(2n+1)^2 \pi^2 + 1\right]\right\} = 1 - 2f(\lambda_k). \quad (A26)$$

Finally, (A17) is found again if one antidifferentiates (A26) with respect to $\lambda_k$. The $\kappa$ term is then incorporated in the constant due to this antidifferentiation. The method can be extended to the complex case.

**Appendix B. Fermi-Dirac distributions in the PNJL and $\mu$PNJL models**

In the NJL model, the Fermi-Dirac distributions are identical for the $N_c$ colors of the quarks (resp. antiquarks) that have the same flavor. For example in the expression of the condensate $\langle\langle \psi_f^+ \psi_f \rangle\rangle$, it explains the factor $N_c$ in (49). In the ($\mu$)PNJL models, because of the replacement $\tilde{\mu}_f \to \tilde{\mu}_f - iA_4$, this property is no longer verified. In practice, one considers instead the "modified" Fermi-Dirac distributions $f_\Phi^\pm$. As in [69], I propose to use the relation, inspired from [63],

$$f_\Phi^\pm(E_f \mp \mu_f) = \pm \frac{1}{\beta} \frac{\partial}{\partial \mu_f}\left[\frac{1}{N_c} \text{Tr}_c \ln\left(Z_\Phi^\pm\right)\right], \quad (B1)$$

where $Z_\Phi^\pm$ correspond to the partition functions of the quarks (plus sign) and antiquarks (minus sign). They are extracted from the ($\mu$)PNJL expression of $\Omega_M$ in (52), i.e.

$$Z_\Phi^+(E_f) = 1 + L^\dagger e^{-\beta(E_f - \mu_f)} \text{ and } Z_\Phi^-(E_f) = 1 + L e^{-\beta(E_f + \mu_f)}. \tag{B2}$$

The use of the equation (B1) with these expressions of $Z_\Phi^\pm$ directly gives the writing of the ($\mu$)PNJL Fermi-Dirac statistics equations (53) and (54), with $L = \exp(i\beta A_4)$ and $L^\dagger = \exp(-i\beta A_4)$. If the trace over the colors is detailed, $f_\Phi^\pm$ are then written as

$$f_\Phi^\pm(E_f \mp \mu_f) = \frac{1}{N_c} \sum_{j=1}^{N_c} \frac{1}{1 + \exp\{\beta[E_f \mp (\mu_f - iA_{4(jj)})]\}} = \frac{1}{N_c} \sum_{j=1}^{N_c} f(E_f \mp (\mu_f - iA_{4(jj)})), \tag{B3}$$

where $f(z) = (e^{\beta z} + 1)^{-1}$ is the "standard" Fermi-Dirac distribution extended to the complex plane. This form [63] underlines that the "modified" ($\mu$)PNJL Fermi-Dirac distributions $f_\Phi^\pm$ correspond to an averaging of the $N_c$ distributions of the quarks/antiquarks that have the same flavor, i.e. an averaging over the colors. Another form of $f_\Phi^\pm$ is obtained following the description performed, e.g., in [72]. Firstly, one writes

$$\text{Tr}_c \ln(Z_\Phi^\pm) = \ln\left\{\prod_{j=1}^{N_c} 1 + \exp[-\beta(E_f \mp \mu_f)]\exp[\mp iA_{4(jj)}]\right\}. \tag{B4}$$

Then, one develops the product present in the logarithm. For $Z_\Phi^+$, it leads to

$$\begin{aligned}
\text{Tr}_c \ln(Z_\Phi^+) = \\
\ln\Big\{ & 1 + \exp[-\beta(E_f - \mu_f)]\big[\exp(-i\beta A_{4(11)}) + \exp(-i\beta A_{4(22)}) + \exp(-i\beta A_{4(33)})\big] \\
& + \exp[-2\beta(E_f - \mu_f)]\big[\exp(-i\beta A_{4(22)} - i\beta A_{4(33)}) \\
& + \exp(-i\beta A_{4(11)} - i\beta A_{4(33)}) + \exp(-i\beta A_{4(11)} - i\beta A_{4(22)})\big] \\
& + \exp[-3\beta(E_f - \mu_f)]\big[\exp(-i\beta A_{4(11)} - i\beta A_{4(22)} - i\beta A_{4(33)})\big]\Big\}
\end{aligned} \tag{B5}$$

In the framework of an $SU(3)_c$ description, e.g., with (9), it can be found that

$$\exp(-i\beta A_{4(ii)} - i\beta A_{4(jj)}) = \exp(i\beta A_{4(kk)}), \text{ with } i \neq j \neq k \tag{B6}$$

and

$$\exp(-i\beta A_{4(11)} - i\beta A_{4(22)} - i\beta A_{4(33)}) = 1. \tag{B7}$$

Furthermore, the Polyakov field $\Phi$ and its conjugate $\bar{\Phi}$, defined equation (6), can be written in the form $\Phi = \frac{1}{N_c}\sum_{j=1}^{N_c} \exp(i\beta A_{4(jj)})$ and $\bar{\Phi} = \frac{1}{N_c}\sum_{j=1}^{N_c} \exp(-i\beta A_{4(jj)})$ in the mean field approximation. Using these relations and the equations (B6) and (B7) in (B5), it is possible to express $\text{Tr}_c \ln(Z_\Phi^\pm)$ exclusively as a function of $\beta$, $E_f$, $\mu_f$, $\Phi$ and $\bar{\Phi}$, i.e.

$$\text{Tr}_c \ln(Z_\Phi^+) = \ln\Big(1 + 3\{\bar{\Phi} + \Phi\exp[-\beta(E_f - \mu_f)]\}\exp[-\beta(E_f - \mu_f)] + \exp[-3\beta(E_f - \mu_f)]\Big). \tag{B8}$$

Finally, with the equation (B1), one finds again the expression of $f_\Phi^+$ mentioned in the literature, as in [58, 93], and used in the ($\mu$)PNJL numerical calculations,

$$f_\Phi^+(E_f - \mu_f) = \frac{\{\bar{\Phi} + 2\Phi\exp[-\beta(E_f - \mu_f)]\}\exp[-\beta(E_f - \mu_f)] + \exp[-3\beta(E_f - \mu_f)]}{1 + 3\{\bar{\Phi} + \Phi\exp[-\beta(E_f - \mu_f)]\}\exp[-\beta(E_f - \mu_f)] + \exp[-3\beta(E_f - \mu_f)]}. \tag{B9}$$

The expression of $f_\Phi^-$ is found in the same way, or with the property

$$f_\Phi^-\left(E_f + \mu_f\right) = f_\Phi^+\left(E_f - \mu_f\right)\Big|_{\substack{\Phi \leftrightarrow \bar\Phi \\ \mu_f \leftrightarrow -\mu_f}} . \tag{B10}$$

**Appendix C. Establishment of the equation** (72)

*C.1. General solutions of a quartic polynomial*

The roots $\lambda_k$ of a quartic polynomial $x^4 + Ax^3 + Bx^2 + Cx + D$ are written as

$$\lambda_{1,2} = \frac{-A}{4} - \frac{\sqrt{Z}}{2} \pm \frac{1}{2}\sqrt{W - \frac{T}{4\sqrt{Z}}} \text{ and } \lambda_{3,4} = \frac{-A}{4} + \frac{\sqrt{Z}}{2} \pm \frac{1}{2}\sqrt{W + \frac{T}{4\sqrt{Z}}}, \tag{C1}$$

where

$$Z = \frac{A^2}{4} - \frac{2B}{3} + Y, \ W = \frac{A^2}{2} - \frac{4B}{3} - Y, \ T = -A^3 + 4AB - 8C,$$

$$Y = \frac{2^{1/3}\Omega}{3X} + \frac{X}{3 \times 2^{1/3}}, \ X = \left(\Xi + \sqrt{-4\Omega^3 + \Xi^2}\right)^{1/3}, \tag{C2}$$

and with

$$\Omega = B^2 - 3AC + 12D \text{ and } \Xi = 2B^3 - 9ABC + 27C^2 + 27A^2 D - 72BD. \tag{C3}$$

The roots satisfy the property

$$\sum_{k=1}^{4} \lambda_k = -A. \tag{C4}$$

*C.2. Calculation of the determinants*

The first determinant used in (70) can be written as

$$\det\left[\not{p} + \gamma_0 \mu_u - m_u + \delta_d\left(\not{p} - \gamma_0 \mu_d - m_d\right)\right] = \left[\frac{(i\omega_n)^4 + A(i\omega_n)^3 + B(i\omega_n)^2 + Ci\omega_n + D}{(i\omega_n - \mu_d)^2 - E_d^2}\right]^2, \tag{C5}$$

with

$$\begin{cases} A = 2(\mu_u - \mu_d) \\ B = -E_u^2 - E_d^2 - 2|\Delta_{ud}|^2 + \mu_u^2 + \mu_d^2 - 4\mu_u\mu_d \\ C = 2\left[\mu_d\left(E_u^2 - \mu_u^2 + |\Delta_{ud}|^2\right) - \mu_u\left(E_d^2 - \mu_d^2 + |\Delta_{ud}|^2\right)\right] \\ D = E_d^2 E_u^2 + \mu_u^2 \mu_d^2 + |\Delta_{ud}|^2\left[2\mu_u\mu_d - (m_u - m_d)^2 + |\Delta_{ud}|^2\right] + E_d^2\left(|\Delta_{ud}|^2 - \mu_u^2\right) + E_u^2\left(|\Delta_{ud}|^2 - \mu_d^2\right) \end{cases} \tag{C6}$$

The four roots of the numerator visible in (C5) are noted as $\lambda_k$. Their analytical expression is given in the first part of this appendix C. In the same way, it can be shown that the second determinant of (70) can be written as

$$\det\left[\not{p} + \gamma_0 \mu_d - m_d + \delta_u\left(\not{p} - \gamma_0 \mu_u - m_u\right)\right] = \left[\frac{(i\omega_n)^4 - A(i\omega_n)^3 + B(i\omega_n)^2 - Ci\omega_n + D}{(i\omega_n - \mu_u)^2 - E_u^2}\right]^2. \tag{C7}$$

The product of the two quartic polynomials present in the numerators of (C5) and (C7) forms a polynomial of degree 8. This one is an even function. It implies that its eight roots are $\pm\lambda_k$. Therefore, it authorizes the factorization $(i\omega_n)^2 - \lambda_k^2$ that is required to be able to use (23).